\documentclass[fleqn,usenatbib]{mnras}

\usepackage{newtxtext,newtxmath}

\usepackage[T1]{fontenc}
\usepackage{ae,aecompl}

\usepackage{graphicx}	
\DeclareGraphicsExtensions{.ps}
\usepackage[utf8]{inputenc}
\usepackage{multirow}
\usepackage{hyperref}
\usepackage{natbib}
\usepackage{array}
\usepackage{verbatim}
\usepackage{float}
\usepackage{epsfig}
\usepackage{subfigure}
\usepackage{color}

\graphicspath{ {Figures/} }

\newcommand{\mum}{$\mu$m}
\newcommand{\ArII}{\textrm{Ar~{\textsc{ii}}}}
\newcommand{\ArIII}{\textrm{Ar~{\textsc{iii}}}}
\newcommand{\HII}{\textrm{H~{\textsc{ii}}}}
\newcommand{\HI}{\textrm{H~{\textsc{i}}}}

\newcommand{\NeII}{\textrm{Ne~{\textsc{ii}}}}

\newcommand{\SIV}{\textrm{S~{\textsc{iv}}}}

\newcommand{\OI}{\textrm{O~{\textsc{i}}}}

\title[PAHs in Orion]{Characterizing the PAH Emission in the Orion Bar}

\author[C. Knight et al.]{
C. Knight$^{1}$, E. Peeters$^{1,2,3}$, A.~G.~G.~M.~Tielens$^{4,5}$, W.~D.~Vacca$^{6}$
\\
$^{1}$Department of Physics and Astronomy, University of Western Ontario, London, ON N6A 3K7, Canada; collinknight11@gmail.com \\
$^{2}$ Institute for Earth and Space Exploration, University of Western Ontario, London, ON, N6A 3K7, Canada\\
$^{3}$Carl Sagan Center, SETI Institute, 189 N. Bernardo Avenue, Suite 100, Mountain View, CA 94043, USA\\
$^{4}$Leiden Observatory, PO Box 9513, 2300 RA Leiden, The Netherlands\\
$^{5}$Department of Astronomy, University of Maryland, MD 20742, USA\\
$^{6}$SOFIA-USRA, NASA Ames Research Center, MS N232-12, Moffett Field, CA 94035-1000, USA\\
}

\date{Accepted XXX. Received YYY; in original form ZZZ}

\pubyear{2021}

\begin{document}
\label{firstpage}
\pagerange{\pageref{firstpage}--\pageref{lastpage}}
\maketitle

\begin{abstract}

We present 5--14~$\mu$m spectra at two different positions across the Orion Bar photodissociation region (PDR) obtained with the Infrared Spectrograph onboard the Spitzer Space Telescope and 3.3~$\mu$m PAH observations obtained with the Stratospheric Observatory for Infrared Astronomy (SOFIA). We aim to characterize emission from Polycyclic Aromatic Hydrocarbon (PAH), dust, atomic and molecular hydrogen, argon, sulfur, and neon as a function of distance from the primary illuminating source. We find that all the major PAH bands peak between the ionization front and the PDR front, as traced by H$_{2}$, while variations between these bands become more pronounced moving away from this peak into the face-on PDRs behind the PDR front and at the backside of the \HII\, region. While the relative PAH intensities are consistent with established PAH characteristics, we report unusual behaviours and attribute these to the PDR viewing angle and the strength of the FUV radiation field impinging on the PDRs. We determine the average PAH size which varies across the Orion Bar. We discuss subtle differences seen between the cationic PAH bands and highlight the photo-chemical evolution of carbonaceous species in this PDR environment. We find that PAHs are a good tracer of environmental properties such as the strength of the FUV radiation field and the PAH ionization parameter.
\end{abstract}




\section{Introduction}
\label{intro}

Mid-infrared (MIR) observations throughout the interstellar medium (ISM) of our Galaxy as well as external galaxies show strong emission features at 3.3, 6.2, 7.7, 8.6, 11.2, and 12.7~$\mu$m attributed to the infrared fluorescence of polycyclic aromatic hydrocarbons (PAHs). These molecules absorb far-ultraviolet (FUV) photons causing electronic excitation \citep[i.e.][]{all89}, which is rapidly converted into vibrational excitation that is radiated away as MIR emission as these PAH species cool. Since their discovery by \cite{gil73}, these bands have been observed in a wide variety of sources including \HII\, regions, young stellar objects (YSOs), post-AGB stars, planetary nebulae (PNe), reflection nebulae (RNe), galaxies as well as the diffuse ISM \citep[e.g.][]{hon01,ver01,pee02,ber07,boe12,sha16,sto16}. 

PAHs and related species account for up to 15\% of the cosmic carbon inventory \citep{all89} and play a key role in the physical and chemical processes in these environments. For instance, PAHs have been shown to be useful tracers of star formation rates \citep[e.g.][]{pee04,cal07,marag18}, they are the dominant heating source in the neutral ISM via photoelectric ejection \citep{bak94}, and are essential to the ionization balance through photoionization and recombination processes \citep{lep88}. Thus, studying these PAH emission features can yield a wealth of knowledge towards our understanding of the important role these molecules have in the physical and chemical processes that occur within the ISM.

The PAH emission features show variations in relative intensities, peak position, and band shape in different Galactic and extragalactic environments as well as within extended sources \citep[e.g.][]{hon01,pee02,gal08}. The main driver for variations in PAH band intensities is the charge state of the underlying population. The 6.2, 7.7, and 8.6~$\mu$m bands are strong in ionic PAHs, whereas the 3.3 and 11.2~$\mu$m bands are more prevalent within neutral PAHs \citep[e.g.][]{all89,hud94,bak94, all99}. 
Generally, PAH bands attributed to the same ionization state tend to be well correlated. For instance, there is a tight relationship between the 6.2 and 7.7~$\mu$m bands found in a wide variety of MIR bright sources \citep[e.g][]{gal08,boe14a,sto17,pee17,marag18}. However, the above relationship does not hold for all astronomical sources. Indeed, it has been found to break down on small spatial scales within the giant star-forming region N66 in the Large Magellanic Cloud \citep{whe13} and towards the center of ultra-compact \HII\, regions within the Galactic massive star-forming region W49A \citep{sto14}. An investigation of PAH emission features in a much closer \HII\, region with similar radiation field properties could provide an explanation for this anomaly by availing of the much higher spatial resolution as set by the observing instrument by virtue of proximity.  

To this end, we consider the prototypical nearby star-forming region, the Orion Nebula (M42), located at a mere distance of 414~$\pm$~7~pc \citep{men07}. Within this nebula lies the Orion Bar, which has long been known to be a source of strong MIR emission \citep[e.g.][]{ait79,sell81, tie85b,bre89a,geb89,sell90,tie93,gia94,ces00,rub11,boe12,har12,sal16,pab19}. Due to the edge--on, stratified nature of this photo-dissociation region (PDR), it is considered to be the benchmark for modelling these environments \citep[e.g.][]{tie85b,tie93}. Furthermore, the edge--on morphology is a key facilitator in our understanding of PDRs in that it allows us to clearly delineate the boundaries between the ionized cavity surrounding a stellar source of strong UV radiation, the neutral PDR where freely flying PAH species are abundant, and the cold molecular cloud that tends to encompass these PDRs \citep[e.g.][]{tau94,tie93,wal00,wer13,cua15,goi15}.

In this study, we examine PAH emission features along with prominent MIR atomic and molecular emission lines towards the Orion Bar using spectroscopic observations from the Spitzer Space Telescope, along with supplementary data previously obtained from FLITECAM on board the Stratospheric Observatory for Infrared Astronomy (SOFIA). In Section~\ref{ob}, we give an overview of the general morphology and physical properties of the Orion Bar. In Section~\ref{data}, we present our spectroscopic observations as well as the data reduction methodology. We describe how the continuum and feature fluxes are measured in Section~\ref{analysis}. In Section~\ref{results}, we describe our primary results in the form of cross cuts of individual emission components and corresponding emission ratios with respect to distance from the primary illuminating source and correlations between these features. We discuss these results with respect to the environmental conditions and the properties of the PAH population in Section~\ref{discussion}. Finally, a summary of this work is provided in Section~\ref{conclusion}.

\begin{figure}
\begin{center}
\includegraphics[width=8.cm]{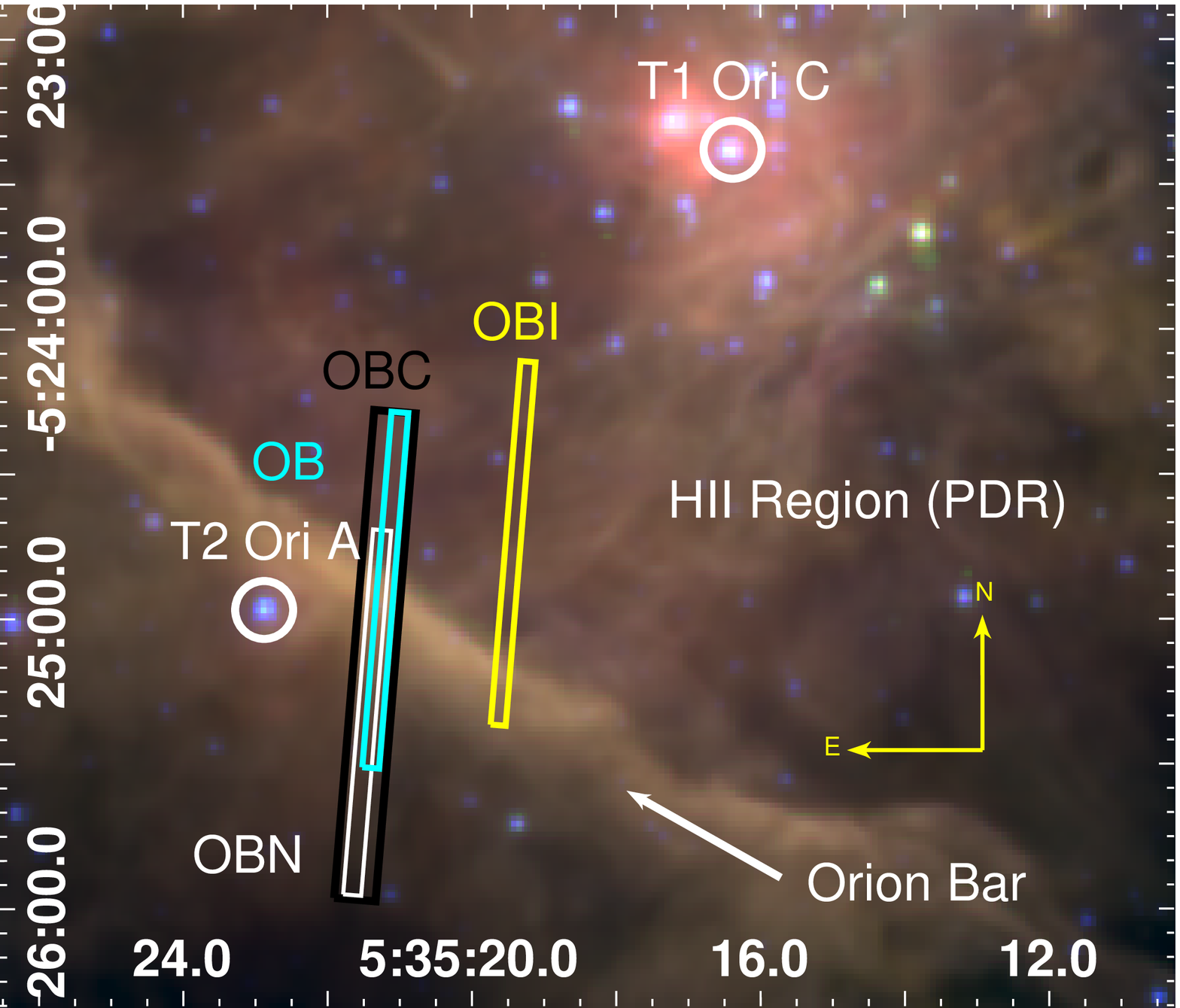}
\end{center}
\caption{Mosaic of the Orion Nebula with IRAC~3.6~$\mu$m in blue, IRAC~5.8~$\mu$m in green, and IRAC~8.0~$\mu$m in red \citep{meg12}. IRS~SL apertures are referred to as `Orion Bar' (OB; cyan), Orion Bar Neutral' (OBN; white), and `Orion Bar ionized' (OBI; yellow). We combine OB and OBN into a single aperture `Orion Bar Combined' (OBC; black) as detailed in Section~\ref{reduc_IRS}. The position of TI Ori C ($\theta^{1}$~Ori~C) and T2 Ori A ($\theta^{2}$~Ori~A) are indicated by white circles. We note that each image is shown in a square root scaling.}
\label{Orion_irac}
\end{figure}

\begin{figure}
\begin{center}
\includegraphics[width=8.cm]{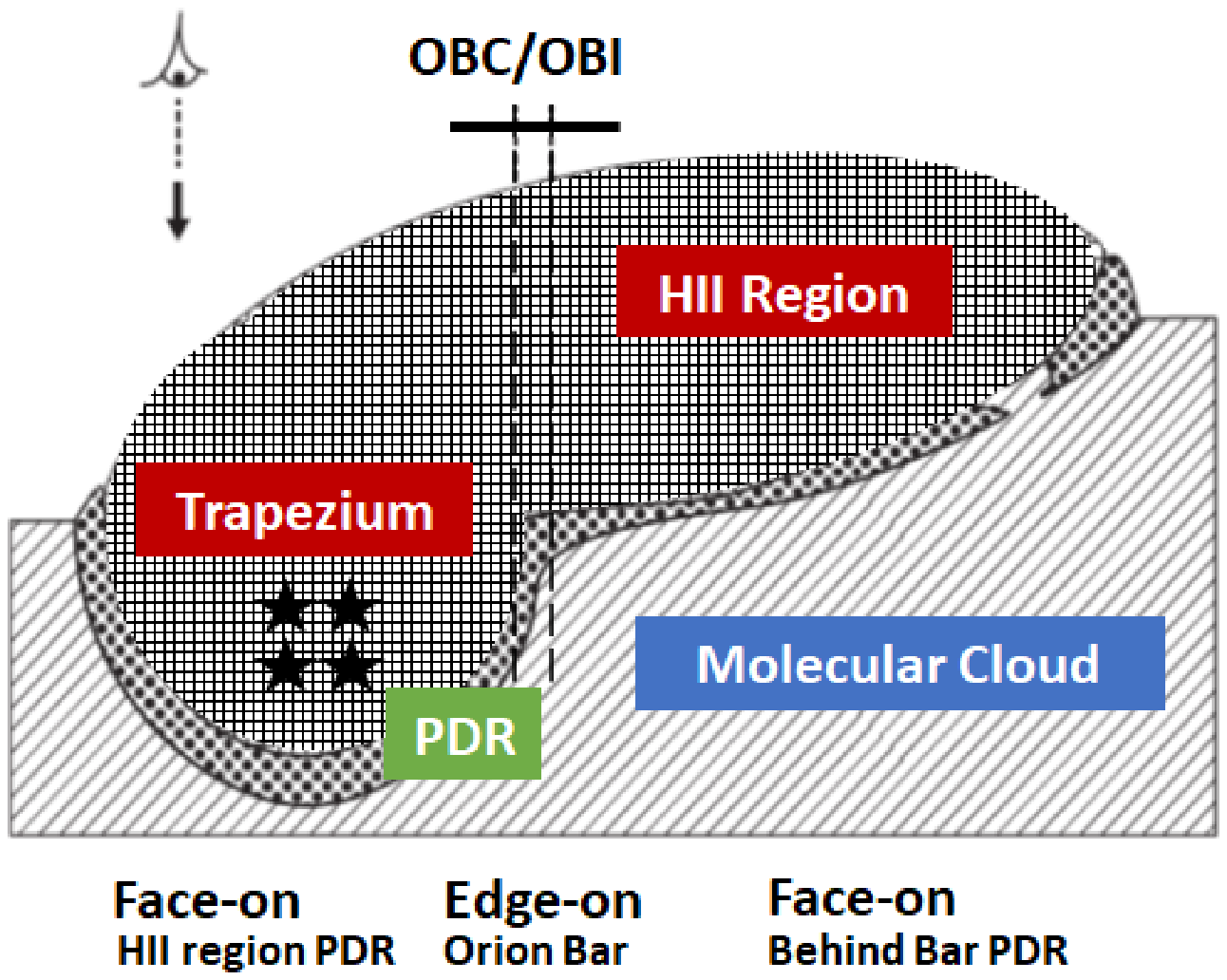}
\end{center}
\caption{Schematic representation of the \HII\, region surrounding the Trapezium stars and the PDR bounding the Orion molecular cloud (OMC-1, not to scale). The line-of-sight view is perpendicular to the IRS~SL apertures which are oriented across the Orion Bar and are indicated here as OBC/OBI (see Section~\ref{obs} for observation details). The PDR viewing angles (face--on or edge--on) and the nomenclature used in this paper for the 3 different regions are indicated below the representation.  Adapted from \citet{boe12} with permission from the authors. \textcopyright AAS. Reproduced with permission.}
\label{Orion_schematic}
\end{figure}

\section{Orion Bar}
\label{ob}
In the Orion Nebula, the primary illuminating source of the PDR is the brightest member of the Trapezium cluster, $\theta^{1}$ Ori C, an O6V type star with T$_{\textrm{eff}}$~=~38,950~K \citep[e.g.][]{ode17}. In Figure~\ref{Orion_irac}, we show a zoomed-in mosaic of the Orion Nebula using MIR imaging. This image demonstrates where the MIR bright gas and dust forms the PDR boundary between the large ionized cavity centered on the Trapezium cluster and the surrounding molecular cloud. The Orion Bar is part of this PDR boundary and is an edge--on, compressed shell \citep{sal16}. The outer boundary of the Orion Nebula is referred to as the Orion Veil, a large expanding shell of neutral gas driven by stellar winds expanding radially from the Trapezium Cluster \citep{pab19}. The stratified edge--on morphology of the Orion Bar and its proximity makes it an ideal probe of a PDR environment as we can investigate the photo--processing of the gas and dust with distance to the illuminating source \citep[e.g.][]{ces00,goi15, Knight:21}. In contrast, in a face--on PDR morphology, the entire processing history is mixed along the line of sight. This PDR morphology makes it significantly more difficult to infer how the gas and dust chemistry is driven by the stellar radiation field. Figure~\ref{Orion_schematic} shows a schematic representation of the Orion Nebula. Face--on PDR emission is seen towards the \HII\, region surrounding the Trapezium cluster (left side in Figure~\ref{Orion_schematic}), which originates from the PDR on the surface of the Orion molecular cloud (OMC--1), as well as behind the edge--on PDR front outwards toward the Veil (right side in Figure~\ref{Orion_schematic}). Henceforth, we refer to the face--on PDR towards the Orion \HII\, region surrounding the Trapezium cluster as the \HII\, region (PDR) and the face--on PDR behind the Orion Bar outwards toward the Veil as the face--on PDR behind the PDR front to distinguish it from the edge--on PDR, the Orion Bar.

 For the Orion Bar, \citet[][]{mar98} reported gas densities of 3--6~$\times$~10$^{4}$~cm$^{-3}$ and a FUV radiation field strength, $G_0$\footnote{In units of the Habing field \citep[1.3~$\times$~10$^{-4}$ ~erg~cm$^{-2}$~s$^{-1}$~sr$^{-1}$, ][]{hab68}.}, of 2.6~$\times$~10$^{4}$~times that of the average interstellar value at the ionization front. As we consider observations that are, in part, positioned behind the PDR front, it is worth noting that it has been suggested that $\theta^{2}$ Ori A, an O9.6V type star with an effective temperature of 34,600~K, is the primary source of UV radiation on the far side of the Bar \citep{ode17}.

\section{Observations and Data Reduction}
\label{data}

\subsection{Observations}
\label{obs}

\subsubsection{Spitzer}
\label{spit}

Spectroscopic observations were obtained with the short-low (SL) staring mode of the Infrared Spectrograph \citep[IRS,][]{hou04} on board the Spitzer Space Telescope \citep{wer04a}. This data set consists of three pointings with slits that transverse the Orion Bar at different locations (PID: 45, PI: Thomas Roellig, Figure~\ref{Orion_irac}). We assign the following nomenclature for these three pointings based on how much (part of) the aperture is in front of the Orion Bar towards the illuminating source,  $\theta^{1}$ Ori C. From closest to farthest from $\theta^{1}$ Ori C, these pointings are referred to as: `Orion Bar ionized' (OBI), `Orion Bar' (OB), and `Orion Bar neutral' (OBN). A summary of our observations is given in Table~\ref{table:1}. 

The SL mode has an effective wavelength range of 5.2--14.5~$\mu$m and a spectral resolution of 60 to 128 over three orders of diffraction: SL1, SL2, and SL3. The pixel size of the SL mode is 1.8$^{\prime\prime}$, with a slit width of 3.6$^{\prime\prime}$ and a slit length of 57$^{\prime\prime}$.

\begin{table}
\caption{\label{table:1}{} Log of Observations.}
\begin{center}
\begin{tabular}{ c  c  c  c }
\hline\hline
 &  Orion Bar & Orion Bar Neutral & Orion Bar Ionized\\ 

\hline\\[-5pt]

\multicolumn{4}{c}{}\\
map $\alpha^{1}$ & 5:35:27.5 & 5:35:27.7 & 5:35:25.7 \\
map $\delta^{1}$ & -5:30:48 & -5:31:14 & -5:30:39\\
AORs$^{2}$ & 4117760 & 4118016 & 4118272 \\[2pt]
\hline\\[-5pt]

\hline\\[-10pt]

\end{tabular}

\end{center}
$^{1}$ $\alpha$, $\delta$ (J2000) are the central coordinates of each map. $\alpha$ has units of hours, minutes, and seconds and $\delta$ has units of degrees, arc minutes, and arc seconds;$^{2}$ AOR is Astronomical Observation Request Identifier.
\end{table}

\subsubsection{SOFIA}
\label{sofia}

We include SOFIA-FLITECAM observations of the Orion Bar in the `PAH' filter \citep[PID: 04 0058, PI: A. Tielens;][]{Knight:21}. This filter has an effective wavelength of 3.302 $\mu$m and a bandwidth of 0.115 $\mu$m. The FLITECAM instrument has a 1024 pixel $\times$ 1024 pixel InSb detector that covers a
8$^\prime$~$\times$~8$^\prime$ area on the sky with 0.475$^{\prime\prime}$ ~$\times$~0.475$^{\prime\prime}$~pixels.

\subsection{Data Reduction}

\subsubsection{Spitzer}
\label{reduc_IRS}

The IRS-SL raw data were processed by the {\it Spitzer} Science Center with the S18.18 pipeline version. The resulting bcd products are further processed with {\it cubism} \citep{smi07b}. Specifically, we set {\it cubism}'s {\it wavsamp} to 0.04--0.96 and applied {\it cubism}'s automatic bad pixel generation with $\sigma_{TRIM}$~=~7 and Minbad-fraction~=~0.50 and 0.75 for global and record bad pixels respectively. Remaining bad pixels were subsequently removed manually. 

Spectra are extracted in an aperture of 2~$\times$~2~pixels moving along the slit in one-pixel steps. As a consequence, adjacent pixels are not independent. We found small mismatches in absolute flux levels between the SL1 and SL2 of  2--16\% and $<$~5\% between SL1 and SL3. To remedy this, the SL3 data were scaled to the SL1 data followed by a scaling of the SL2 data to the combined SL1 and scaled SL3 data. Subsequently, the SL1 and SL2 orders were combined into a single spectral cube for each pointing.

Due to the considerable spatial overlap of the OB and OBN apertures (see Figure~\ref{Orion_irac}), we combine both slits into one extended aperture. We take pixels corresponding to the OB slit where the pointings overlap as it has a higher SNR in overlapping pixels. We refer to this combined aperture as `Orion Bar Combined' (OBC) for the remainder of the text. 

\subsubsection{SOFIA}
\label{reduc_SOFIA}

We refer the reader to \citet{Knight:21} for details on the data reduction of the SOFIA-FLITECAM observations. We regrid the FLITECAM 3.3~$\mu$m image to each of the three IRS~SL apertures and applied a 2~$\times$~2 binning of the FLITECAM data to be consistent with our analysis of the IRS~SL data. We convert the 3.3~$\mu$m observations from units of integrated (over the filter) surface brightness (W~m$^{-2}$~$\mu$m$^{-1}$~sr$^{-1}$) to average surface brightness (in units of W~m$^{-2}$~sr$^{-1}$) following the method employed in \citet{Knight:21}. Specifically, we multiply by the bandwidth of the 3.3~$\mu$m filter of $\sim$~0.1~$\mu$m, which assumes emission within the filter can be approximated by a nominal flat spectrum. 

Five ISO-SWS spectra are available across the Orion Bar \citep[see Table~\ref{table:sofia} and Figure 3 from][]{Knight:21}. The 3.3~$\mu$m PAH emission accounts for $\sim$~72\% of the flux in the FLITECAM 3.3~$\mu$m filter \citep{Knight:21}. The three PDRs (\HII\ region (PDR), edge--on PDR, and the face--on PDR behind the PDR front) have only slightly different PAH contributions (range of 6.9\%) with the edge--on PDR exhibiting the highest values (Table~\ref{table:sofia}). No correction factor is applied in this paper to account for this as such small variations do not influence our conclusions.

\begin{table}
\caption{\label{table:sofia}{} PAH contribution to SOFIA 3.3 $\mu$m observations.}
    \begin{center}
    \begin{tabular}{c c c c c}
     Name$^1$ & TDT$^2$ & distance$^3$ & PDR & PAH \\
     & & ($^{\prime\prime}$) & & fraction \\
     \\[-7pt]
     \hline
     \\[-7pt]
     D8 & 69501409 & 81.4 & \HII\, region (PDR) & 69.5\\
     Br$\gamma$ & 69502108 & 106.0 & \HII\, region (PDR) & 69.8\\
     D5 & 83101507 & 118.3 & Orion Bar & 73.6\\
     H2S1  &  69501806 & 130.5 & Orion Bar & 76.4\\
     D2    &  69502005 & 155.8 & Behind Bar & 71.2\\
    \end{tabular}
    \end{center}
    $^1$ Name given to the observation in the ISO archive; $^2$ TDT numbers uniquely identifying the ISO observation; $^3$ Distance between the illuminating source and the center of the 14$^{\prime\prime}$~x~20$^{\prime\prime}$ aperture. 
\end{table}

\subsection{Spectra}
\label{spectra}

\begin{figure*}
\begin{center}
\resizebox{\hsize}{!}{%
\includegraphics[clip,trim =0cm 0cm 0.25cm 0.5cm,width=8.5cm]{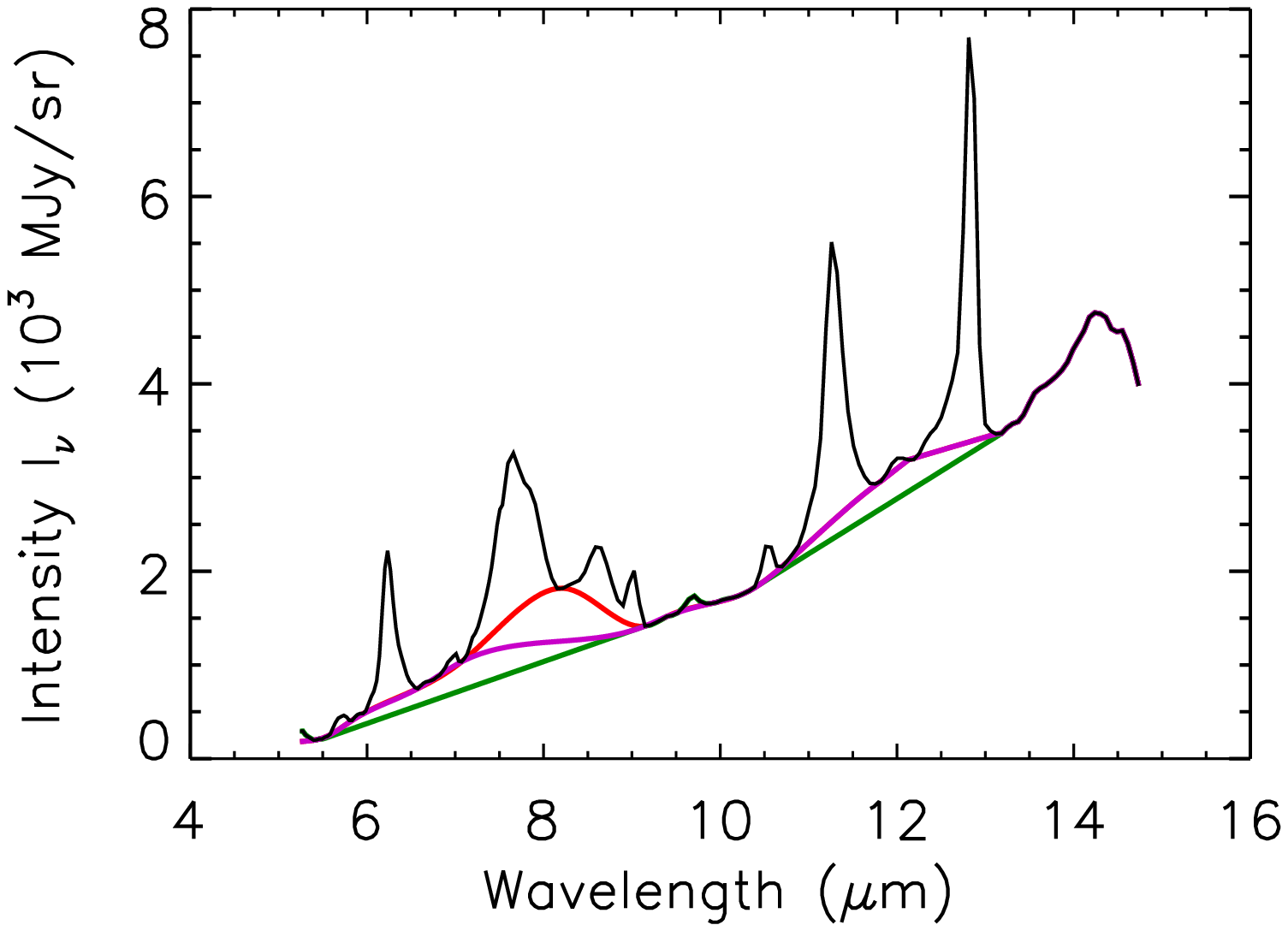}
\includegraphics[clip,trim =0cm 0cm 0.25cm 0.5cm,width=8.5cm]{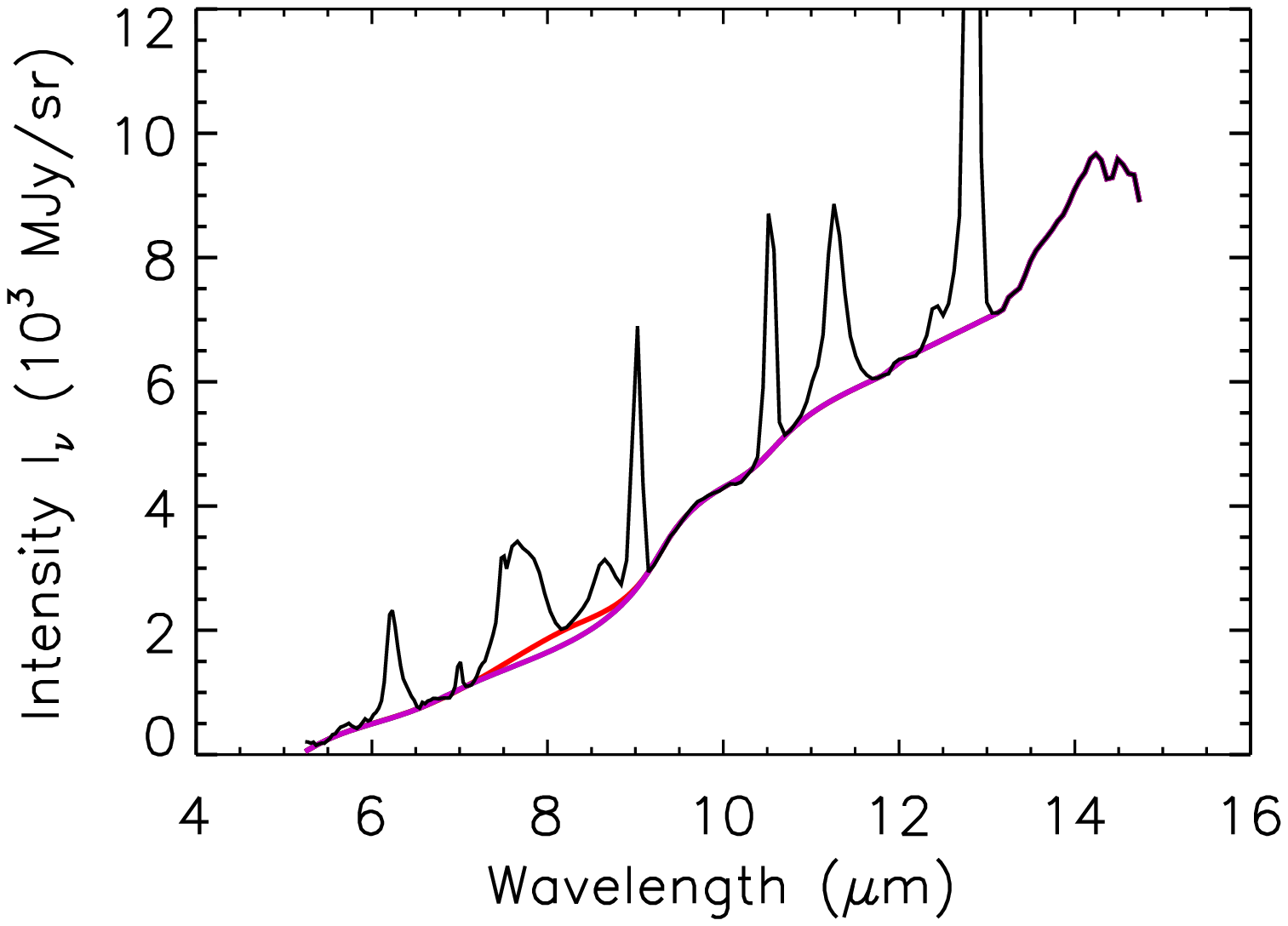}}
\end{center}
\caption{Typical IRS~SL spectra from the Orion Bar PDR (left) and the \HII\, region PDR (right) are shown. The red line traces the local spline continuum (LS), the magenta line traces the global spline continuum (GS), and the green line traces the underlying dust continuum (PL). See Section~\ref{cont} for more details on the continuum fitting procedure.}
\label{orion_spectra}
\end{figure*}

Typical spectra observed towards the Orion Bar are displayed in Figure~\ref{orion_spectra}. Comparison of these spectra demonstrates how the slope of continuum rises with increasing proximity to the illuminating source. Emission features discernible above the dust continuum include the major PAH bands at 6.2, 7.7, 8.6, 11.2, and 12.7~$\mu$m as well as weaker PAH bands at 5.7, 6.0, 11.0, 12.0, and 13.5~$\mu$m. These PAH bands are usually on top of broad emission plateaus at 5--10 and 10--15~$\mu$m (separate from the dust continuum, detailed in Section~\ref{cont}). Additionally, other atomic and molecular lines were detected such as the 6.98~$\mu$m [\ArII] line, the 7.46~$\mu$m Pfund~$\alpha$ line, the 8.99~$\mu$m [\ArIII] line, the 9.7~$\mu$m H$_{2}$ S(3) line, the 10.5~$\mu$m [\SIV] line, the 12.37~$\mu$m \HI\, recombination line, and the 12.8~$\mu$m [\NeII] line. In the ionized gas, the underlying dust continuum is much steeper and displays broad silicate emission at $\sim$~10~$\mu$m \citep{ces00}. 

\section{Data Analysis}
\label{analysis}

\subsection{Continuum Fitting}
\label{cont} 

In order to separate the PAH emission features from the underlying continuum, we make use of the spline decomposition method \citep[e.g][]{vker00,hon01,pee02,van04,boe12,sto14,sto16, sha15,sha16,pee17} to define a local spline (LS), a global spline (GS), and the underlying dust (PL) continuum (Figure~\ref{orion_spectra}). For the LS continuum, we use anchor points at 5.37, 5.52, 5.83, 6.54, 7.07, 8.25, 9.15, 9.40, 9.89, 10.33, 10.76, and 11.82~$\mu$m. In order to better fit the continuum underneath the 12.7~$\mu$m complex, we extend our spline fits as two straight lines from 11.82 to 12.1~$\mu$m and 12.1 to 13.2~$\mu$m respectively.  We do not fit the spectra beyond 13.2~$\mu$m due to the abrupt change in slope at the end of the spectra. The GS continuum fitting uses the same anchor points as the LS, except for the removal of the 8.25~$\mu$m anchor point. The difference between these two continua (LS and GS) is referred to as the 8~$\mu$m bump \citep[e.g.][]{pee17}. The dust continuum consist of a straight line between anchor points at 5.5~$\mu$m and 10.1~$\mu$m as well as a straight line between 10.4~$\mu$m and 13.2~$\mu$m. 

We find two very different shapes in the underlying dust continuum which is related to the position with respect to the illuminating source (Figure~\ref{orion_spectra}). Spectra obtained at positions closest to the star have a much steeper rise in continuum emission towards longer wavelengths (Figure~\ref{orion_spectra}, right panel). All other spectra located behind the Orion Bar ionization front (IF, see Section~\ref{proj} for details) have a much shallower rise in continuum emission. In the case of the spectra in front of the IF, we do not detect significant plateau emission and the steep slope of the underlying dust continuum does not allow to apply the dust continuum method described earlier. Hence, the GS fit is used to characterize this dust continuum (which is represented by the PL continuum for spectra located behind the IF). This is similar to what was found for \HII\, region spectra by \cite{sto17}. 

\subsection{Flux Measurement}
\label{fluxes}

The fluxes of the major PAH bands are determined through integrating the LS continuum subtracted spectra over the wavelength range of the feature. However, in the case of the 6.2, 11.2, and 12.7~$\mu$m features, another method is needed due to blending with weaker PAH features or atomic emission lines. Similar to \cite{pee17}, a two Gaussian fit of the 6.0 and (blue wing of the) 6.2~$\mu$m PAH bands was done with peak positions/FWHM of 6.02/0.12~$\mu$m and 6.232/0.156~$\mu$m respectively. We determine these values by allowing them to vary during the initial fitting procedure and subsequently take the average values over all the spectra. The 6.2~$\mu$m band flux is determined by subtracting the 6.0~$\mu$m Gaussian from the integrated flux of the LS subtracted spectra taken over the wavelength range spanning the 6.0~$\mu$m and 6.2~$\mu$m bands. We use a similar decomposition method to obtain the 11.2~$\mu$m emission feature flux. The 2 Gaussian fit of the 11.0 and (blue wing of the) 11.2~$\mu$m PAH bands has peak positions/FWHM of 11.003/0.15~$\mu$m and 11.262/0.227~$\mu$m respectively.

The 12.7~$\mu$m PAH band is significantly blended with the 12.8~$\mu$m [\NeII] line in all observations and, in some cases, with a weak 12.37~$\mu$m \HI\, recombination line. To differentiate between these emission features, the decomposition method used in \citet{sto14}, \citet{sha15}, and \citet{sto16} is employed. We use the NGC~2023 12.7~$\mu$m line profile in the Southern Ridge PDR front detailed in \cite{pee17} as a template for the 12.7~$\mu$m band. We simultaneously fit two Gaussian functions to the 12.37~$\mu$m \HI\, recombination line and the 12.8~$\mu$m [\NeII] line, along with the 12.7~$\mu$m template, which is scaled to align with the spectra in the 12.4 to 12.7~$\mu$m range. The 12.7~$\mu$m band flux is obtained by integrating the continuum subtracted spectra from 12.15 to 13.2~$\mu$m and subtracting the 12.37~$\mu$m \HI\, and the 12.8~$\mu$m [\NeII] fluxes determined from the Gaussian fits. We find an average peak position/FWHM of 12.829/0.13~$\mu$m for the Gaussian fitted to the 12.8~$\mu$m [\NeII] line for all of our spectra. We also note that the 12.37~$\mu$m \HI\, line flux is detected at the 3~$\sigma$ level or higher only in spectra closest to the illuminating source. Thus for most spectra, the 12.37~$\mu$m \HI\, recombination line does not influence the measurement of the 12.7~$\mu$m PAH strength.

A Gaussian decomposition is performed to extract individual components within the 7 to 9~$\mu$m spectral range, similar to \cite{pee17,sto17}. Taking the GS continuum subtracted spectra, 6 Gaussians are simultaneously fitted to the prominent features within this range: 4 PAH Gaussian components at 7.6, 7.8, 8.2, and 8.6~$\mu$m, the 8.99~$\mu$m [\ArIII] line, as well as the 7.46~$\mu$m Pfund~$\alpha$ line (see Appendix~\ref{Fitparameters} for details). Figure~\ref{orion_7to9_spectra} shows examples of this decomposition. The fit is unable to match the sharpness of the 7.6~$\mu$m peak due to the chosen FWHM (and they thus overshoot around 7.8~$\mu$m). 

Aside from the [\NeII] 12.8~$\mu$m line, lines that are isolated upon LS continuum subtraction are fit using a Gaussian profile. These include the 6.98~$\mu$m [\ArII] line, the 9.7~$\mu$m H$_{2}$ line, and the 10.5~$\mu$m [\SIV] line. 

The signal-to-noise ratio of the PAH emission features is estimated as $ \textrm{SNR} = \textrm{F}/ (rms~\times~\sqrt{\textrm{N}}~\times~\Delta \lambda) $ where F is the feature's flux (in W~m$^{-2}$~sr$^{-1}$), {\it rms} is the rms noise, N is the number of spectral wavelength bins within the feature, and $\Delta \lambda$ is the wavelength bin size determined from the spectral resolution. The rms noise is determined from featureless portions of the spectra between 5.36--5.52, 9.2--9.4, and 9.95--10.3~$\mu$m. For atomic and H$_{2}$ lines, the signal-to-noise is the ratio of the peak line flux to the underlying rms noise.

\begin{figure*}
\begin{center}
\resizebox{\hsize}{!}{%
\includegraphics[clip,trim =3cm 1.cm 1cm 1.cm,width=8.5cm]{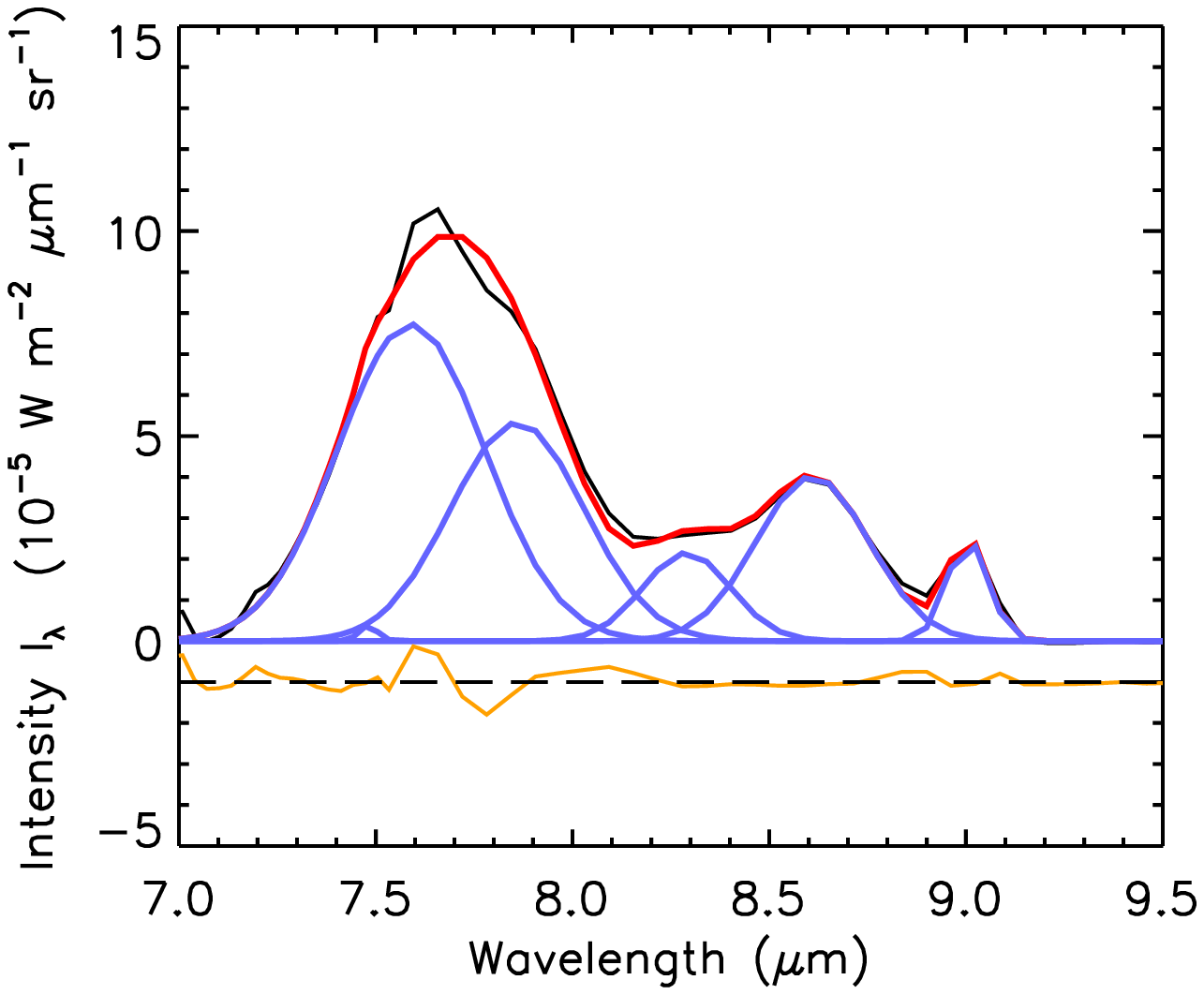}
\includegraphics[clip,trim =3cm 1cm 1cm 1cm,width=8.5cm]{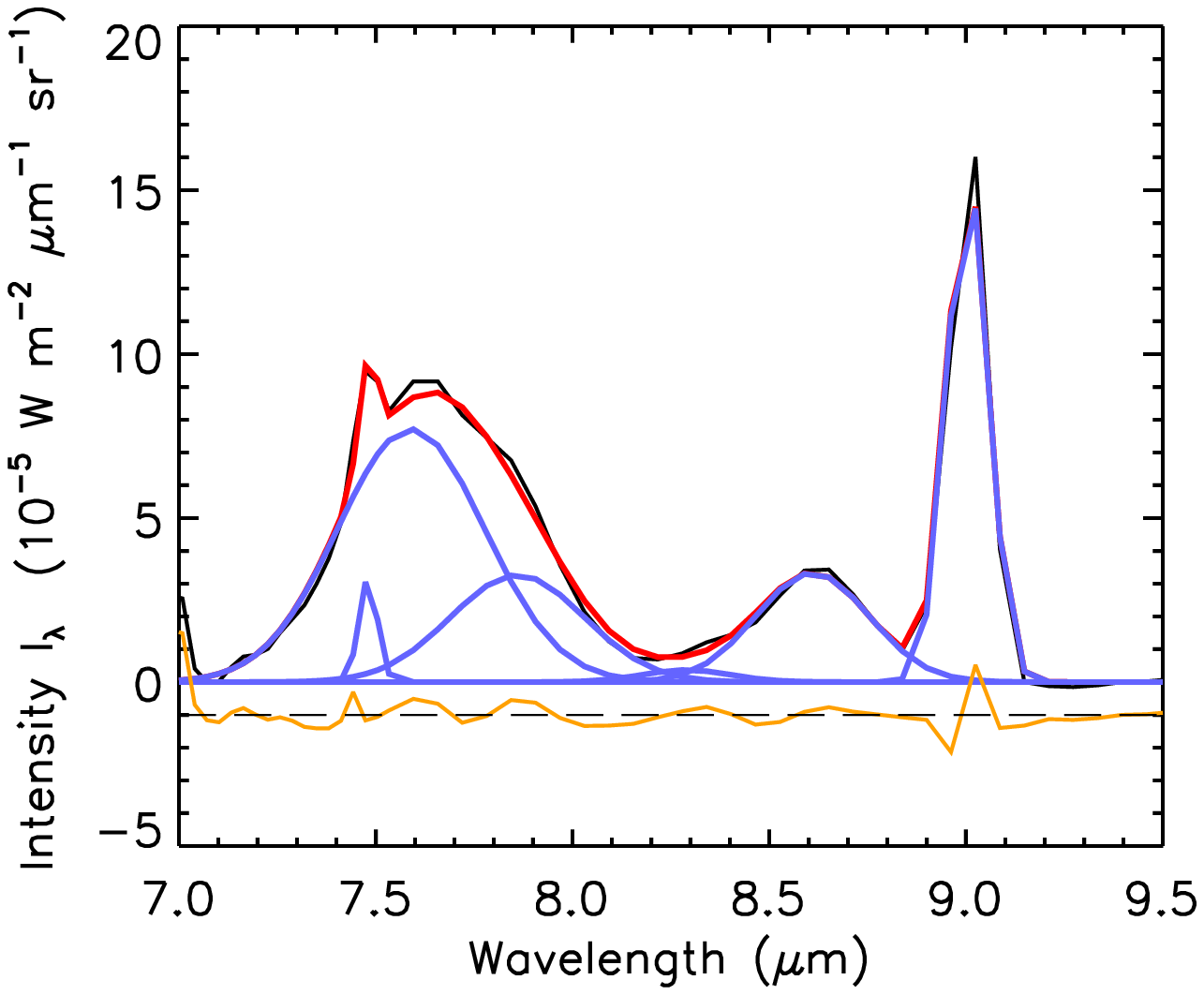}}
\end{center}
\caption{The 7 to 9~$\mu$m decomposition at positions behind ({\it Left}) and in front ({\it Right}) of the Orion Bar with respect to the illuminating source. The GS continuum subtracted spectra are displayed in black, individual Gaussian components are shown in blue, and the combined fit in red. The residuals (shown in yellow) and black dashed lines are offset by 1~$\times$~10$^{-5}$~W~m$^{-2}$~$\mu$m$^{-1}$~sr$^{-1}$ for clarity. }
\label{orion_7to9_spectra}
\end{figure*}

\section{Results}
\label{results}
 
In this section, we investigate the relationships between individual PAH emission bands, atomic spectral lines, the 9.7~$\mu$m~H$_{2}$ emission line, the underlying plateaus, and the dust continuum emission within our pointings across the Orion Bar. We use two separate methods to analyse these spectral features, namely cross cuts and correlation plots. Our cross cuts (or linear projections) allow us to measure how these spectral features as well as their ratios vary i) with distance to the illuminating source, and ii) relative to the changing environmental conditions across the Orion Bar.

\begin{figure*}
\begin{center}
\resizebox{0.935\hsize}{!}{%
\includegraphics[clip,trim =0cm 1.5cm 0cm 0cm,width=7.2cm]{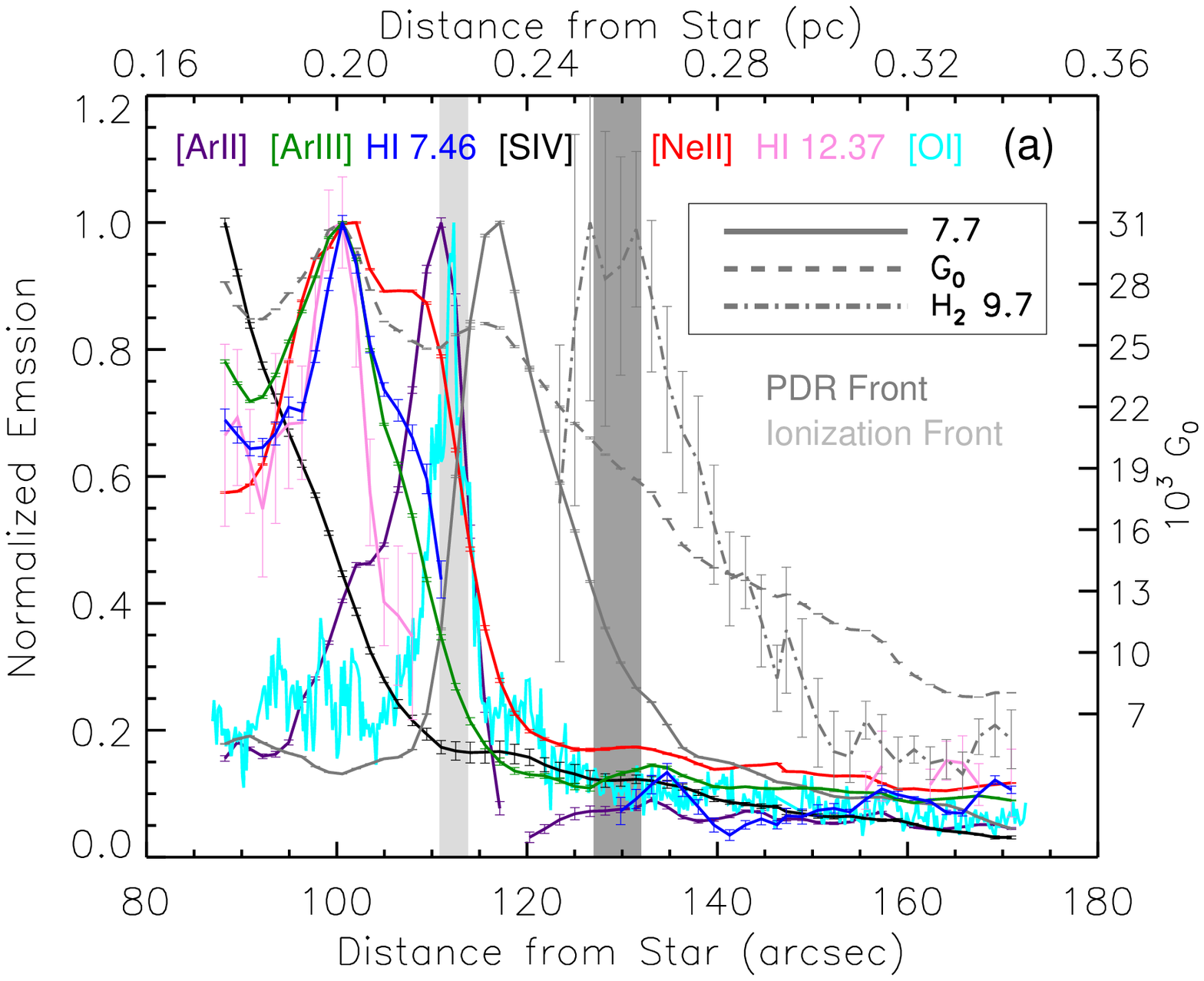}
\includegraphics[clip,trim =0cm 1.5cm 0cm 0.0cm,width=7.2cm]{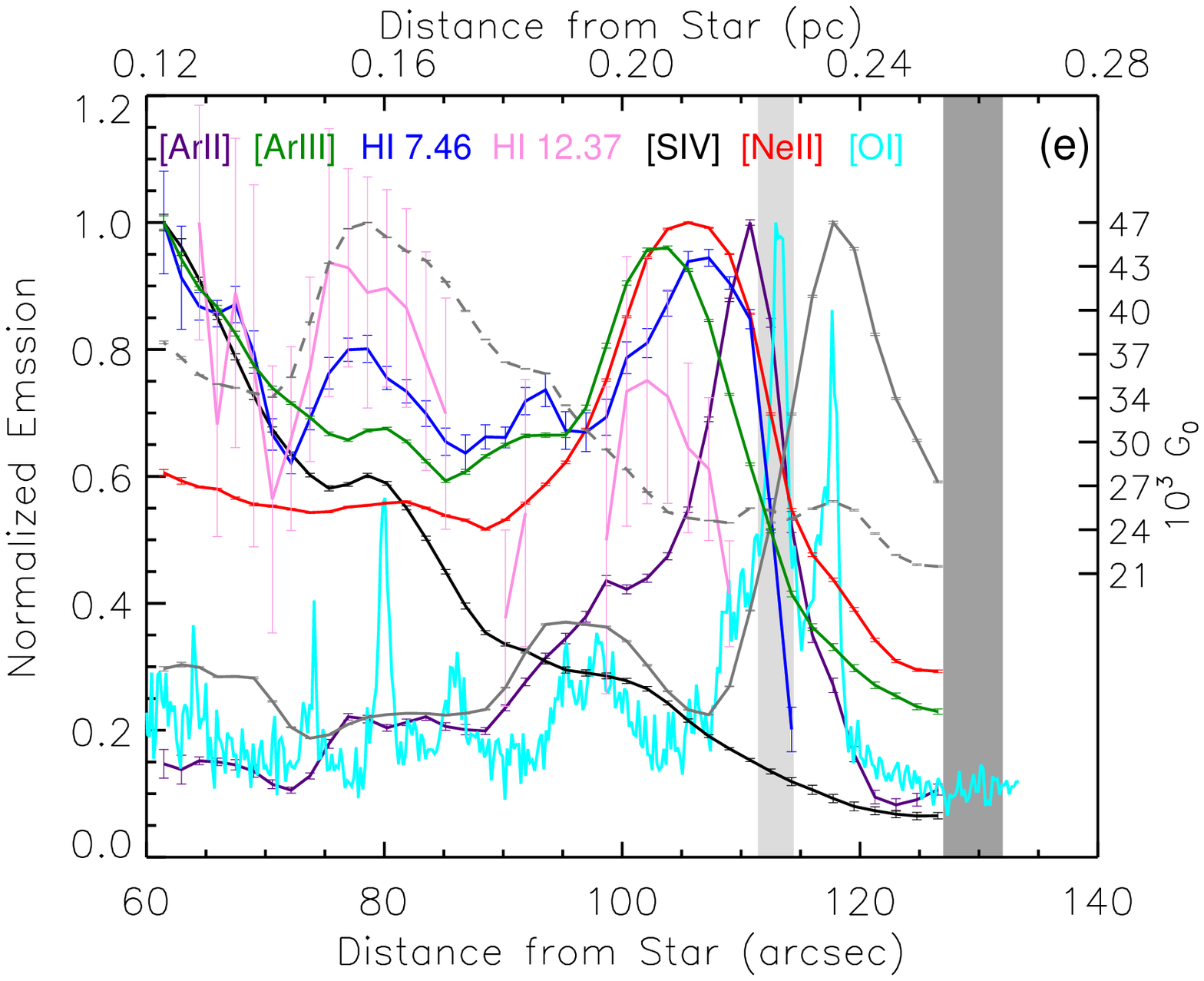}
}
\resizebox{0.935\hsize}{!}{%
\includegraphics[clip,trim =0cm 1.5cm 0cm 2.2cm,width=7.2cm]{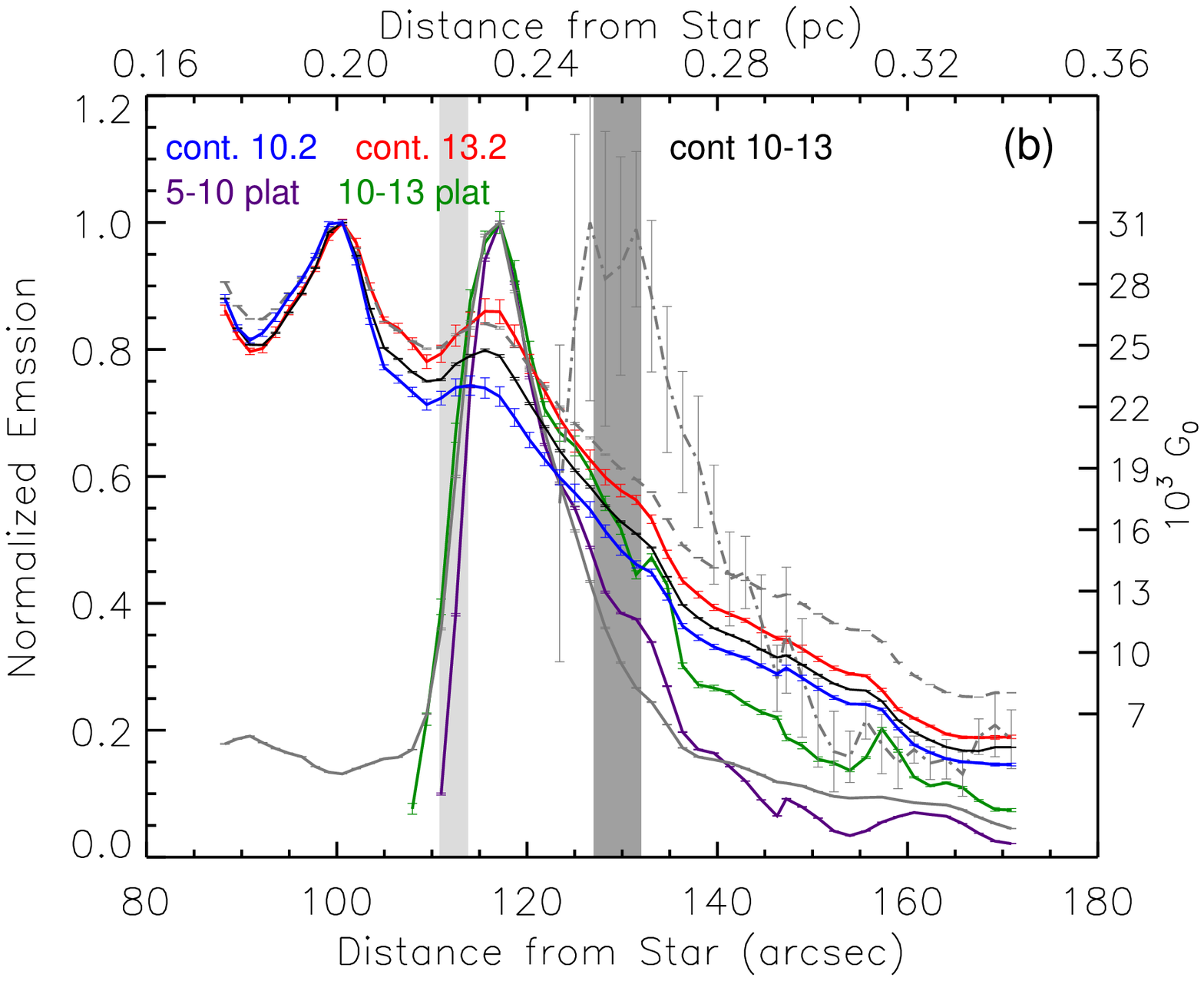}
\includegraphics[clip,trim =0cm 1.5cm 0cm 2.2cm,width=7.2cm]{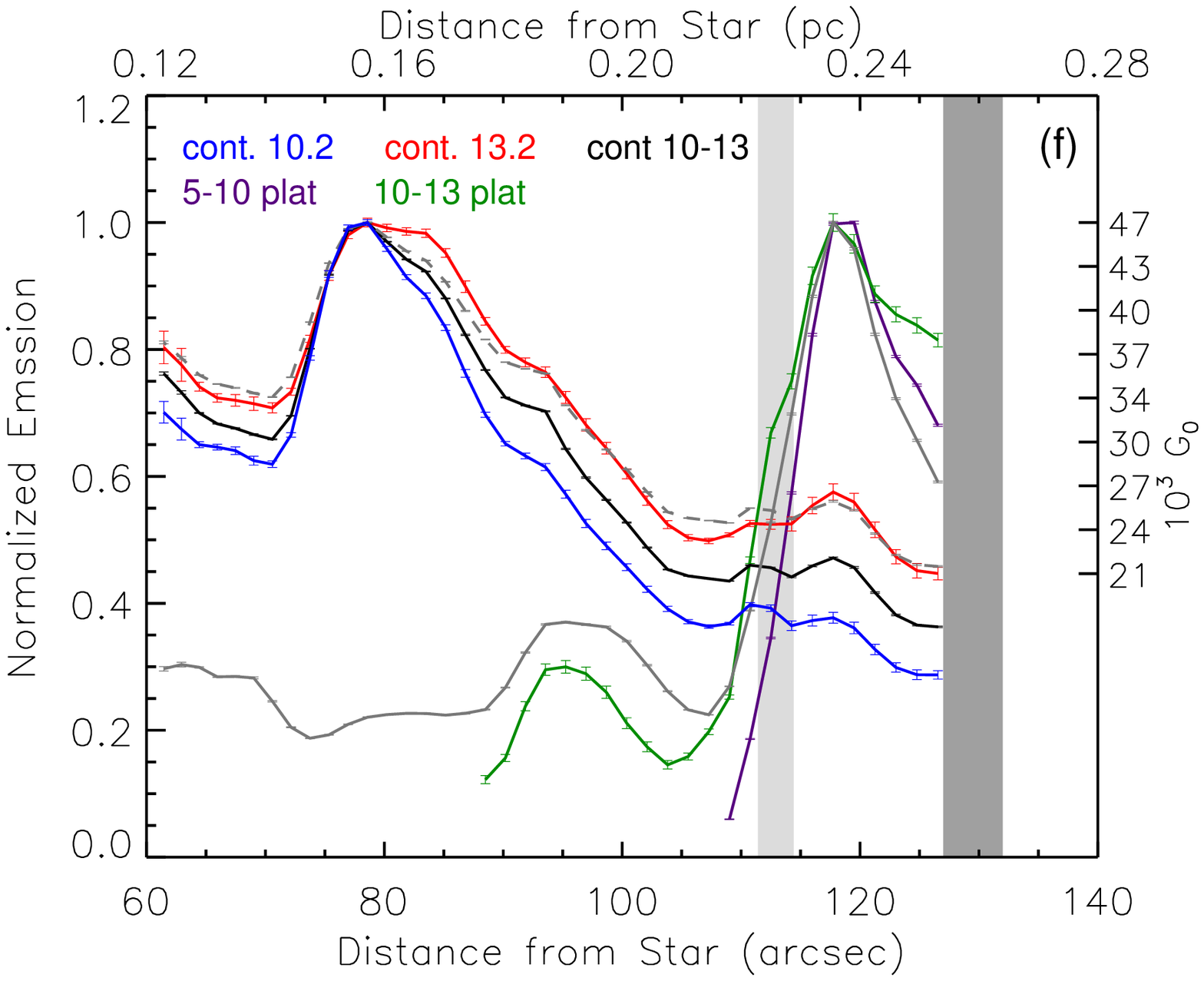}
}
\resizebox{0.935\hsize}{!}{%
\includegraphics[clip,trim =0cm 1.5cm 0cm 2.2cm,width=7.2cm]{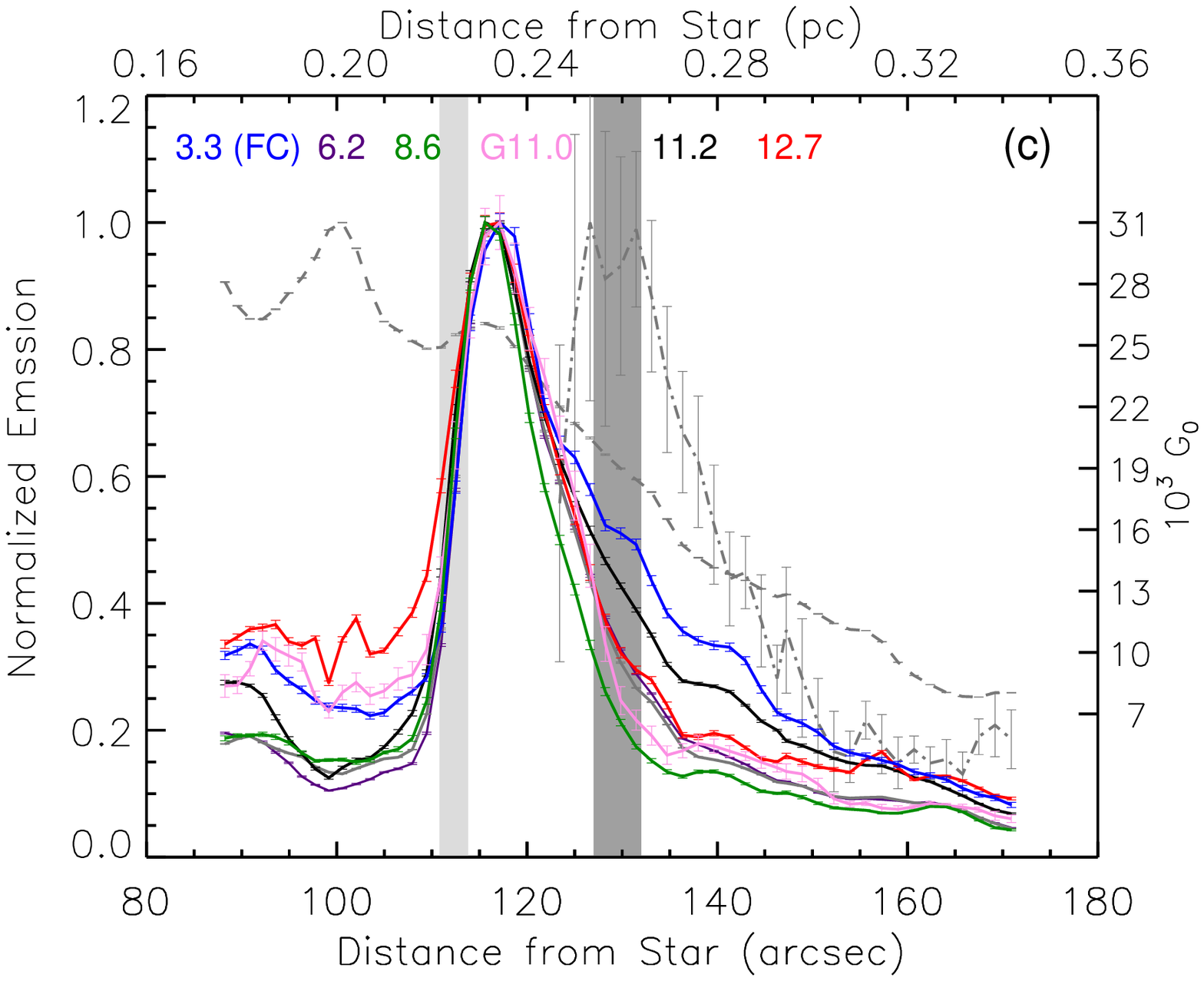}
\includegraphics[clip,trim =0cm 1.5cm 0cm 2.2cm,width=7.2cm]{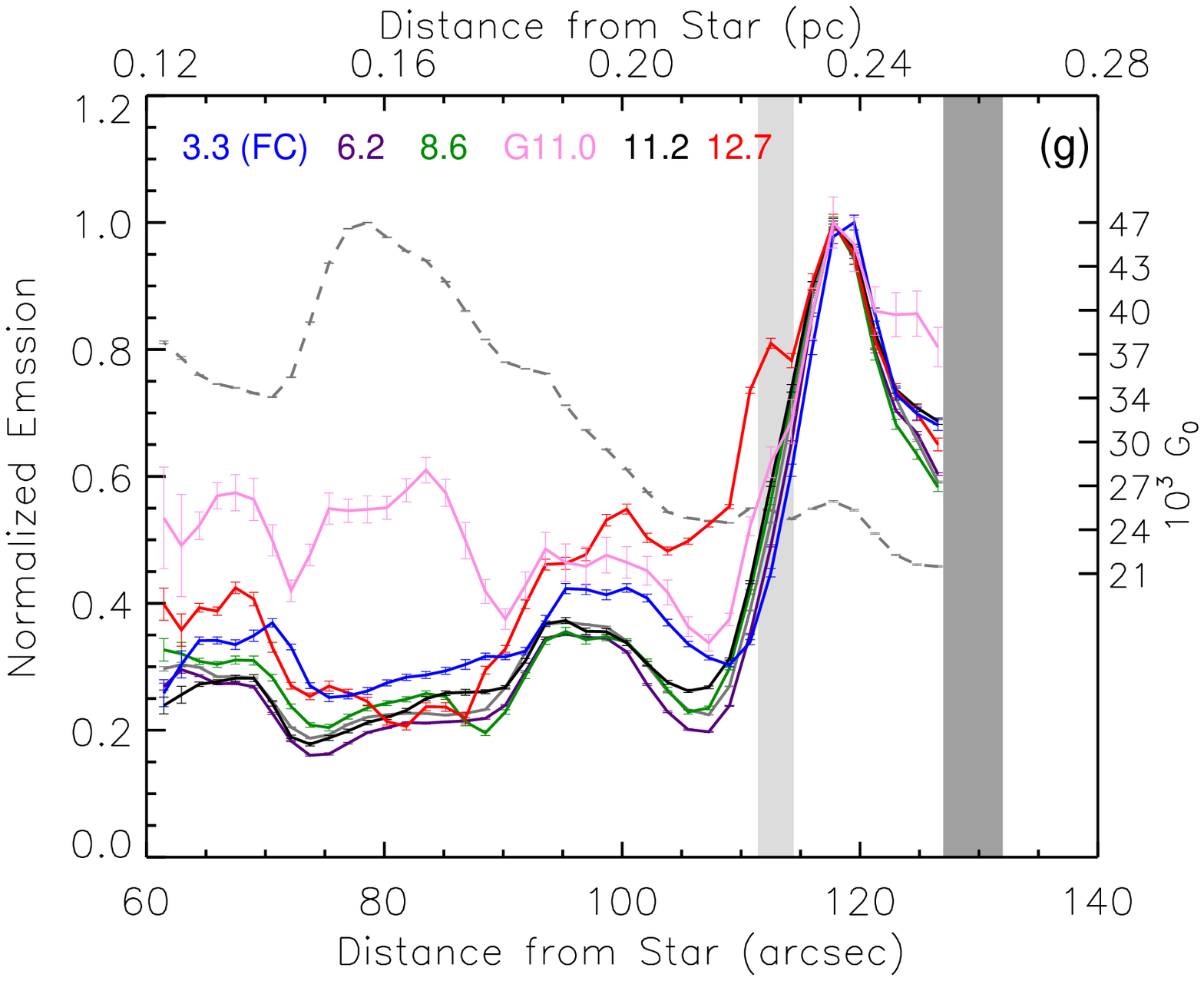}
}
\resizebox{0.945\hsize}{!}{%
\includegraphics[clip,trim =0cm 0cm 0cm 2.2cm,width=8cm]{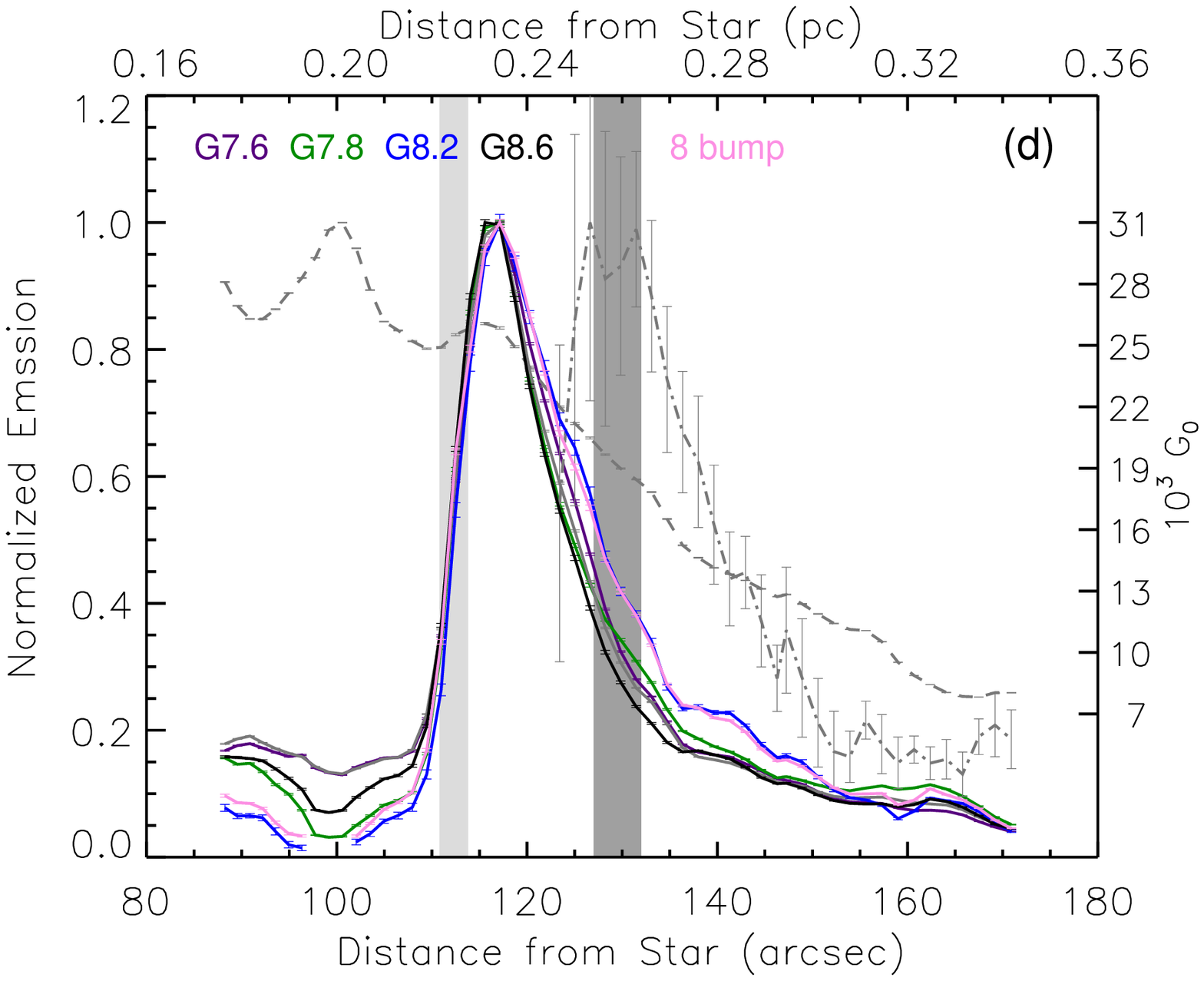}
\includegraphics[clip,trim =0cm 0cm 0cm 2.2cm,width=8cm]{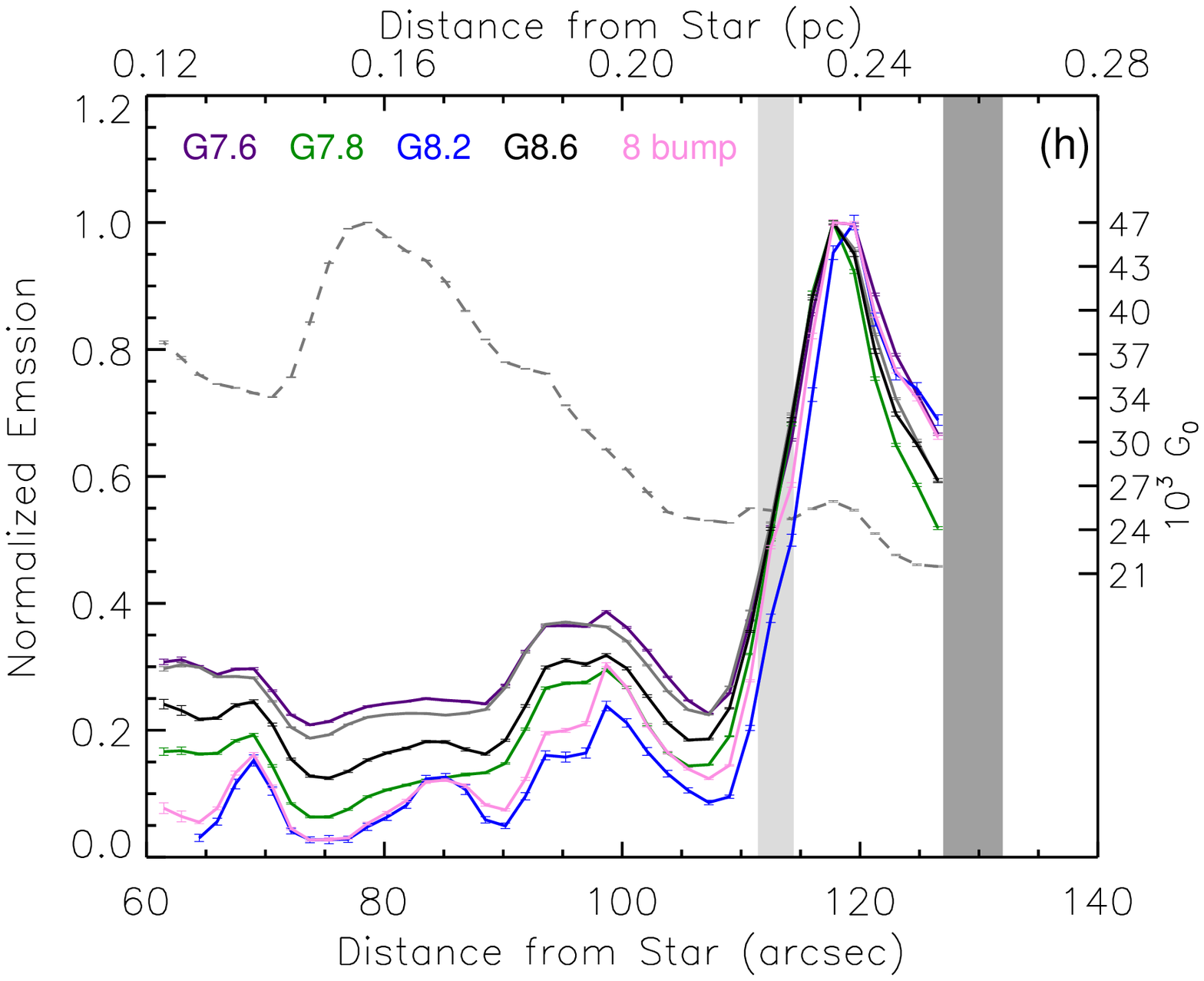}
}
\end{center}
\caption{Orion Bar combined (OBC, left) and Orion Bar ionized (OBI, right) cross cuts normalized to the peak values for each emission feature. The dark grey shaded region corresponds to the location of the Orion Bar PDR front as defined by the peak of the 9.7~$\mu$m H$_{2}$ line. The light grey shaded region corresponds to the ionization front as defined by the [OI] 6300~\AA \ line peak given in panels (a) and (e) for the OBC and OBI apertures respectively. G$_{0}$ cross cut values are shown on the right y-axis in units of 10$^{3}$ Habings (see Section~\ref{PDR} for derivation).}
\label{orion_line_profiles1}
\end{figure*}

\begin{figure*}
\begin{center}
\resizebox{0.935\hsize}{!}{
\includegraphics[clip,trim =0cm 1.5cm 0cm 0.0cm,width=7.5cm]{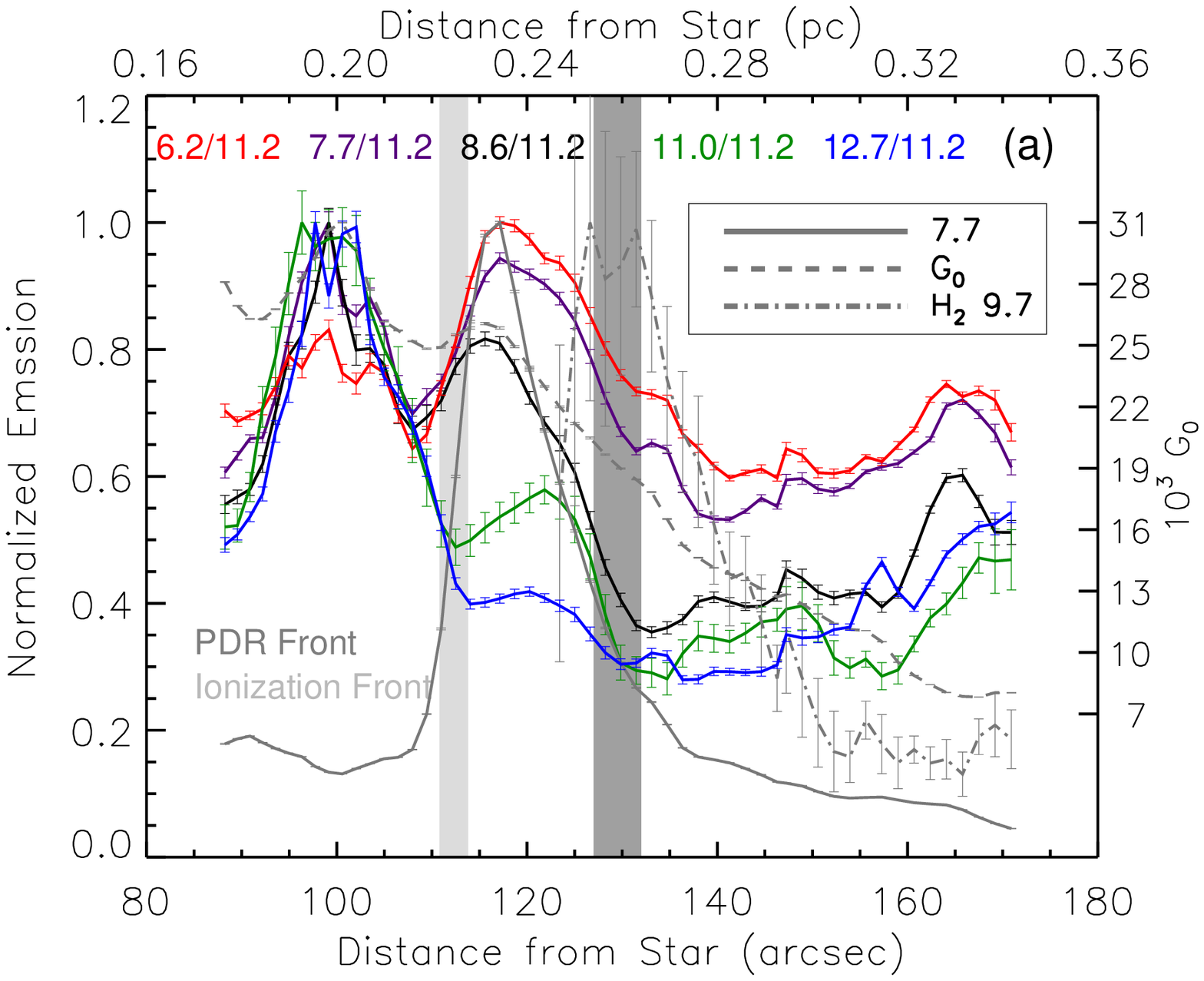}
\includegraphics[clip,trim =0cm 1.5cm 0cm 0.0cm,width=7.5cm]{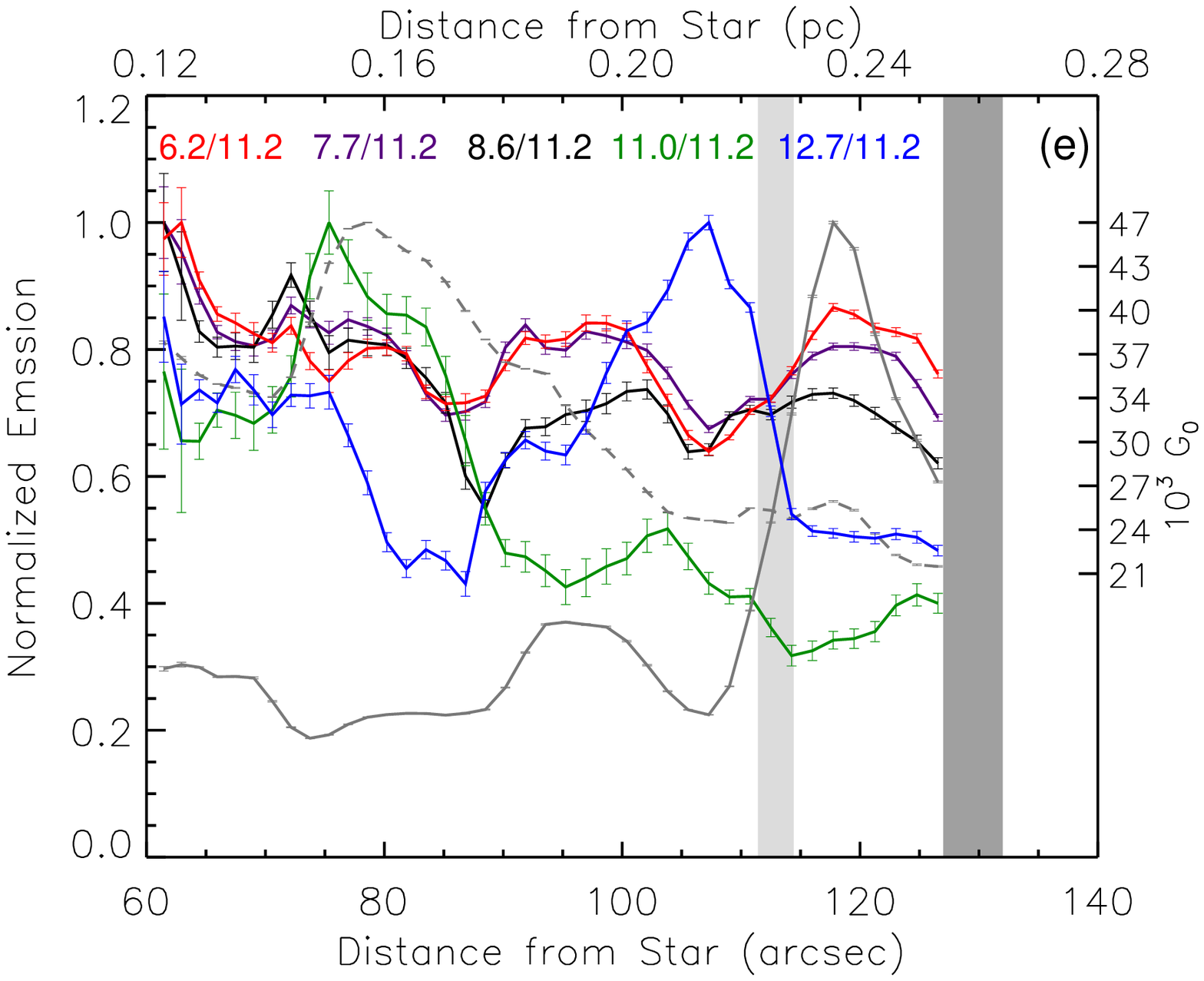}
}
\resizebox{0.935\hsize}{!}{%
\includegraphics[clip,trim =0cm 1.5cm 0cm 2.2cm,width=7.5cm]{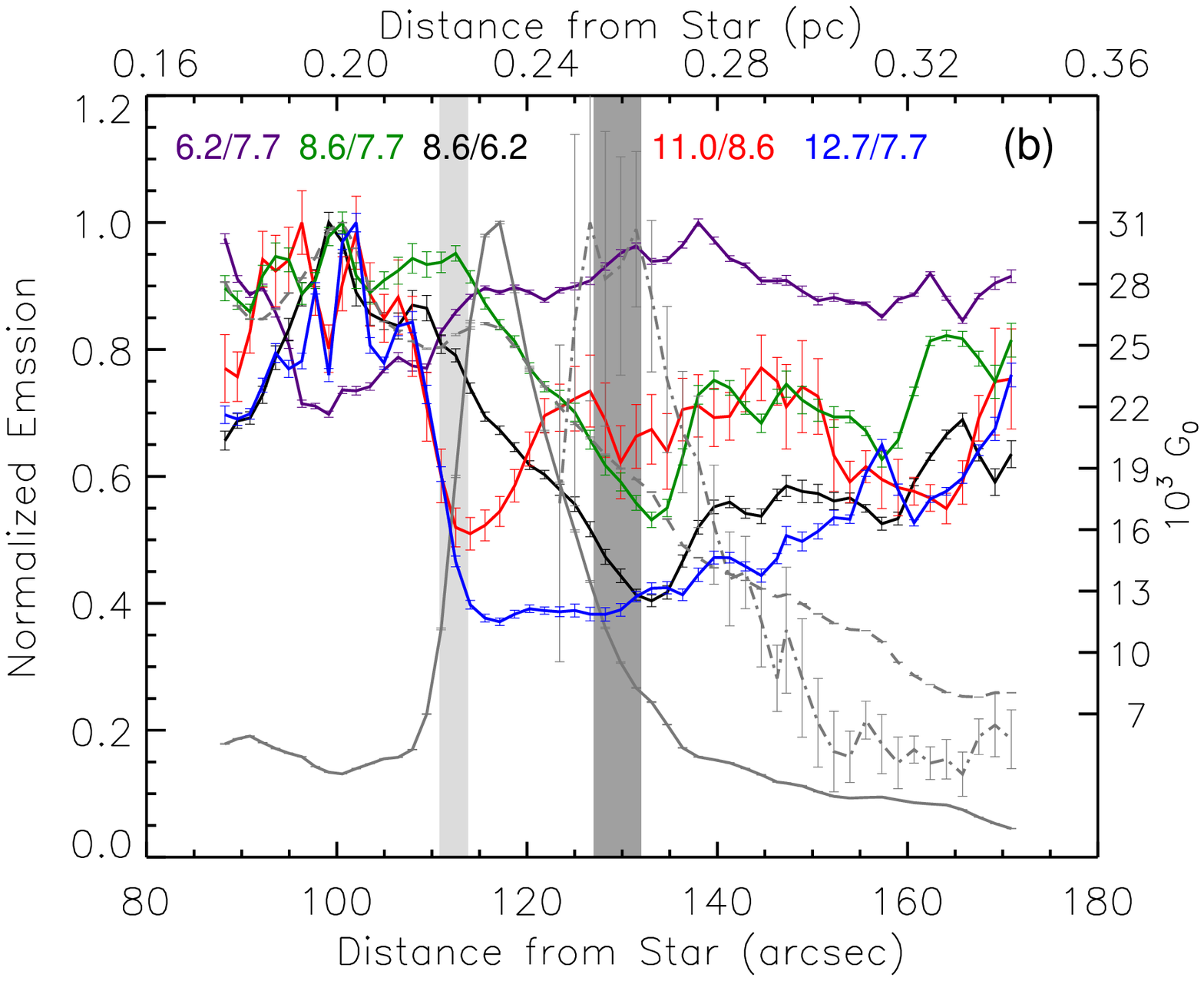}
\includegraphics[clip,trim =0cm 1.5cm 0cm 2.2cm,width=7.5cm]{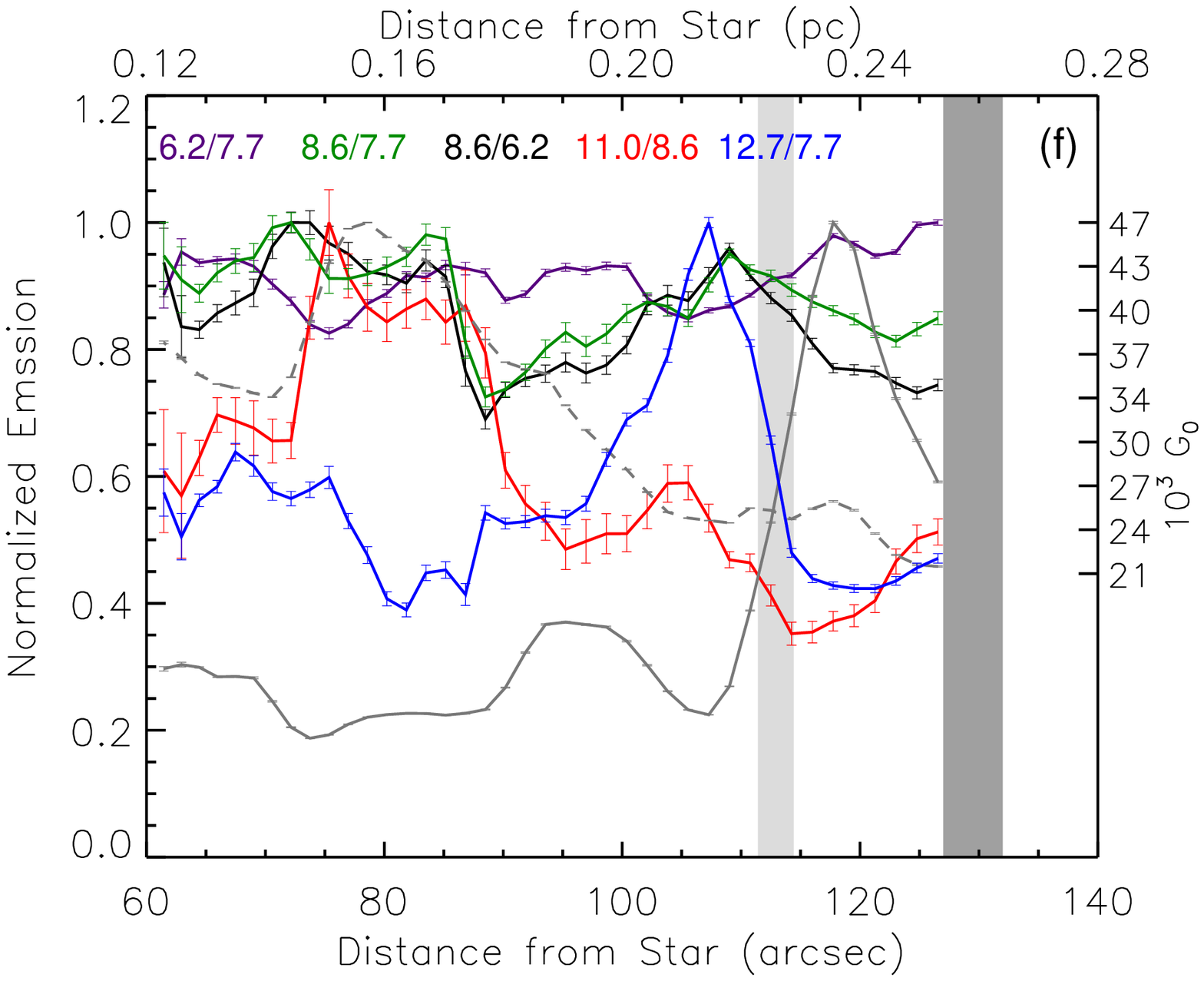}
}
\resizebox{0.935\hsize}{!}{%
\includegraphics[clip,trim =0cm 1.5cm 0cm 2.2cm,width=7.5cm]{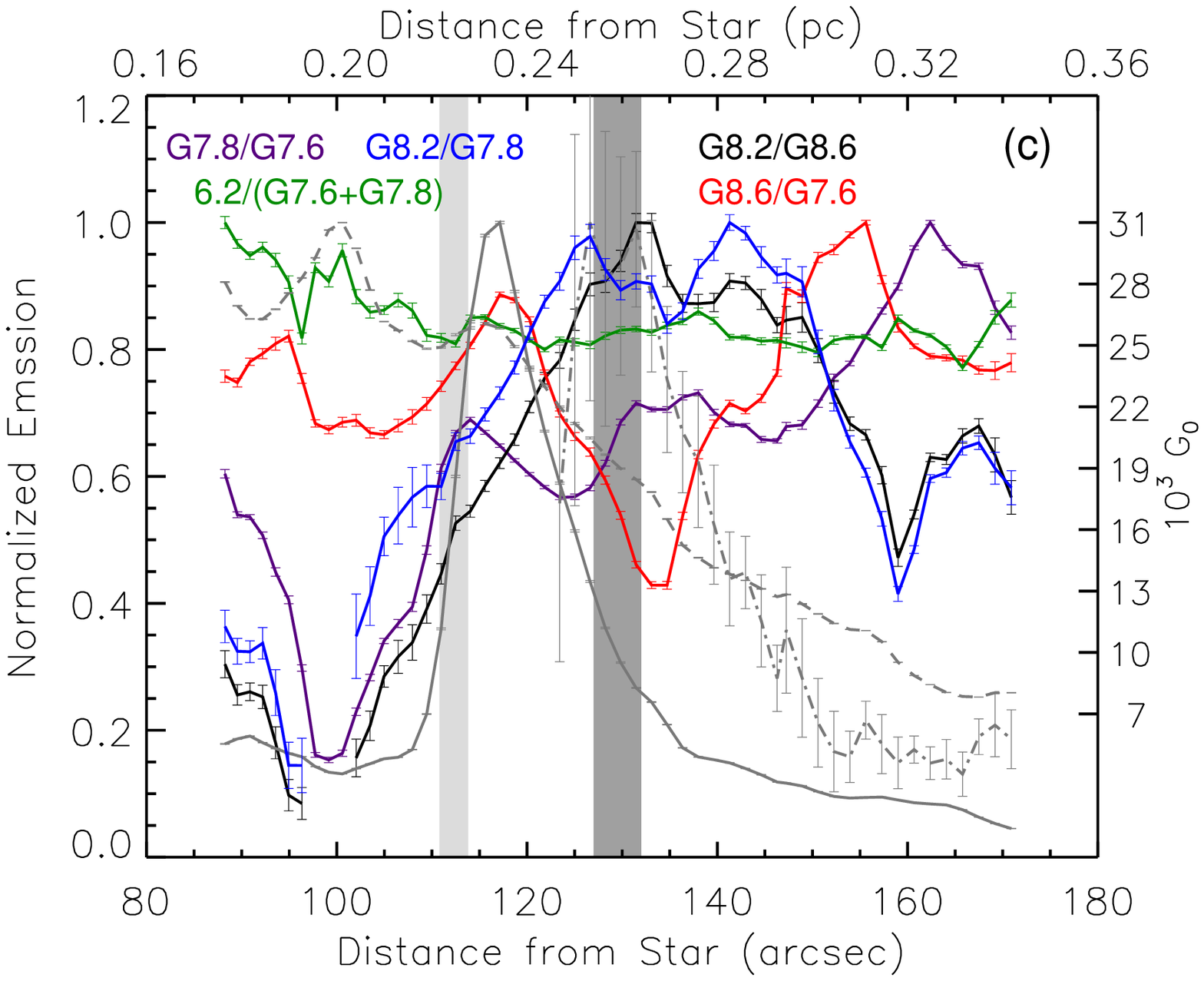}
\includegraphics[clip,trim =0cm 1.5cm 0cm 2.2cm,width=7.5cm]{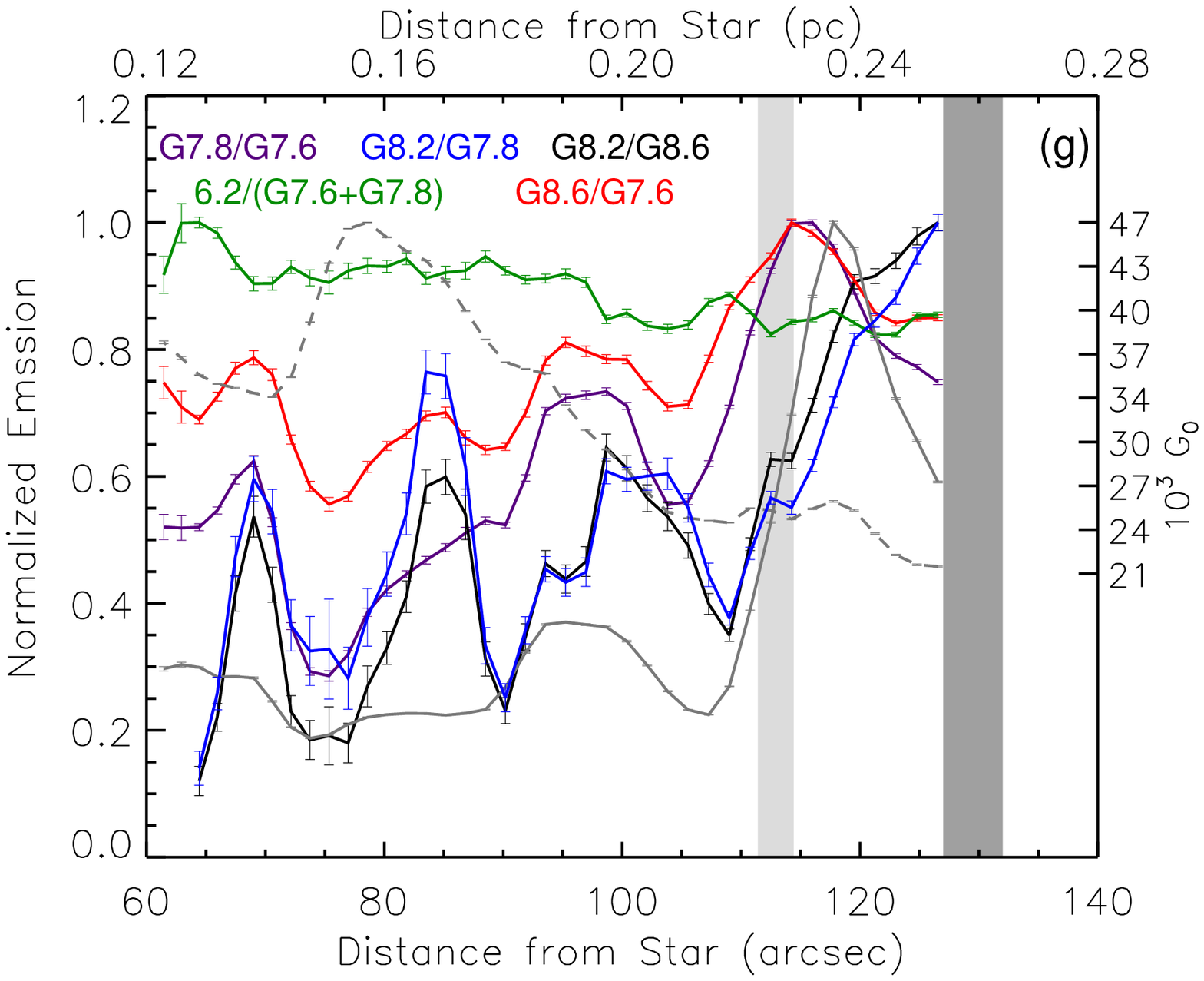}
}
\resizebox{0.935\hsize}{!}{%
\includegraphics[clip,trim =0cm 0cm 0cm 2.2cm,width=7.5cm]{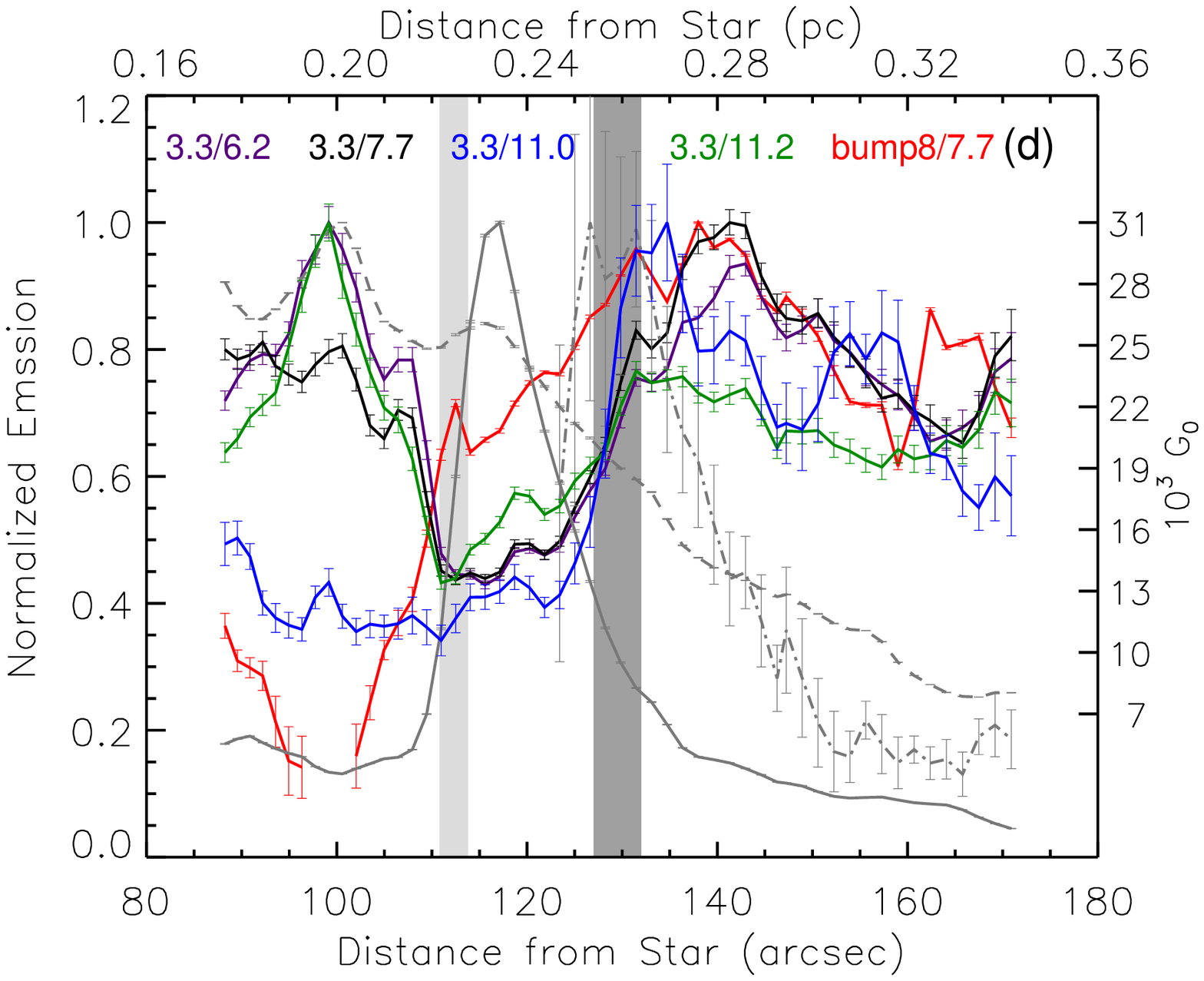}
\includegraphics[clip,trim =0cm 0cm 0cm 2.2cm,width=7.5cm]{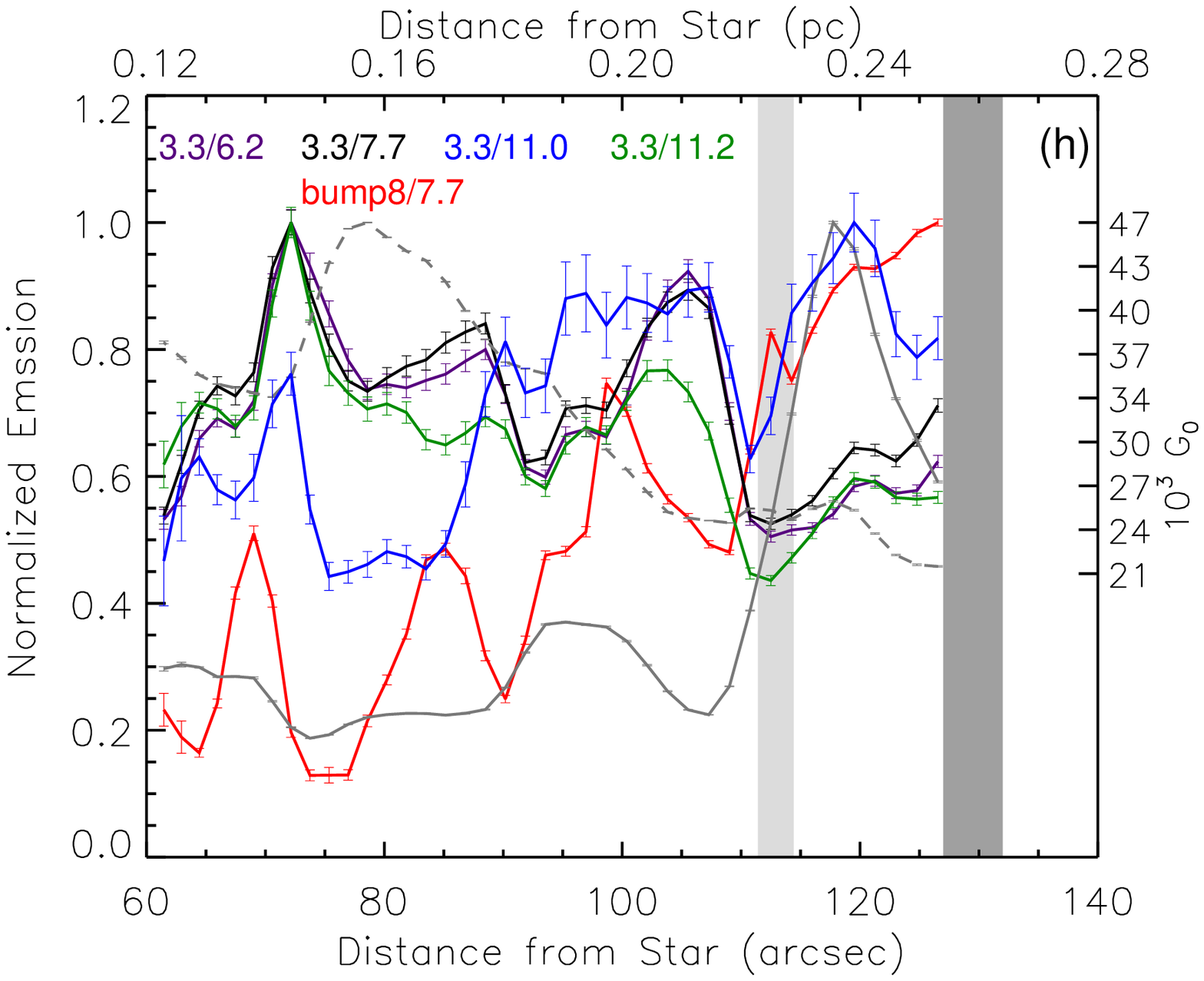}
}
\end{center}
\caption{Orion Bar combined (OBC, left) and Orion Bar ionized (OBI, right) emission ratio cross cuts normalized to the peak values for each ratio. The dark grey shaded region corresponds to the location of the Orion Bar PDR front as defined by the peak of the 9.7~$\mu$m H$_{2}$ line. The light grey shaded region corresponds to the ionization front as defined by the [OI] 6300~\AA\ line peak given in Figure~\ref{orion_line_profiles1} panels (a) and (e) for the OBC and OBI apertures respectively. G$_{0}$ cross cut values are shown on the right y-axis in units of 10$^{3}$ Habings (see Section~\ref{PDR calc} for derivation). }
\label{orion_line_profiles3}
\end{figure*}

\subsection{Cross cuts}
\label{proj}
Figures~\ref{orion_line_profiles1} and~\ref{orion_line_profiles3} show cross cuts of the intensity of emission features and their ratios for the Orion Bar Combined and Orion Bar Ionized apertures\footnote{We give a summary of all of the emission components for which we have derived cross cuts and their normalization factors in Appendix~\ref{ob irs components}.}. We normalize these cross cuts to their maximum value within each respective aperture.
Only fluxes and emission ratios equal or larger than 3~$\sigma$ are presented here. We make use of the following groupings for the remainder of this section based on the relative position to the Orion Bar IF in each aperture, i.e in front of and behind the IF (Figure~\ref{Orion_schematic}). We refer to pixels in front of the IF as the \HII\, region (PDR) where we find the steep underlying dust continuum coinciding with the ionized cavity surrounding the Trapezium cluster. We further refer to the region between the Orion Bar IF and PDR front, encompassing the PAH peak, as the edge--on PDR and beyond the edge--on PDR front as `behind the PDR front'. We emphasize that the edge--on PDR dominates the PAH emission up to 5.5$^{\prime\prime}$ beyond the PDR front (see Section~\ref{PDR morph} for details). The latter transition is used in Figure~\ref{Orion_schematic} and in Section~\ref{discussion}.  

We find that all of the major PAH bands peak at the same distance from the illuminating source in both apertures between the edge--on PDR front, as defined by the 9.7~$\mu$m H$_{2}$ peak, and the edge--on IF, as defined by the [\OI] 6300~\AA \ line peak obtained from MUSE IFU spectroscopic observations of the Orion Nebula \citep{wei15}. The distance between the peak of the PAH emission and H$_{2}$ emission (at 117$^{\prime\prime}$ and 130$^{\prime\prime}$) from the illuminating source respectively) is 13.0$^{\prime\prime}$~$\pm$~3.6$^{\prime\prime}$, which agrees with the distance found between the 3.3~$\mu$m peak and the 2.122~$\mu$m H$_{2}$ peak of $\sim$~12~$\pm$~2$^{\prime\prime}$ in \cite{tie93}. The  edge--on IF is located at 112.5$^{\prime\prime}$~$\pm$~1.5$^{\prime\prime}$, $\sim$~4$^{\prime\prime}$ in front of the PAH emission peak, in agreement with cross cuts presented in \cite{sal16}.
In this section, we will first discuss in detail the cross cuts along the OBC aperture, followed by a discussion on the observed differences and similarities between the OBC and OBI aperture.

\subsubsection{Atomic Lines}
\label{atom}

In Figure~\ref{orion_line_profiles1} (a), the cross cuts for each of the atomic emission lines peak inward of the IF towards the illuminating source in the following order: [\ArII], [\NeII], [\ArIII], Pfund~$\alpha$, and [\SIV] (which does not show a peak but a steady rise towards the star). We also include the MUSE [\OI] 6300~\AA \ cross cut for reference. Note that the order in which these atomic emission lines peak (aside from Pfund~$\alpha$) towards the star is directly related to the ionization potential of each respective species. The relative emission of these lines sharply drops from their peak emission moving away from the Trapezium cluster but they are still detected well beyond the IF.

\subsubsection{Dust Emission}
\label{dust}

The dust continuum emission measured at 10.2 and 13.2~$\mu$m as well as the integrated 10--13.2~$\mu$m continuum emission generally agree with each other (Figure~\ref{orion_line_profiles1} (b)). These continua all have a strong peak at 100$^{\prime\prime}$ from the illuminating star, coinciding with the peak emission of [\ArIII] and Pfund~$\alpha$. A secondary (local) continuum maximum, not seen in the [\ArIII] and Pfund~$\alpha$ emission, is found where the PAH emission peaks. These continuum measurements show a gradual decrease moving further away from the Trapezium cluster past the PAH emission peak, the PDR front, and beyond. We note that \cite{fel93} identified a bar--like emission structure in the ionized gas which is located in front of the Orion Bar. The location of the peak in the dust continuum emission coincides with this `ionized gas Bar'. 

\subsubsection{PAH Emission Features}
\label{PAHs}

All PAH bands, the Gaussian PAH components from the 7--9~$\mu$m decomposition, the 5--10 and the 10--13~$\mu$m PAH plateaus show the same peak position within the Orion Bar PDR at 117$^{\prime\prime}$ from the illuminating star (see Figure~\ref{orion_line_profiles1} (b), (c), and (d)). Moving towards the illuminating source, all PAH bands display a rapid decrease in emission strength but remain detected throughout (except for the G8.2 component and 8~$\mu$m bump). In contrast, behind the PDR front, a more gradual decline occurs for all PAH emission features. Additionally, significant variations in the relative intensities of the PAH emission features become evident upon moving away from their shared peak position. In particular, the intensity at which the PAH features level off in the \HII\, region (PDR) varies and does not seem to be solely governed by the ionization state of the feature's carrier. Behind the PDR front, the drop in PAH band intensity (relative to the peak emission) varies with the 3.3, 11.2, 12.7, 6.2, 7.7, 11.0, and 8.6~$\mu$m bands in decreasing order respectively\footnote{We note that the ISO-SWS D5 observation is centered on the PAH emission peak. Hence, the normalized flux of the 3.3 $\mu$m emission in the \HII\ region (PDR) and behind the PDR front will further decrease by respectively $\sim4\%$ and $\sim2.5\%$ when accounting for the variable PAH contribution to the SOFIA filter (see Table~\ref{table:sofia}). Likewise, at the PDR front, the normalized flux of the 3.3 $\mu$m emission may increase by 2-3\%. Such changes do not influence our major conclusions.}. In other words, the PAH bands that are attributed to neutral species have a less pronounced decrease in relative flux in this region. Similarly, we find that the 7--9~$\mu$m Gaussian components show a decrease in relative intensity (with respect to the peak intensity) towards the illuminating source in the following order: G7.6, G8.6, G7.8, and G8.2~\mum\,  components. We note that the G8.2 component is very weak or absent in the \HII\, region (PDR). Behind the PDR front, these components show a decrease in relative intensity in the reverse order to what is found in the  \HII\, region (PDR): i.e. G8.2, G7.8, G7.6, and G8.6~$\mu$m. We note that the G7.6~$\mu$m component and LS derived 7.7~$\mu$m band have very similar spatial cross cuts, reflecting the dominance of the G7.6~\mum\, component to the 7.7~$\mu$m complex.

The 5--10 and 10--13~$\mu$m plateaus cross cuts are very similar to the PAH bands within the edge--on Orion Bar PDR and have a gradual decline moving behind the PDR front (Figure~\ref{orion_line_profiles1} (b)). We note the rapid drop in these cross cuts at the IF to the point where they are no longer detected in the  \HII\, region (PDR). Similarly, the 8~$\mu$m bump has a cross cut that is comparable with other PAH features, most notably the G8.2~$\mu$m component as it is derived from essentially the exact same spectral region (Figure~\ref{orion_line_profiles1} (d)). 

\subsubsection{PAH Emission Ratios}
\label{PAH ratios}

In Figure~\ref{orion_line_profiles3} (a), the 6.2/11.2, 7.7/11.2, 8.6/11.2, and 11.0/11.2 cross cuts are very comparable: these ratios show a broad maximum at the PAH peak and at the dust continuum peak, which coincides with the peak emission of [\ArIII] and Pfund~$\alpha$. The relative strength of the maxima of these ratios at the dust continuum peak with their maxima at the PAH peak varies significantly and can be organized in decreasing order from 11.0/11.2 to 8.6/11.2, 7.7/11.2, and finally to 6.2/11.2. 

In Figure~\ref{orion_line_profiles3} (b), the 6.2/7.7 ratio shows little variation behind the IF. However, within the  \HII\, region (PDR), the 6.2/7.7 ratio decreases to a minimum roughly co--spatial with the dust continuum peak. We note that the 6.2/(G7.6 +G7.8) ratio shows very little variation (see Figure~\ref{orion_line_profiles3} (c)). This arises from the difference in the PAH behaviour being traced by the 7.7~$\mu$m band, for which  the LS continuum is subtracted, and the combined G7.6 and G7.8~$\mu$m components which include emission from the 8~$\mu$m bump which is minimized within the  \HII\, region (PDR, Figure~\ref{orion_spectra}). In contrast, the 8.6/7.7, 8.6/6.2 and 11.0/8.6 ratios are strong in the  \HII\, region (PDR) and weaker behind the IF (Figure~\ref{orion_line_profiles3} (b)).

The 12.7/11.2 and 12.7/7.7 emission ratios show overall similar trends, characterized by a strong peak near the dust continuum peak akin to the 11.0/11.2 ratio (Figure~\ref{orion_line_profiles3} (a) and (b)). However, despite each of these ratios showing a sharp decrease at the IF, these ratios differ with the 11.0/11.2 ratio as they do not show any significant maximum near the PAH peak.

Regarding emission ratios between the 7--9~$\mu$m Gaussian components (Figure~\ref{orion_line_profiles3} (c)), G7.8/G7.6 shows a very pronounced minimum at the dust continuum peak followed by a substantial rise towards the IF and a local minimum within the Bar, followed by a rise moving behind the PDR front. The G8.6/G7.6 ratio fluctuates across the cross cut with local maxima roughly corresponding to the dust emission peak, the PAH emission peak, and at 155$^{\prime\prime}$ from the illuminating source (i.e. behind the PDR front). The G8.2/G7.8, G8.2/G8.6, and the 8~$\mu$m bump/7.7 emission ratios are comparable, with a strong minimum roughly corresponding with the dust continuum peak. Further from the illuminating star, these cross cuts shows a steep rise into the edge--on PDR, which levels off near the edge--on PDR front. Further into the PDR, these ratios proceed to significantly drop again to a sharp local minimum found at $\sim$~160$^{\prime\prime}$ attributable to a `blip' in the G8.2~$\mu$m and 8~$\mu$m bump components here.

Overall, the 3.3/6.2, 3.3/7.7, 3.3/11.2 ratios all show similar trends with a strong peak near the dust continuum peak comparable with the 6--9/11.2 and 11.0/11.2 peaks found here (Figure~\ref{orion_line_profiles3} (d)). However, each of these emission ratios involving the 3.3~$\mu$m have a minimum corresponding to the PAH peak followed by a subsequent rise behind the PDR front. 
The 3.3/11.0 is unique amongst PAH ratios involving the 3.3~$\mu$m PAH emission feature as it lacks a local strong maximum at the dust continuum peak. It shows very little variation in front of the Orion Bar IF and in the edge--on PDR. At the edge--on Orion Bar PDR front, this ratio increases considerably and, within the uncertainties, plateaus behind the PDR front. We note that the 3.3/11.0 behaves very similar to the 3.3/11.2 in the transition to the Orion Bar and beyond.

\subsubsection{Aperture Differences}
\label{ap diff}

The OBI aperture is a slightly different case in comparison to the OBC aperture as it does not intersect with the edge--on Orion Bar PDR front and extends much deeper into the ionized cavity surrounding the Trapezium cluster (i.e. closer to the cluster). The atomic lines show, in general, the same behavior as the OBC aperture with the peaks of the [\NeII], [\ArIII] and \HI\, Pfund~$\alpha$ line being broader, encompassing the shoulder seen in the OBC aperture (Figure~\ref{orion_line_profiles1} (e)). The 10--13.2~$\mu$m continuum cross cut (and the 10.2 and 13.2~$\mu$m continuum emission) is distinctly different in the OBI aperture with a continual rise towards the star and a broad peak much deeper into the  \HII\, region (PDR, Figure~\ref{orion_line_profiles1} (f)). Many of the atomic emission lines show a minor bump that can be associated with the dust continuum peak in the OBI aperture. 

All of the major PAH bands excluding the 12.7~$\mu$m band show very similar cross cuts moving towards the source, with a somewhat broad local maximum centered at $\sim$~98$^{\prime\prime}$, a few arcseconds behind a corresponding small bump in the dust continuum cross cuts (Figure~\ref{orion_line_profiles1} (g) and (h)). Only minor variations in each cross cut are seen towards the star with a slight rise closest to the star. Emission ratio cross cuts between PAH features within the OBI aperture show very similar trends with those found for the OBC aperture in the \HII\, region (PDR) with each emission ratio generally having the same behaviour at the dust continuum peak and within the edge--on Orion Bar PDR (Figure~\ref{orion_line_profiles3} (e)--(h)). The most significant discrepancy between both apertures is the local (weaker) maximum, the `PAH bump', at  $\sim$~98$^{\prime\prime}$ which is only found in the OBI aperture. In addition, the 12.7/11.2 and 12.7/7.7 ratios are notably different in the OBI aperture, with a strong peak in front of the IF and a broad minimum at the dust continuum peak. This peak coincides with the peak of the [\NeII] emission, which may influence the 12.15--13.2~$\mu$m decomposition we applied. Higher spectral resolution data from, for example, the James Webb Space Telescope, will settle this.

\begin{figure*}
\begin{center}
\resizebox{\hsize}{!}{%
\includegraphics[clip,trim =0cm 0cm 0cm 0cm,width=0.33\textwidth]{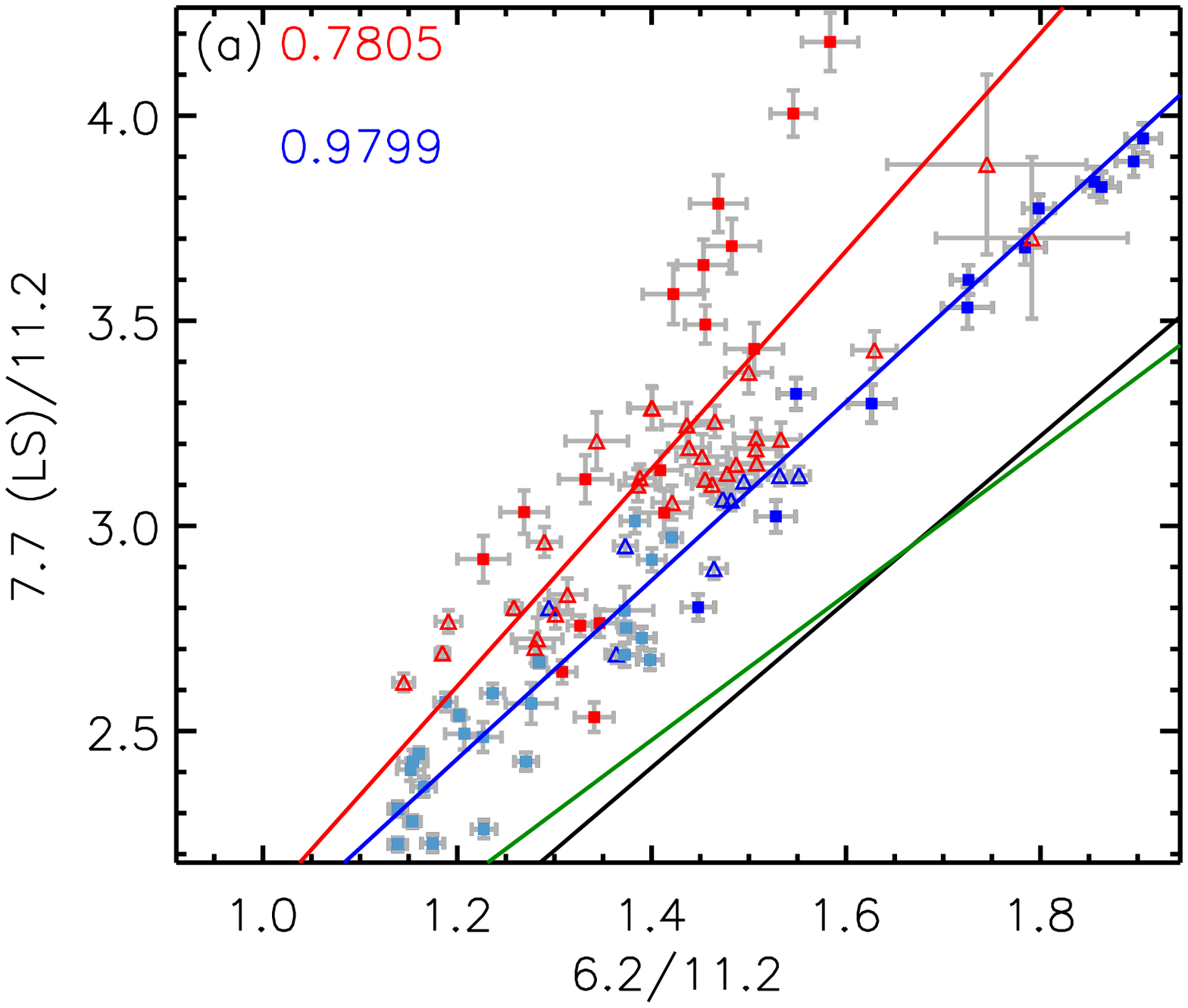}
\includegraphics[clip,trim =0cm 0cm 0cm 0cm,width=0.33\textwidth]{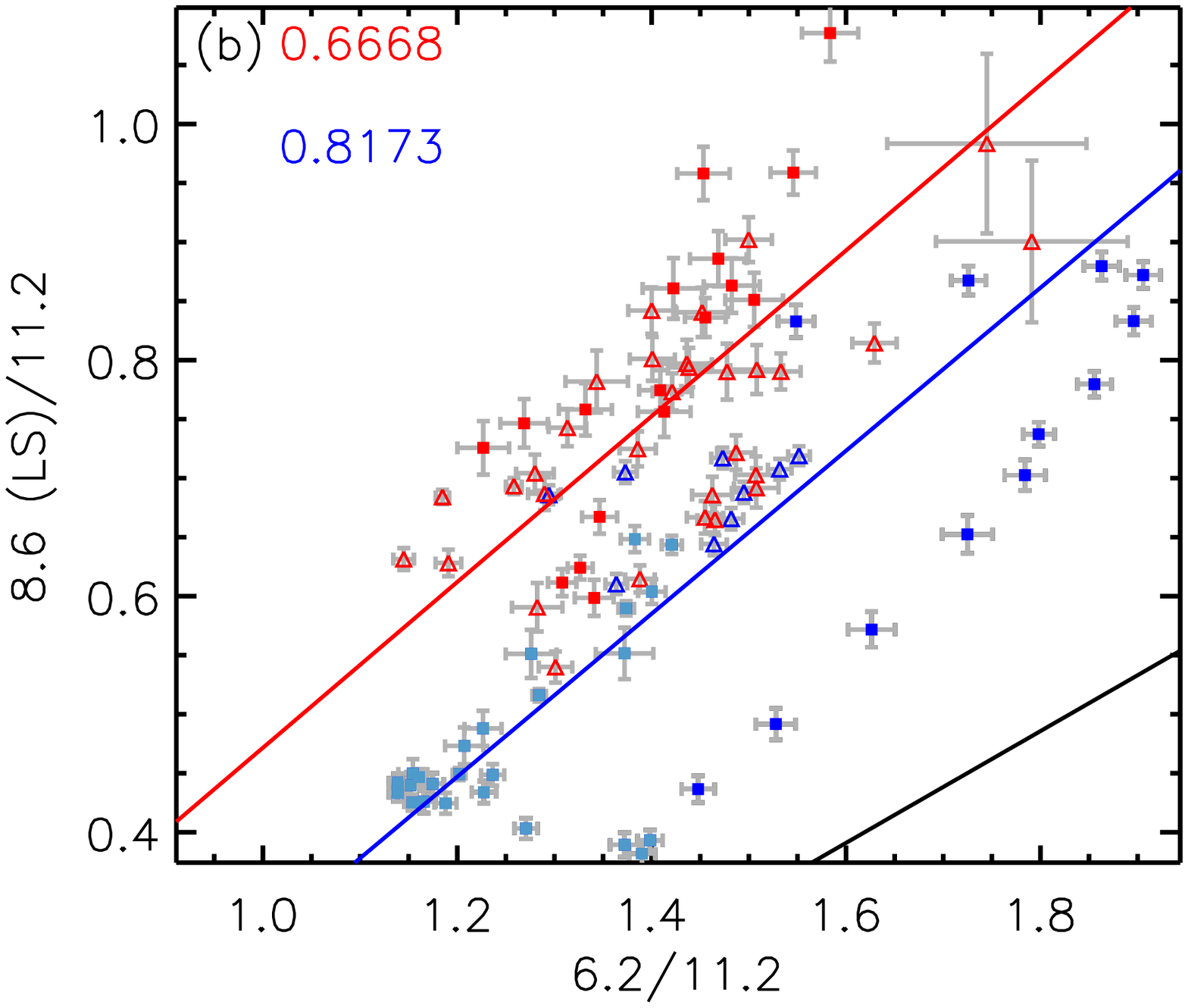}
\includegraphics[clip,trim =0cm 0cm 0cm
 0cm,width=0.33\textwidth]{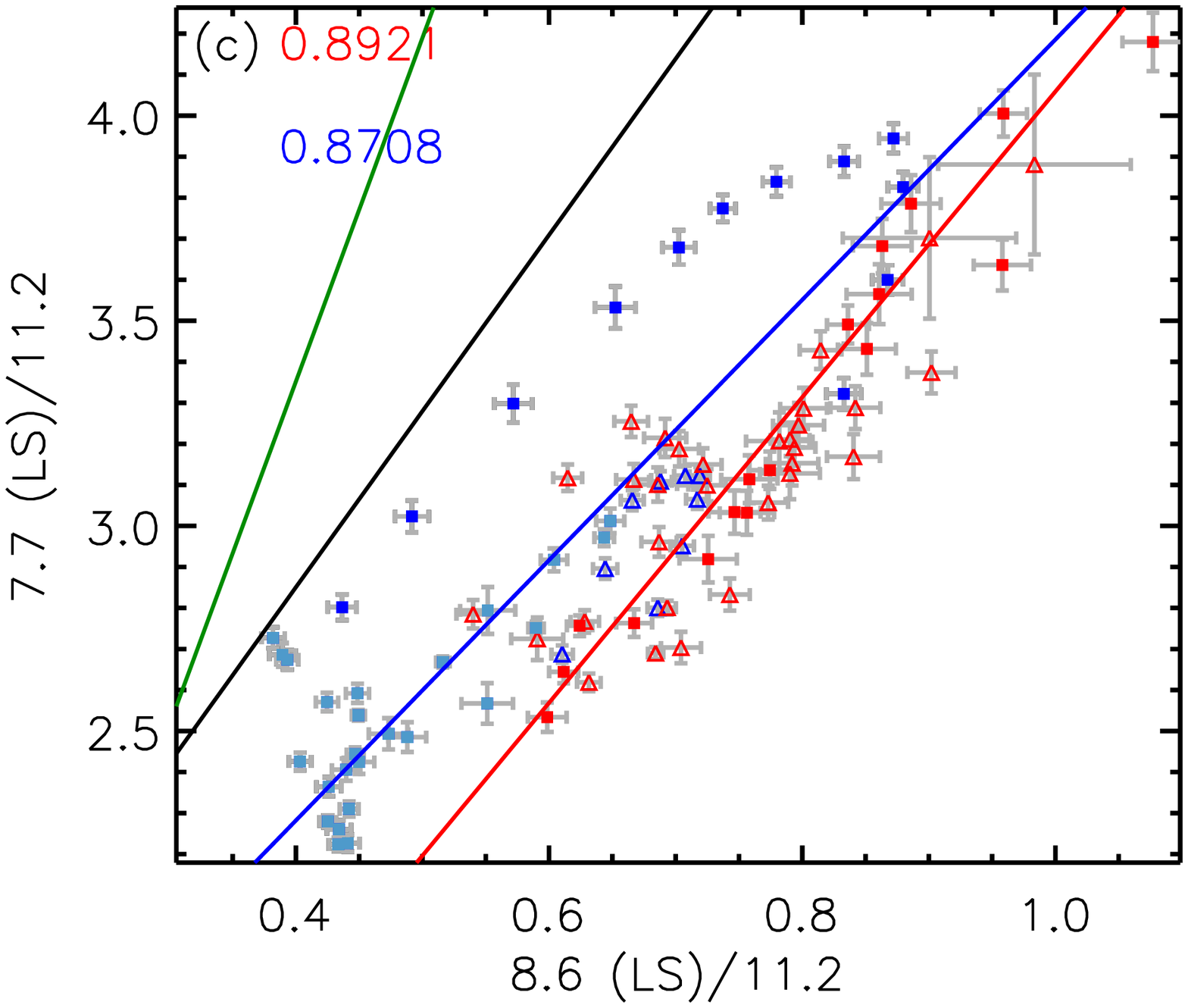}}
\resizebox{\hsize}{!}{%
\includegraphics[clip,trim =0cm 0cm 0cm 0cm,width=0.33\textwidth]{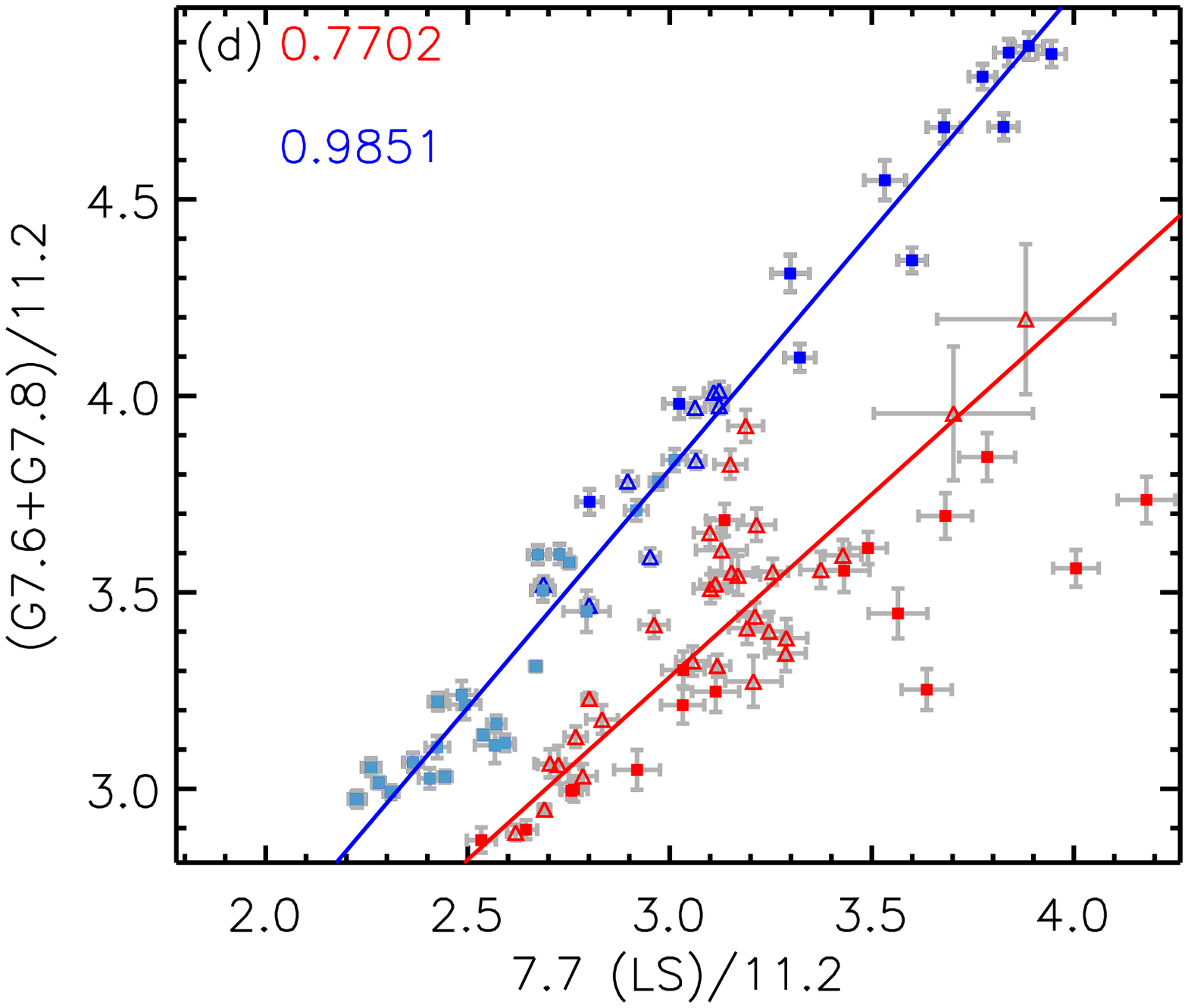}
\includegraphics[clip,trim =0cm 0cm 0cm 0cm,width=0.33\textwidth]{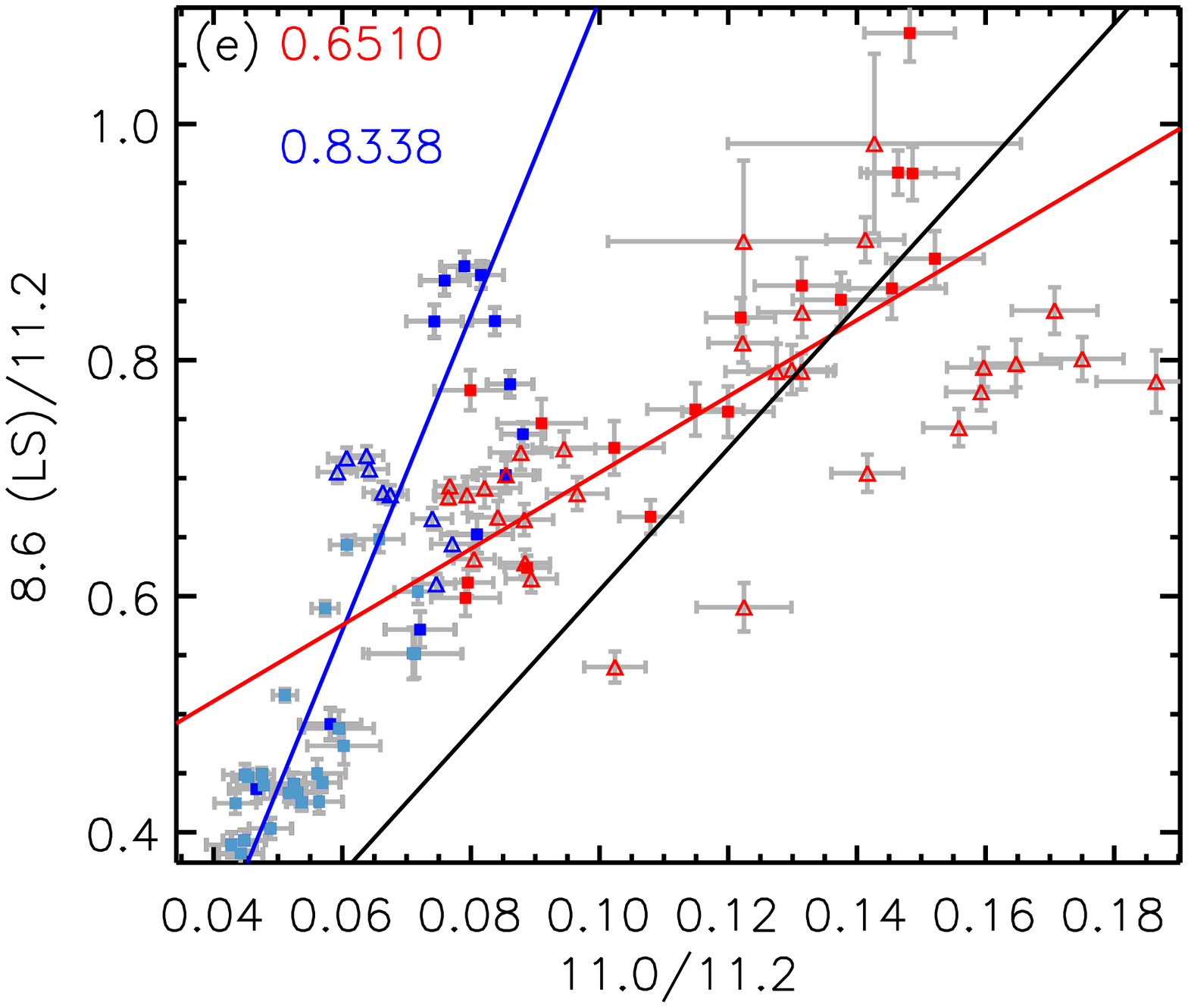}
\includegraphics[clip,trim =0cm 0cm 0cm 0cm,width=0.33\textwidth]{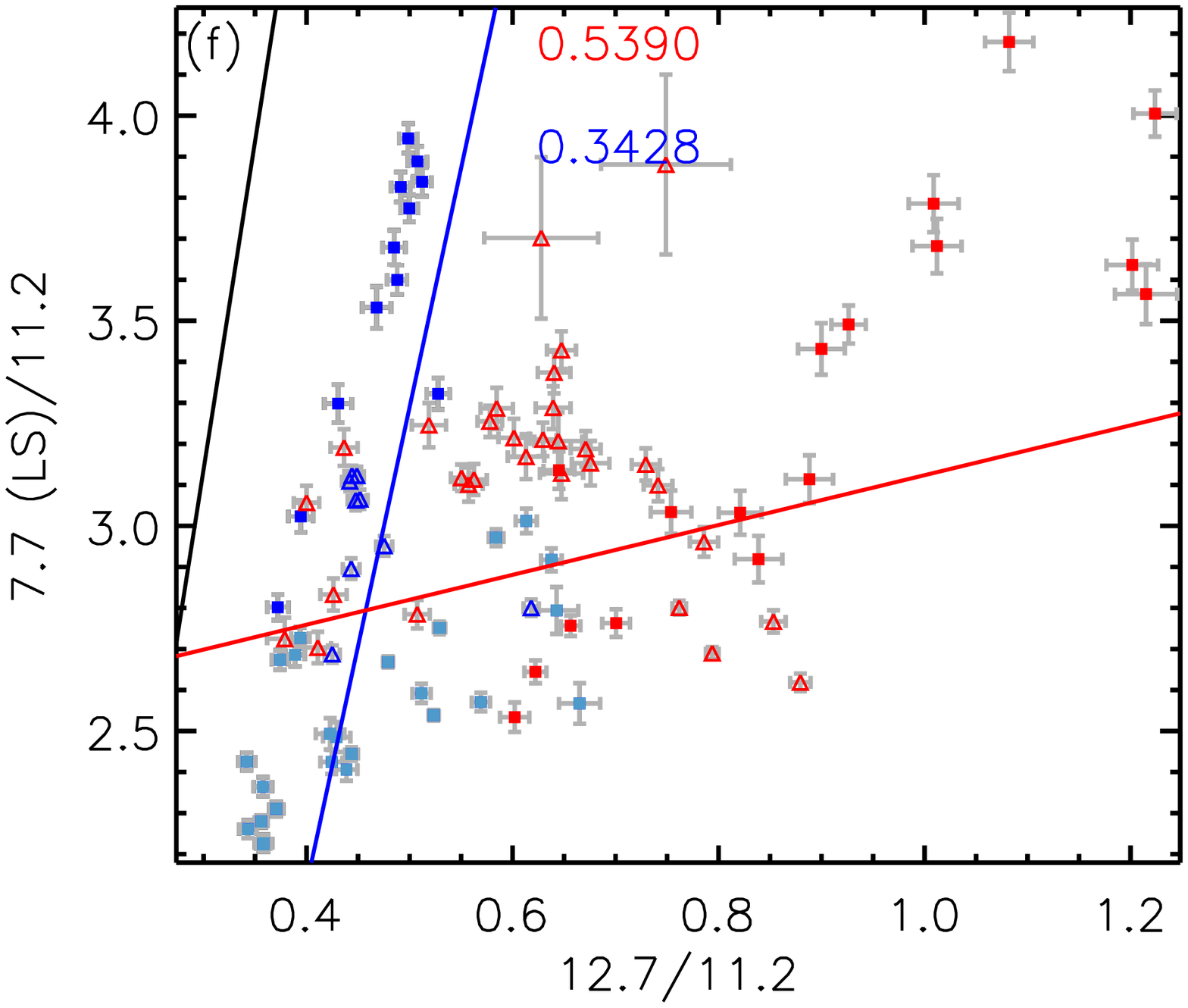}}
\includegraphics[clip,trim =0cm 0cm 0cm
0cm,width=0.33\textwidth]{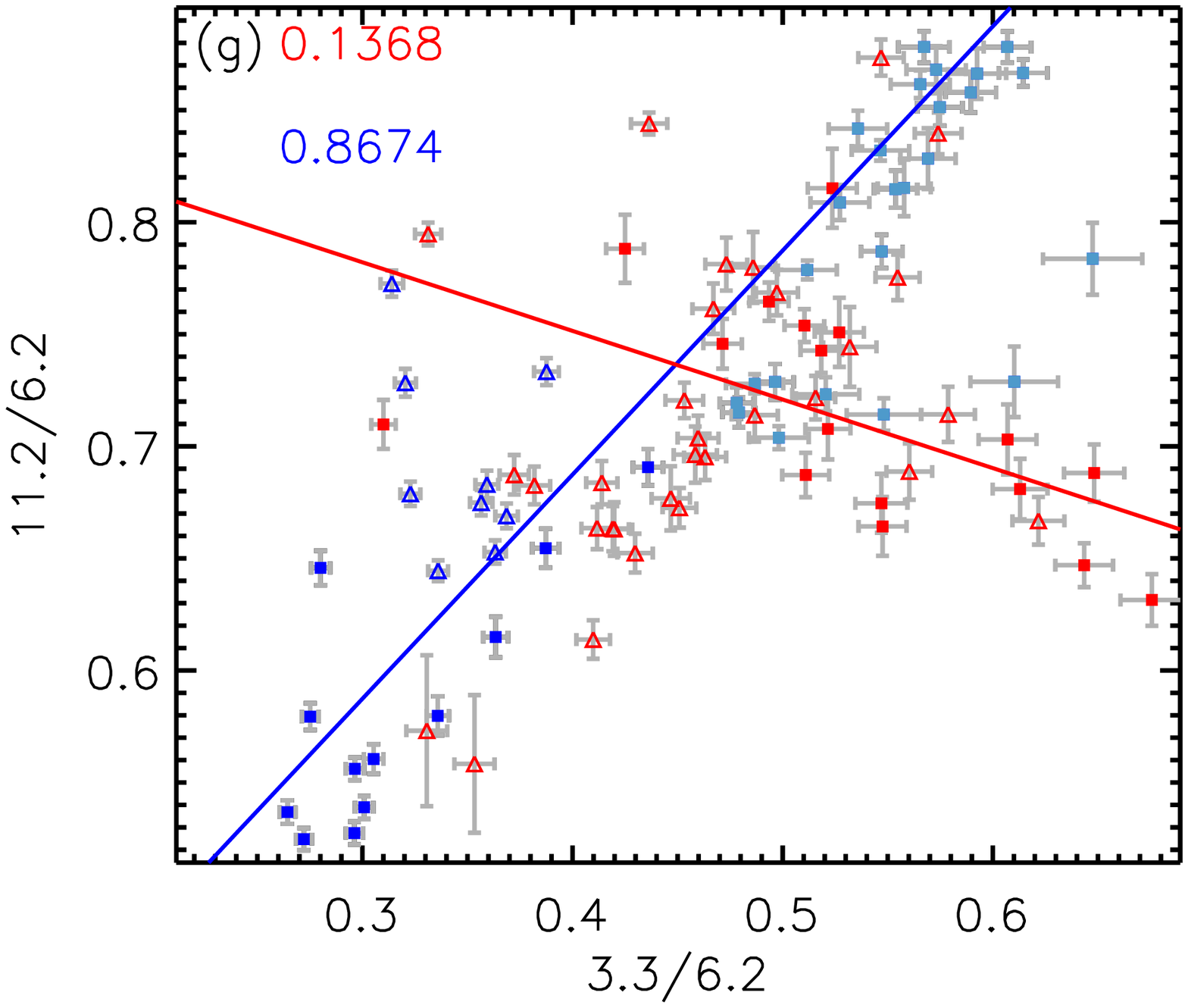}
\end{center}
\caption{Correlation plots within the Orion Bar Combined (OBC, squares) and the Orion Bar Ionized (OBI, triangles) I: ratios from within the Orion Bar PDR are shown in blue, ratios from the PDR spectra behind the Orion Bar PDR front in light blue ( > 131.5$^{\prime\prime}$ from the Trapezium), and ratios from the \HII\, region (PDR) in red (see shaded areas in panel (a) of Figure~\ref{orion_cont_10_14}). Correlation coefficients for the PDR behind the Orion Bar IF (i.e both blue and light blue data points) and the \HII\, region (PDR; red data points) are given in blue and red respectively. Weighted linear fits are shown as solid lines for each respective region given in the same colors as the correlation coefficients. The black and green lines correspond to the respective correlation fits found for NGC~2023 in \citet{pee17} and the Orion Bar using the spline method in \citet{gal08} respectively.}
\label{orion_corr1}
\end{figure*}

\subsection{Correlation plots}
\label{corr}

We investigate potential intensity correlations between major PAH features as well as the 7--9~$\mu$m Gaussian components (Figures~\ref{orion_corr1}, ~\ref{orion_corrD1},~\ref{orion_corrD2}; Appendix~\ref{corr_add}). We separate our data into two groups based on the relative position to the IF: 1) the \HII\, region (PDR, shown in red) and 2) the PDR spectra behind the Orion Bar IF as described in Sections~\ref{ob} and~\ref{proj} and illustrated by the shaded regions in panel (a) of Figure~\ref{orion_cont_10_14}. For comparison, we include the reported correlation fits for the RN NGC~2023 \citep[black line,][]{pee17} and the Orion Bar \citep[green line,][]{gal08}.

We observe modest to strong correlations between the 6.2, 7.7, and 8.6~$\mu$m bands with the degree of correlation depending on the environment (Figure~\ref{orion_corr1} (a, b, c)). Specifically, the 6.2 and 7.7~$\mu$m bands as well as the 6.2 and 8.6~$\mu$m bands are strongly correlated for the PDR spectra behind the Orion Bar IF and significantly weaker correlated within the \HII\, region (PDR). In contrast, the 8.6 vs 7.7~$\mu$m bands correlate similarly strong in both environments.
In addition, a separation is present between data points from the \HII\, region (PDR) and the PDR behind the Orion Bar IF, albeit with some overlap. This separation is most pronounced when comparing the (G7.6 + G7.8)~\mum\, bands directly with the 7.7~$\mu$m band (Figure~\ref{orion_corr1} (d)). Indeed, a bi--modal distribution is present with the  \HII\, region (PDR) data fit being shallower and located below the fit for the PDRs behind the IF. 
We also note that numerous data points from the PDR behind the Orion Bar IF are located well below (for the 6.2 vs 8.6~\mum\, bands) or above (for the 8.6 vs 7.7~\mum\, bands) the line of best fit for the PDR behind the Orion Bar IF.

As the behaviour of the 11.0~$\mu$m in the PDR behind the Orion Bar IF and \HII\, region (PDR) tends to change drastically, correlation plots involving the 11.0 $\mu$m band show two distinct distributions (Figures~\ref{orion_corr1}~(e),~\ref{orion_corrD1}). Of these, the 8.6~$\mu$m shows the best correlation with the 11.0~$\mu$m band (Figure~\ref{orion_corr1} (e)), with strong to moderate correlations in the PDR behind the Orion Bar IF and \HII\, region (PDR) respectively.
We do not find any strong correlations with the 12.7~$\mu$m band (Figures~\ref{orion_corr1}~(f),~\ref{orion_corrD2}) but note that the \HII\, region (PDR) and the PDR behind the Orion Bar IF behave very different again, as demonstrated by the 7.7 and 12.7~$\mu$m bands (Figures~\ref{orion_corr1} (f)). 
Similarly, most PAH features do not show a strong correlation with the 3.3 $\mu$m PAH feature (Figure~\ref{orion_corrD2}) with the exception of the 11.2 $\mu$m PAH feature which correlates very well with the 3.3 $\mu$m PAH feature beyond the Orion Bar PDR front (Figure~\ref{orion_corr1} (g)). The \HII\, region (PDR) exhibit two groups, one following the correlation seen beyond the IF front and one seemingly opposite to it. We note that most ``outliers" from the correlation seen beyond the IF front occur in the OBC slit at the dust emission peak (3.3/6.2~$\gtrsim~5.4$ with $0.63~\gtrsim~11.2/6.2~\gtrsim~0.71$). As the \HI\, recombination line peaks very sharply at this position, expected (enhanced) free-free continuum emission in the SOFIA 3.3 $\mu$m filter will decrease the PAH contribution to this filter. For these data points to agree with the observed correlation beyond the PDR front, a decrease in the PAH contribution to the SOFIA filter of $\sim$30\% is required at the dust emission peak (with respect to the fraction seen at other positions). This is consistent with the \HI\, recombination emission as it decreases by $\sim$30\% (relative to its peak) at the pointings of the two ISO-SWS positions that straddle the dust emission peak (Table~\ref{table:sofia}, Figure~\ref{orion_line_profiles1}a). Hence, the deviations from the correlation is likely due to a varying and uncorrected PAH contribution to the SOFIA 3.3 $\mu$m filter in the \HII\, region (PDR).

\section{Discussion}
\label{discussion}

We presented the behaviour of the emission features observed towards the Orion Bar (Section~\ref{results}). A stratified structure is clearly present between the various atomic lines tracing the ionized gas, the PAH emission, and the H$_2$ emission. While all PAH related components peak at the same distance from the illuminating source, variations in relative strengths are present in both cross cuts. In general, these relative variations are consistent with well established PAH characteristics though we do report unexpected  behaviour. This is best exemplified by enhanced scatter and bi-model distributions in the presented correlation plots. In this section, we investigate potential drivers of this behaviour such as the environmental conditions and properties of the underlying PAH populations. Specifically, we discuss the effect of the PDR viewing angle in Section~\ref{PDR morph}. We investigate the PAH size dominating the PAH emission in the edge--on PDR in Section~\ref{PAH size}, the behaviour of the ionic PAH bands in Section~\ref{spatial seq}, and the characteristics of the dust continuum, silicate, and plateau emission in Section~\ref{dust and plateau}. In Section~\ref{PDR}, we explore the diagnostic power of PAHs as PDR tracers.   

\subsection{The influence of PDR Viewing Angles}
\label{PDR morph}

In Section~\ref{results}, we separated the Orion spectra from both apertures into two groups based on the relative position of each pixel to the edge--on IF of the Orion Bar (as depicted in Figure~\ref{orion_cont_10_14} (a) by the shaded areas). Using this grouping, the correlations found are significantly worse than reported for the Orion Bar and other PDR sources in the literature \citep[e.g.][]{gal08,boe14a,sto16,sto17,pee17}. To investigate the origin of these weaker correlations, we explore the effect of the different PDR viewing angles (i.e. face--on versus edge--on) present within the two slits. Indeed, as discussed in Section~\ref{ob}, we detect PDR emission coming from the direction of the \HII\, region (PDR) which is associated with the face--on PDR on the backside of the \HII\, region, PDR emission associated with the edge--on PDR of the Orion bar, and PDR emission beyond $\sim$136$^{\prime\prime}$ which is associated with the face--on PDR created by illumination of the backside gas by (likely) $\theta^2$ Ori A. In Section~\ref{UV_fit}, we determine the depth of the penetration of UV photons into the Orion Bar as a means to quantify the transition from primarily an edge--on to a face--on PDR orientation. In Section~\ref{corr_morph}, we compare correlations between the 6--9 $\mu$m PAH bands using different pixel groupings based on PDR viewing angles. 
To summarize the ensuing discussion, the edge--on PDR dominates the observed PAH emission up to 19$^{\prime\prime}$ beyond the PAH emission peak. The PAH characteristics of the two edge--on PDRs (one in the OBC slit and one partially probed in the OBI slit) are distinct reflecting the highly structured nature of the PDR interface \citep{Goicoechea:16}. In addition, PAH emission characteristics from the edge--on PDR in the OBC slit are unique indicating the importance of the PDR viewing angle.

\subsubsection{Determining the depth of the edge--on PDR}
\label{UV_fit}

\begin{figure}
\begin{center}
\includegraphics[clip,trim =1cm .8cm 1cm 1cm,width=7.5cm]{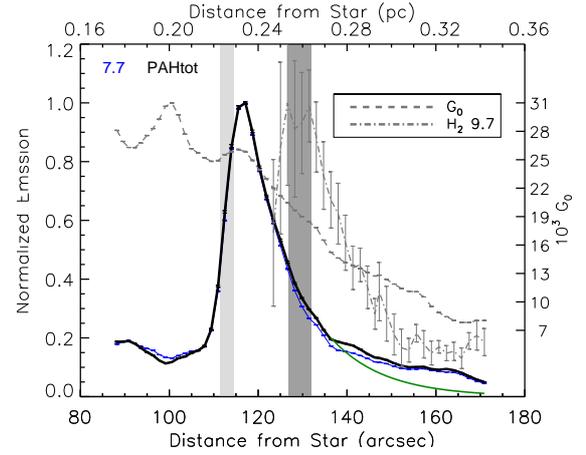}
\end{center}
\caption{The expected decrease in total PAH emission into the Orion Bar PDR due to the decrease in UV photons available for excitation for the Orion Bar Combined aperture (green; see Section~\ref{UV_fit} for details). The normalized total PAH emission and, for reference, the 7.7~$\mu$m emission cross cut are shown in respectively blue and black. G$_{0}$ cross cut values are shown on the right y-axis in units of 10$^{3}$ Habings (see Section~\ref{PDR calc} for derivation). The dark and light grey shaded region correspond to the Orion Bar PDR front and the IF respectively. }
\label{orion_combined_UV_extinction}
\end{figure}

In order to understand the effects of the different PDR viewing angles on the relative behaviour of the PAH emission features in the FOV, we estimate the relative contribution of the edge--on and face--on PDRs to the PAH emission deeper into the Bar and behind the PDR front (i.e. for distances larger than $\sim$ 117$^{\prime\prime}$ from $\theta^1$ Ori C). To this end, we quantify the effect of the extinction of UV photons from dust grains into the edge--on PDR and thus the decrease in available energy for PAH excitation. Following \cite{sal16} and beginning at the PAH peak in the Bar at 117$^{\prime\prime}$, we fit the total PAH emission as a function of distance {\it s} by:

\begin{equation}
F_{\textrm{PAH}}(\textrm{s}) = F_{\textrm{PAH}}(\textrm{peak}) \times e^{\textrm{-k} \ (\textrm{s} - \textrm{s}_{\textrm{peak})}},
\label{uv_ext_eq}
\end{equation}

\noindent where k is a free parameter and the optical depth is defined as $\tau_{\textrm{UV}} = \textrm{-k} \ (\textrm{s} - \textrm{s}_{\textrm{peak}})$. This exponentially decreasing trend traces the observed total PAH emission very well out to about 136$^{\prime\prime}$ (Figure~\ref{orion_combined_UV_extinction}). 
Beyond 136$^{\prime\prime}$, equation~\ref{uv_ext_eq} increasingly deviates from the observed total PAH emission. Taking the ratio of the total PAH emission predicted due to UV extinction and the observed total PAH emission, we find that 86\% of the PAH emission between 117$^{\prime\prime}$ (the PAH peak) and 170$^{\prime\prime}$ (the largest distance from the illuminating source in the OBC slit) can be attributed to the edge--on PDR where the UV photon flux, and thus the PAH excitation, exponentially attenuates with depth. From 136$^{\prime\prime}$ to 170$^{\prime\prime}$, only 45\% of the total emission is accounted for by the edge--on PDR. The 55\% of excess emission thus arises from the face--on PDR, as shown in diagrams depicting the structure of the Orion Nebula \citep[Figure~\ref{Orion_schematic}; e.g.][]{boe12,wer13,pab19}. 
We note that we are unable to quantify the contribution of the face--on PDR to the observed emission in the \HII\, region (PDR).

\subsubsection{Effect on PAH correlations}
\label{corr_morph}

\begin{figure*}
\begin{center}
\resizebox{\hsize}{!}{%
\includegraphics[clip,trim =1cm 1.5cm 1cm 0cm,width=10.5cm]{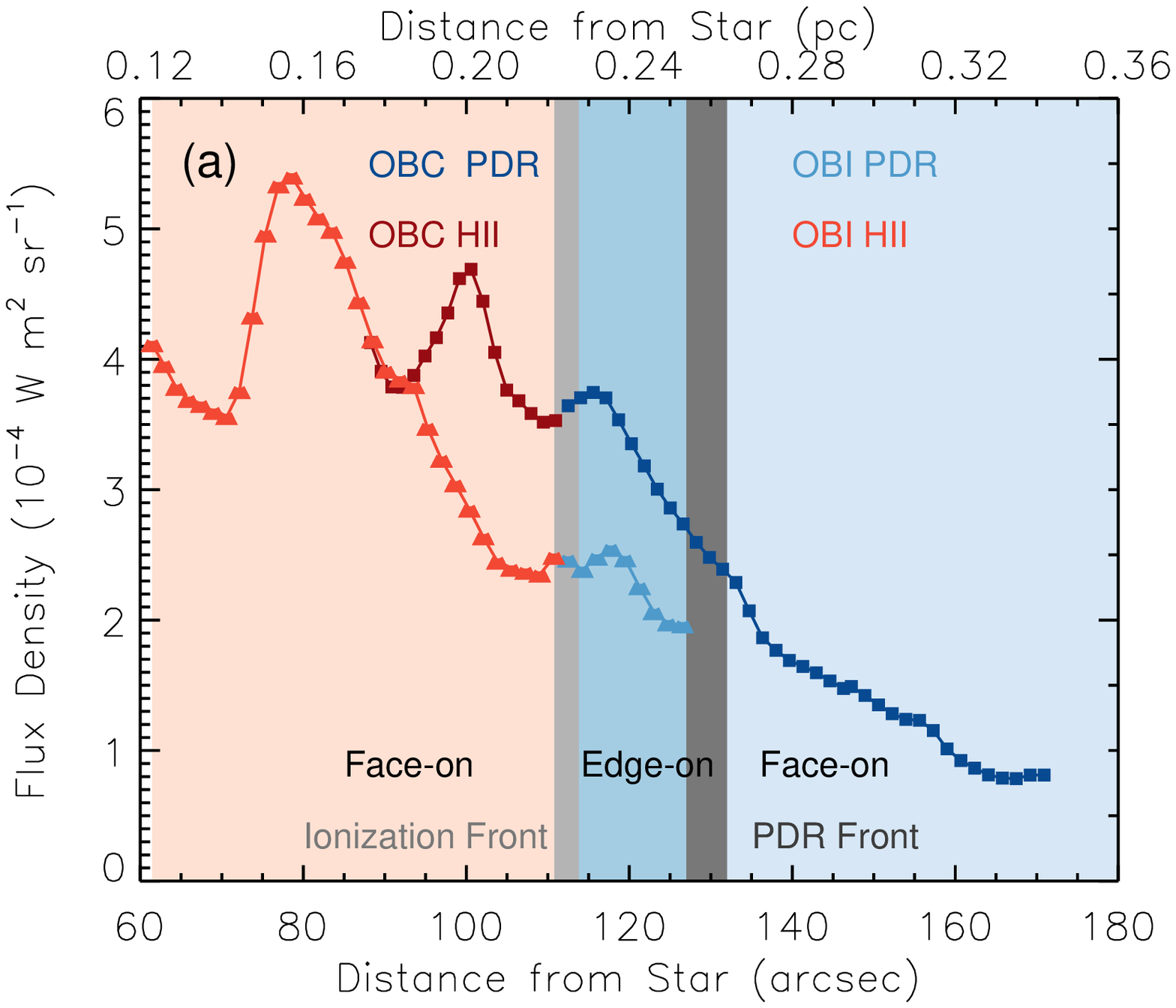}
\includegraphics[clip,trim =1cm 1.5cm 1cm 0cm,width=10.5cm]{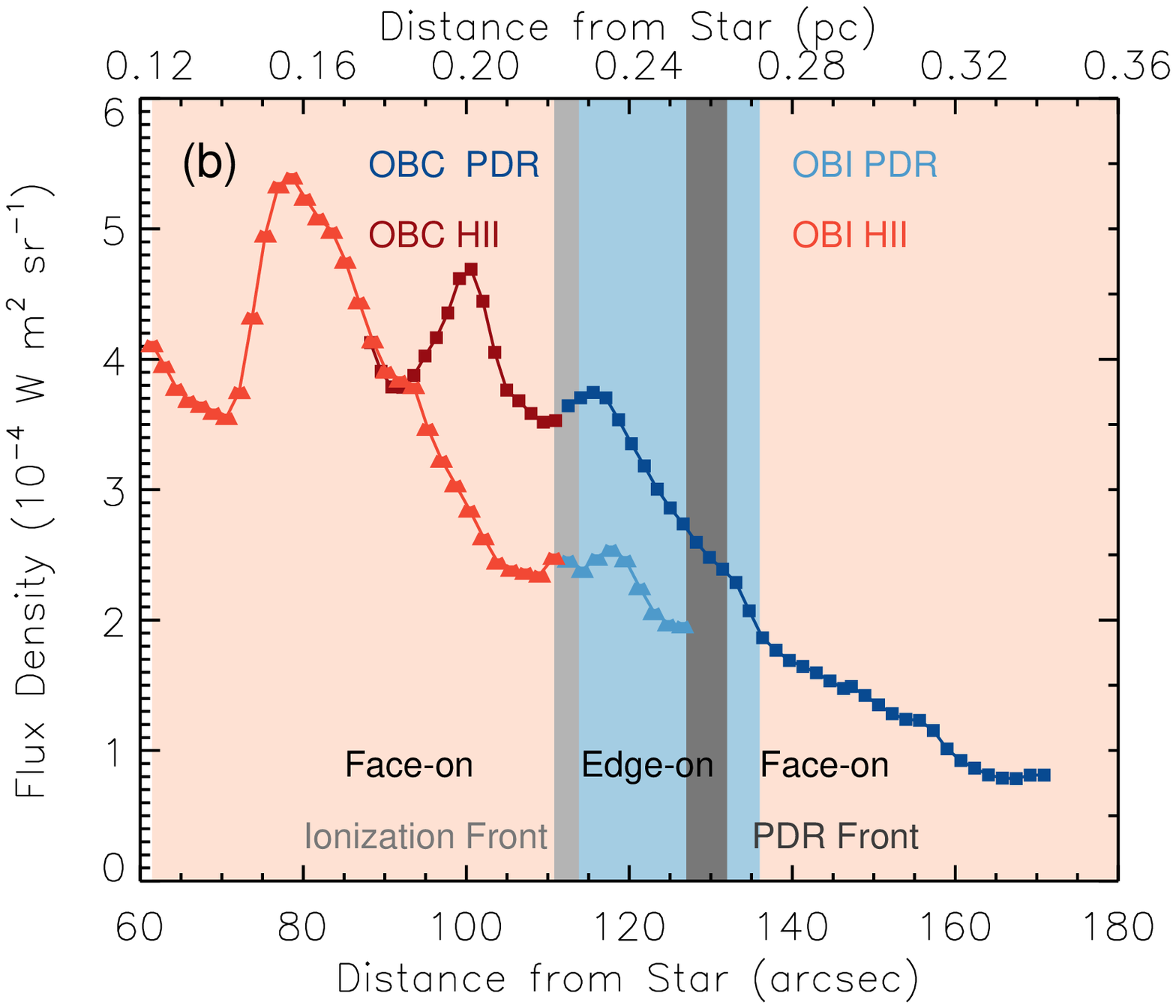}
\includegraphics[clip,trim =1cm 1.5cm 1cm 0cm,width=10.5cm]{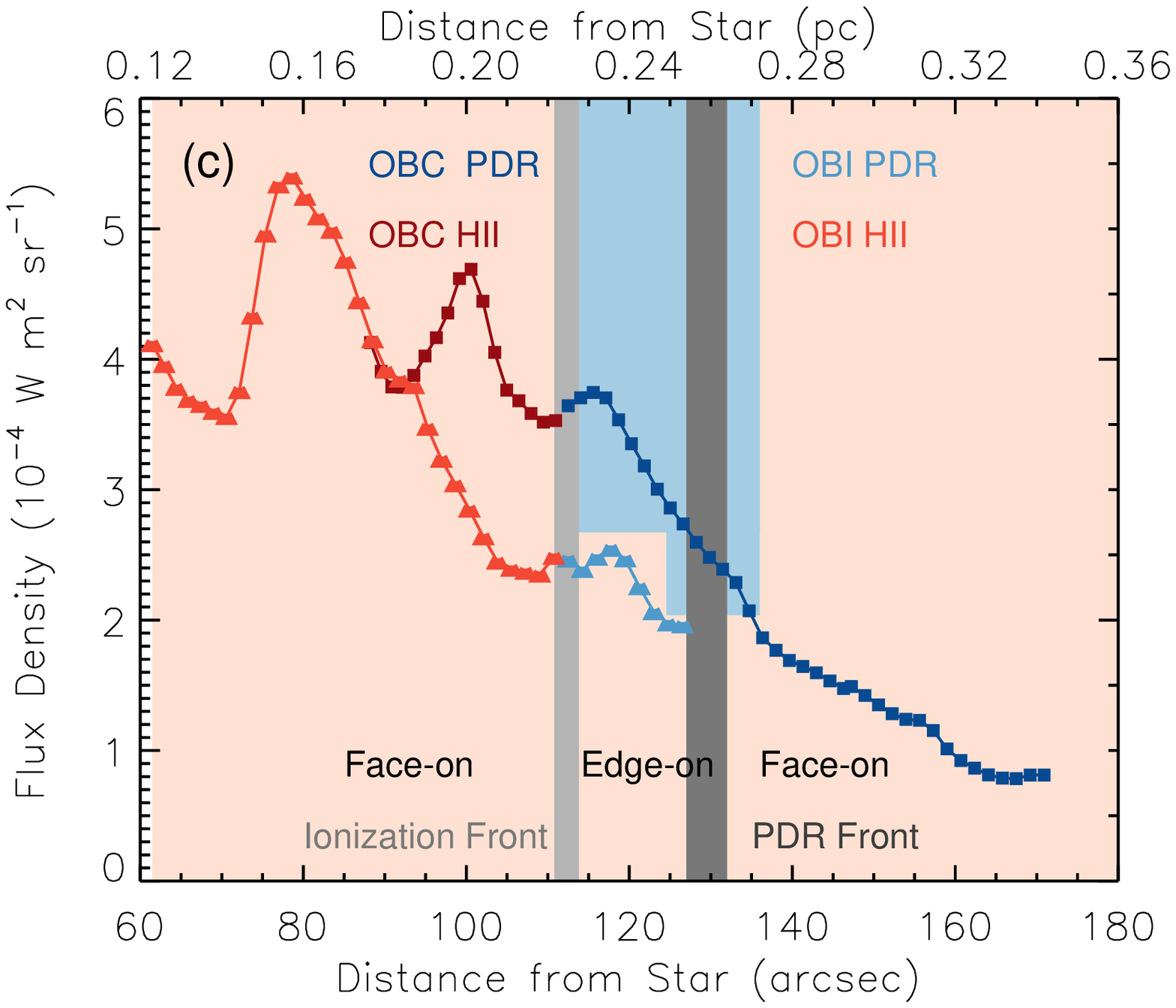}
}
\end{center}
\caption{The 10--13.2~$\mu$m continuum cross cut for Orion Bar combined (squares; dark colors) and Orion Bar ionized (triangles; light colors) with the pixels located beyond the IF and pixels in the \HII\, region (PDR) represented in respectively blue and red symbols. The shaded areas highlight the pixel groupings and thus the color coding used for Figures~\ref{orion_corr1},~\ref{orion_corrD1}, and~\ref{orion_corrD2} (panel a) and Figure~\ref{orion_corr3} (panel b and c). The dark and light grey shaded regions correspond to the Orion Bar PDR front and the IF (see Figure~\ref{orion_line_profiles1}). Face--on and edge--on labels refer to regions to be dominated by said PDR viewing angle as detailed in Sections~\ref{ob} and~\ref{PDR morph}.}
\label{orion_cont_10_14}
\end{figure*}

\begin{figure*}
\begin{center}
\resizebox{\hsize}{!}{%
\includegraphics[clip,trim =0cm 0cm 0cm
 0cm,width=0.25\textwidth]{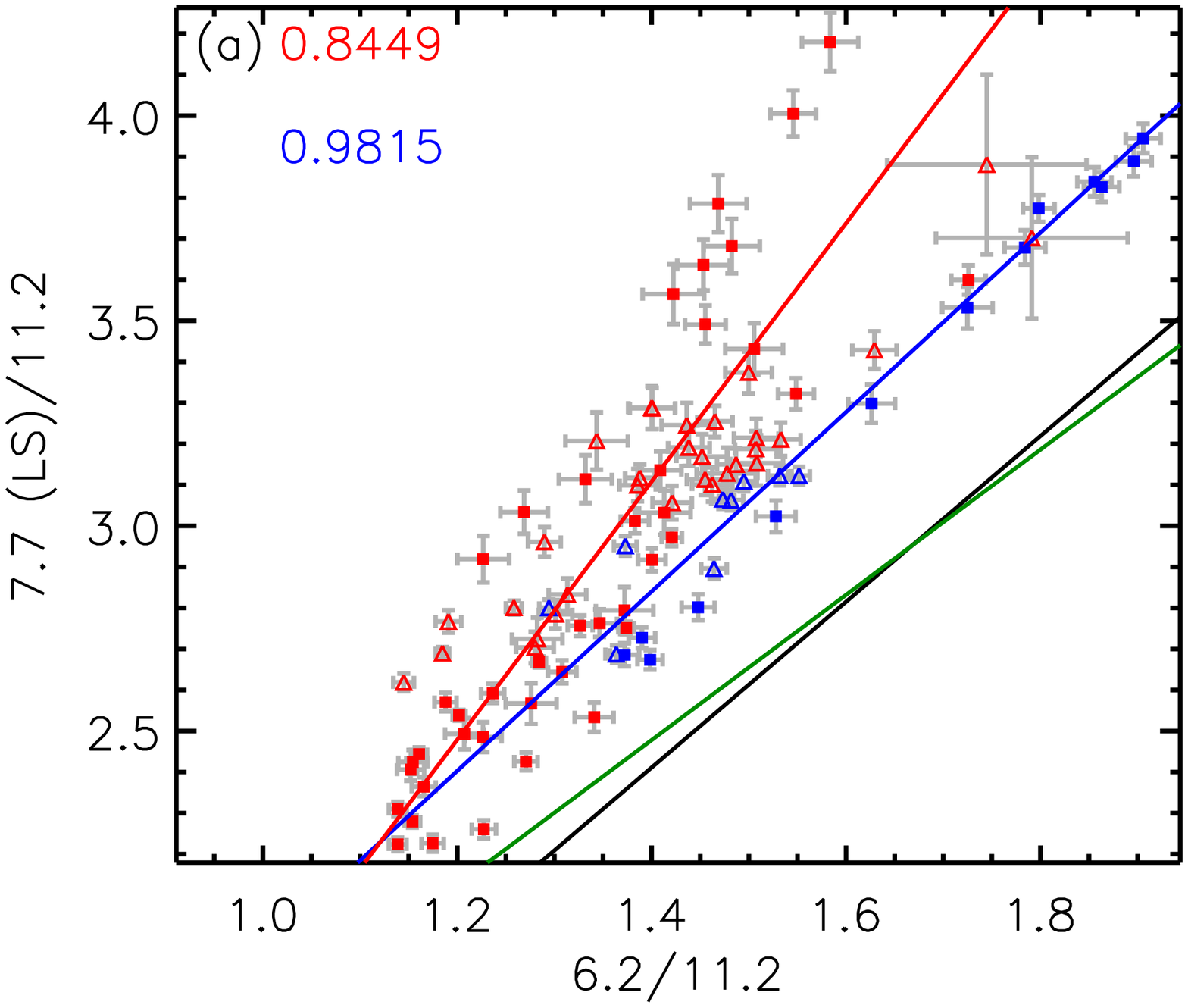}
\includegraphics[clip,trim =0cm 0cm 0cm 0cm,width=0.25\textwidth]{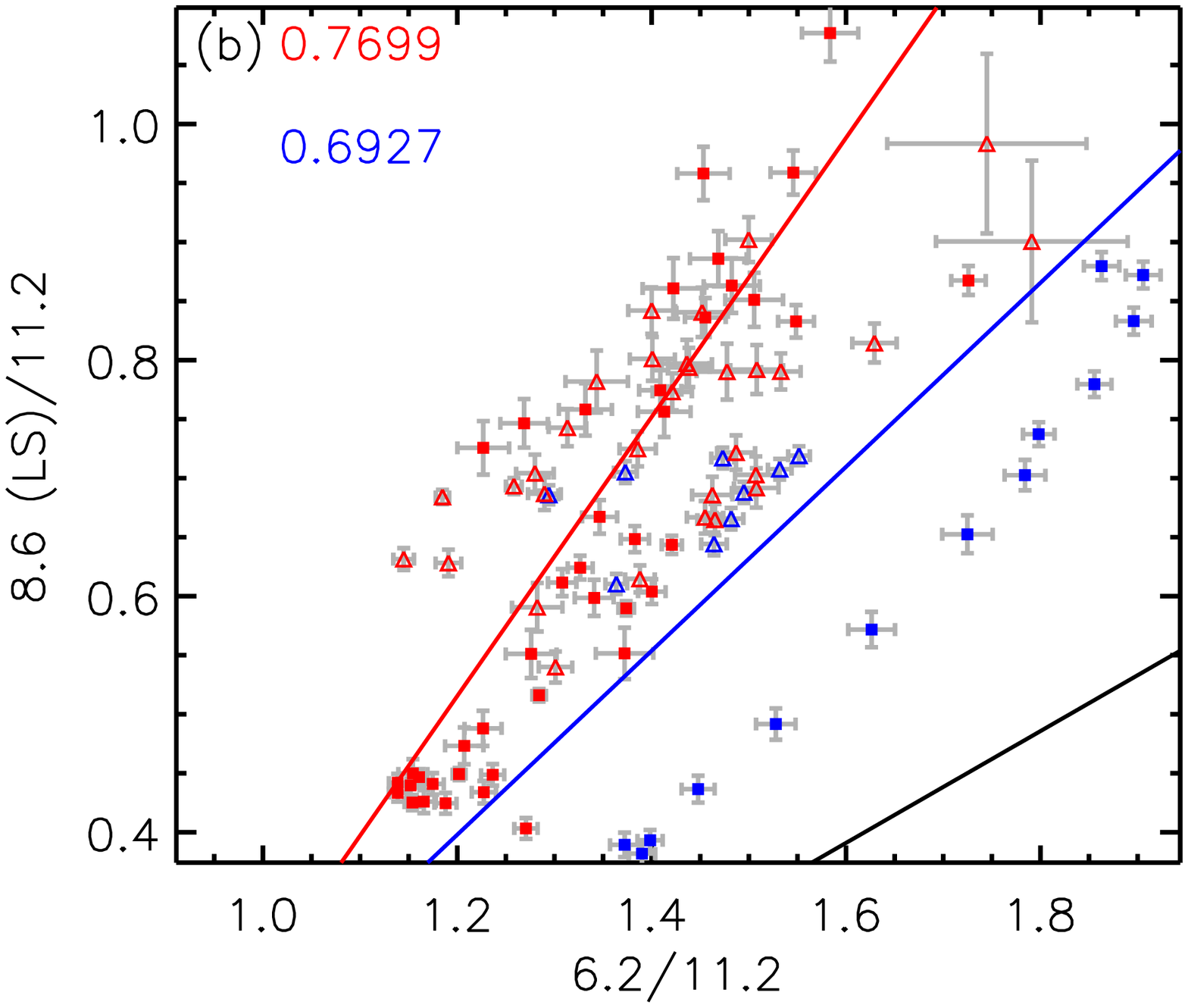}
\includegraphics[clip,trim =0cm 0cm 0cm 0cm,width=0.25\textwidth]{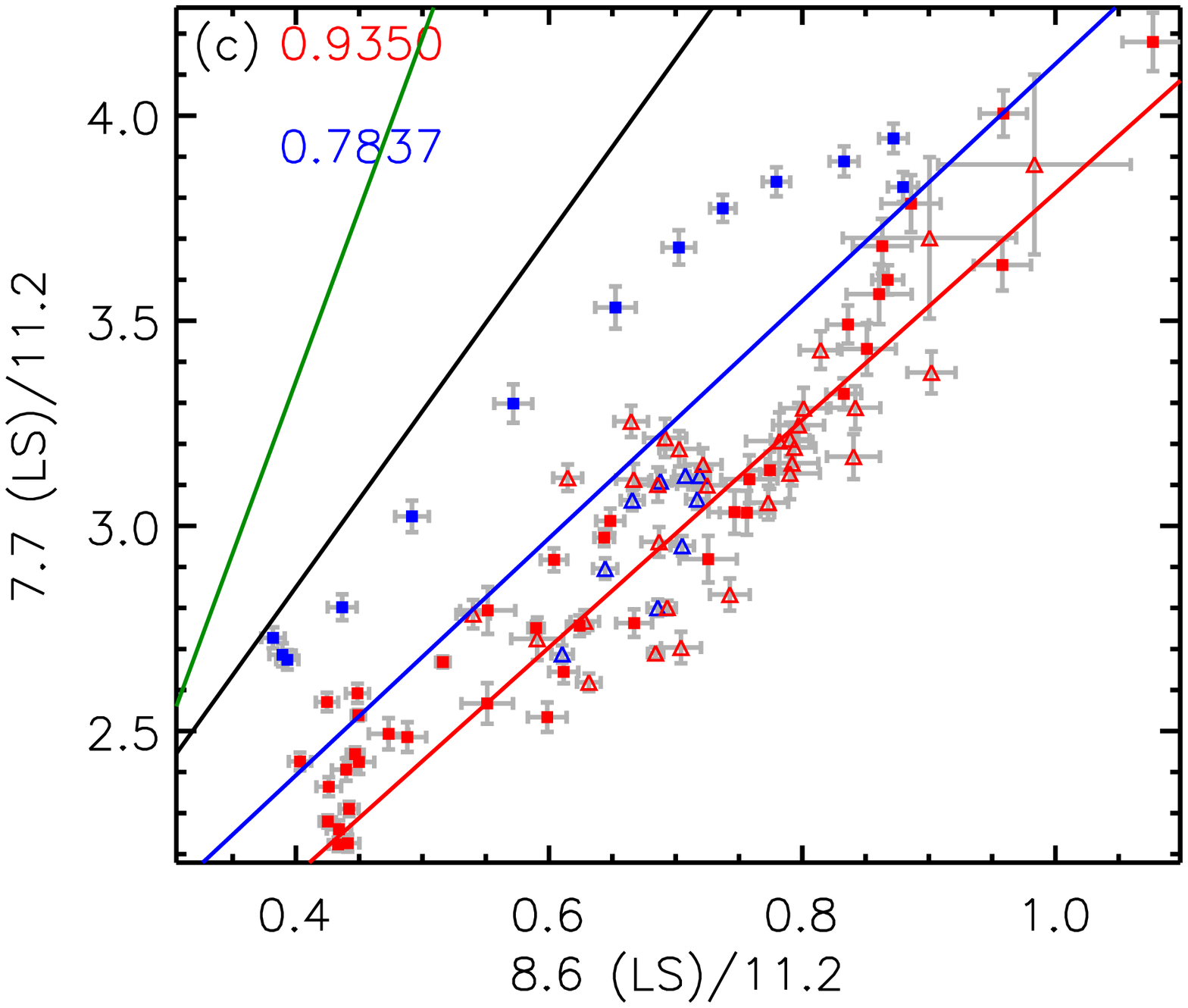}
}
\resizebox{\hsize}{!}{%
\includegraphics[clip,trim =0cm 0cm 0cm
 0cm,width=0.25\textwidth]{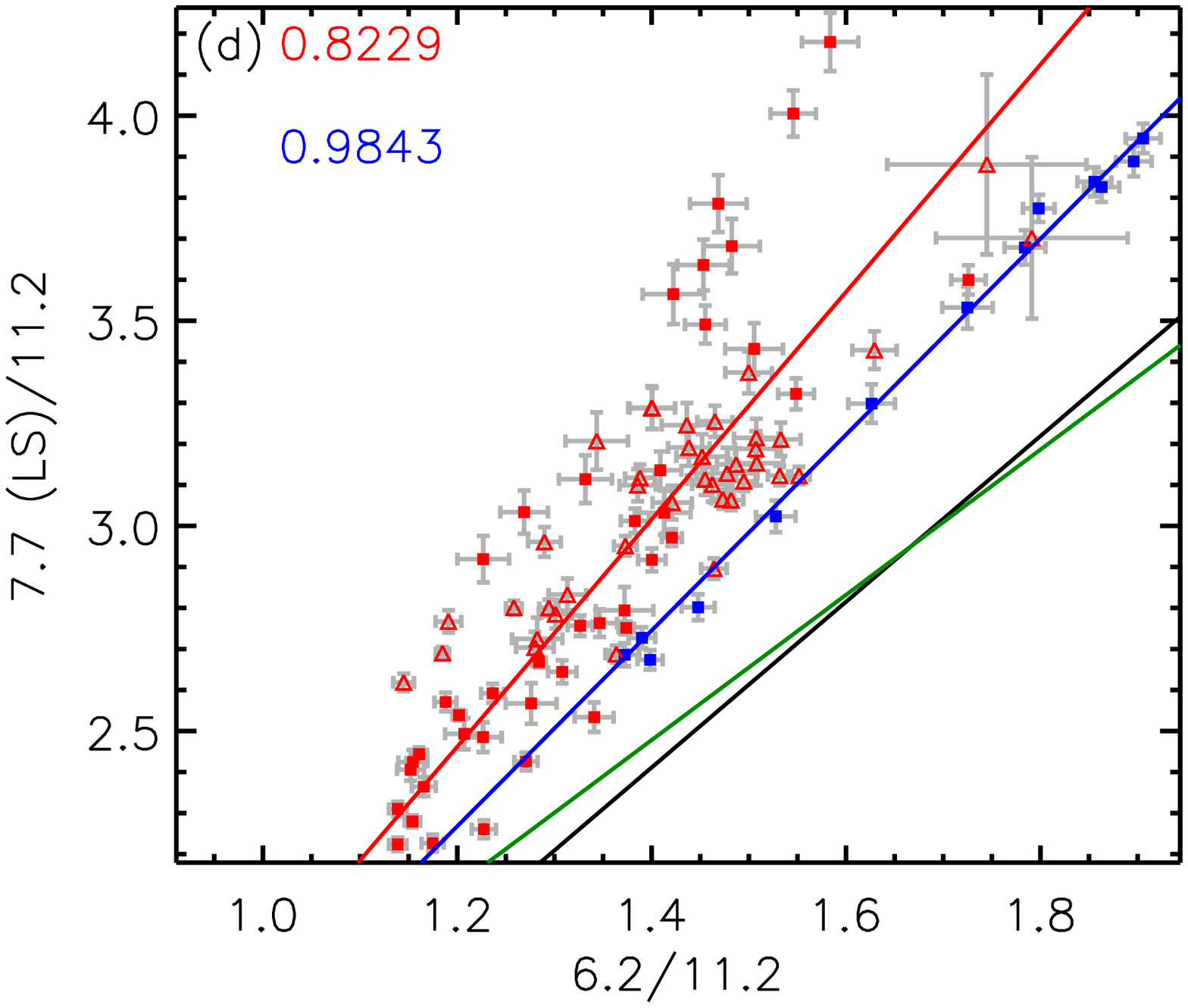}
\includegraphics[clip,trim =0cm 0cm 0cm 0cm,width=0.25\textwidth]{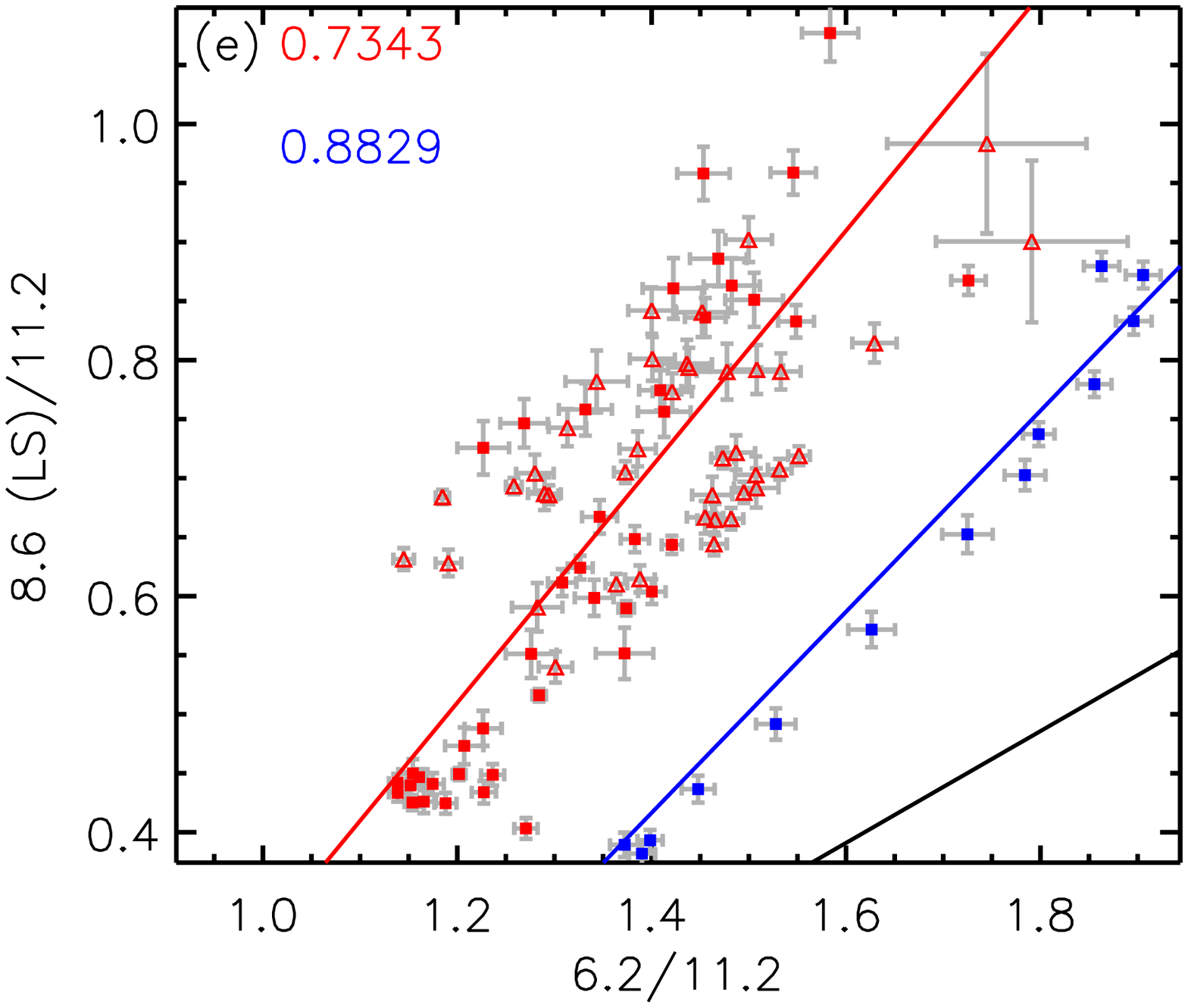}
\includegraphics[clip,trim =0cm 0cm 0cm 0cm,width=0.25\textwidth]{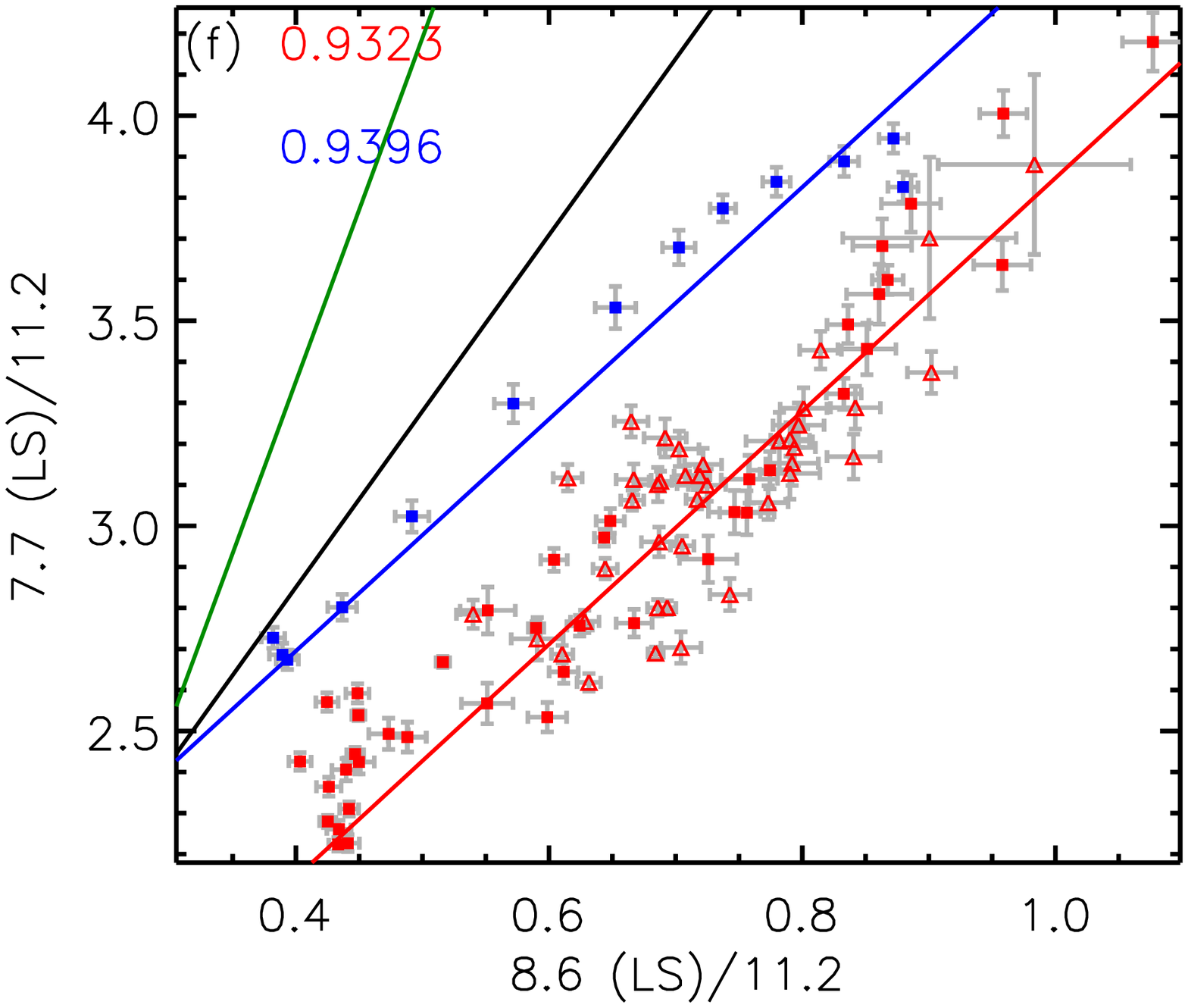}
}
\end{center}
\caption{Correlation plots within the Orion Bar Combined (squares) and the Orion Bar Ionized (triangles) II:  
Edge--on and face--on PDR data points are shown in respectively blue and red in the top row (as depicted in Figure~\ref{orion_cont_10_14} panel b). In the bottom row, OBC edge--on PDR data points are shown in blue while OBI edge--on PDR and all face--on PDR data points are shown in red (as depicted in Figure~\ref{orion_cont_10_14} panel c). 
Correlation coefficients as well as weighted linear fits (solid lines) for each grouping are given in their respective color. The black and green lines correspond to the correlation fits found for NGC~2023 \citep{pee17} and the Orion Bar using the spline method \citep{gal08} respectively.
}
\label{orion_corr3}
\end{figure*}

We now re--visit the observed PAH correlations applying a different grouping: one group includes only data points from the edge--on PDR, bounded by the IF and 136$^{\prime\prime}$ distance from the illuminating source, and the second group representing the remaining data points of i) the face--on PDR and cavity located in front of the IF (i.e. the \HII\, region (PDR)), and ii) the face--on PDR beyond 136$^{\prime\prime}$ (as depicted by the shaded areas in panel b of Figure~\ref{orion_cont_10_14}). The resulting correlations between the 6--9~$\mu$m PAH emission features are shown in Figure~\ref{orion_corr3} (top row). Overall, this new grouping better represents two distinct behaviours in the observed correlations but some overlap between the two groups still remains. In addition, the correlation coefficients are higher considering all face--on PDRs pixels (as traced in red in Figure~\ref{orion_corr3}) in comparison to considering only the \HII\, region (PDR) pixels (as traced in red in Figure~\ref{orion_corr1}). This suggests that the PAH emission at projected stellar distances of more than 136$^{\prime\prime}$ agrees better with the PAH emission arising from within the \HII\, region (PDR) than with the PAH emission from the edge--on PDR of the Orion Bar. 

This employed grouping (i.e. edge--on PDR data points in blue and face--on PDR data points in red) also exhibit some mixing in Figure~\ref{orion_corr3} (top row): i.e. a few blue data points are located with the red data points. We therefore employ a third grouping: one group representing only the OBC edge--on PDR pixels and the second group representing the remaining pixels, including the face--on PDR pixels {\it and} the OBI edge--on PDR pixels (as depicted by the shaded areas in panel c of Figure~\ref{orion_cont_10_14}).  This grouping clearly represent two distinct behaviours in the PAH emission without any confusion (Figure~\ref{orion_corr3}, bottom row). Indeed, the bi--linear trend now clearly separates the OBC edge--on PDR emission from the remainder. The addition of the OBI edge--on PDR emission to the face--on PDR emission results in a slight decrease in its correlation coefficients relative to our second grouping. This suggests that while it is more similar to the PAH emission of a face--on PDR, the PAH emission from the OBI edge--on PDR may still be slightly distinct.
The distinct behaviour of the edge--on PDR in both slits is puzzling. It likely arises from the detailed structure on small spatial scales within the slits. Indeed, in contrast to the transition from the ionized region to the PDR, the molecular emission (HCO$^+$ and CO) from the Orion Bar displays a fragmented ridge of high-density substructures \citep{Goicoechea:16}. Future observations with the James Web Telescope \citep[JWST; ERS ID 1288;][]{pee18}\footnote{https://stsci.edu/jwst/observing-programs/approved-ers-programs/program-1288} will be able to explore the PAH emission on similar smaller spatial scales. 

Based on these Spitzer observations, we conclude that the viewing angle of the PDR influences the observed PAH correlations. In the case of Orion, different PDR viewing angles results in relationships between PAH ratios which are almost parallel but offset from each other (Figure~\ref{orion_corr3}). This is most pronounced in correlations involving the 8.6 \mum\, PAH emission.  Specifically, either there is additional 8.6 \mum\, emission in the face--on PDR compared to the edge--on PDR or there is additional 6.2 and 7.7 \mum\, emission in the edge--on PDR compared to the face--on PDR or both. This extra component seem to be roughly constant for all pixels. We note that also for correlations involving the 8.6 \mum\, PAH emission, two distinct distributions are present within the face--on PDRs and the OBI edge--on PDR pointings (Figure~\ref{orion_corr3} (e)). The higher distribution (i.e. with a larger y-offset) arises from the face--on PDR in front of the Bar (i.e. the \HII\, region (PDR)) where the strength of the UV radiation field impinging on the PDR \citep[G$_0 \sim 10^5$,][]{tie85b} is largest. Unique behaviour for the OBC edge--on PDR is also seen in correlations involving the 11.0 and 12.7 \mum\, bands (Figures~\ref{orion_corr1}, ~\ref{orion_corrD1}, ~\ref{orion_corrD2} ). The observed relationships in PAH ratios differ both in offset and slope between the OBC edge--on PDR and the remaining pixels.

Other studies have reported the presence of bi--linear trends or bifurcation between different PAH emission features \citep[e.g.][]{boe14a, sto16, sto17}. This bifurcation has been attributed to the distinct physical environments of low G$_{0}$ diffuse ISM and high G$_{0}$ \HII\, regions. These reported bi--linear trends involving the 6.2, 7.7, and 8.6 \mum\, PAH bands are however distinct from the results of this paper: these authors observe a difference in slope in these correlations while we mainly find a difference in offset. This is consistent with a different origin of the reported bifurcation (FUV radiation field intensity versus the PDR viewing angle). 
      
\cite{gal08} studied ISOCAM observations of the Orion Bar which included the ionized region, the Orion Bar, and the region behind the PDR front. These observations thus cover all the different regions discussed in this paper. While these authors also find tight correlations between the 6.2 versus 7.7 and the 7.7 versus 8.6~$\mu$m bands, their correlation is offset from our data points (Figure~\ref{orion_corr1}, panels a and c). 
The origin of the offset between the trends in both data sets of the Orion Bar is currently unclear and warrants further investigation. Such offsets have been seen between sources \citep[e.g.][]{sto16, sto17, and18, marag18}. For example, the correlations in the reflection nebula NGC~2023 \citep{pee17} are in many cases displaced from our trends albeit parallel (Figures~\ref{orion_corr1},,~\ref{orion_corr3}). In particular, the NGC~2023 correlations involving the 11.0 and 12.7 \mum\, PAH bands exhibit a similar slope to those from the OBC edge--on PDR. Given that these NGC~2023 observations largely cover edge--on PDRs or filaments, this supports our conclusion that the PDR viewing angle influences the relationships between PAH ratios. However, \cite{sto16} reported that the spherical symmetric reflection nebula NGC~1333 shows correlations that closely mimic those of NGC~2023 which is thus not consistent with our conclusion. This suggests that other parameters contribute to the observed relationships as well. \\

\subsection{Probing PAH size across the edge--on PDR}
\label{PAH size}

The 3.3/11.2 emission ratio is a well known tracer of PAH size \citep[e.g.][]{sch93,ric12,cro16, Maragkoudakis:20}. In this section, we derive the average PAH size across the edge--on PDR and conclude that intense UV fields lead to increased photo-processing of PAHs destroying the smallest PAHs.

\cite{ric12} employed the NASA Ames PAH IR spectroscopic database \citep[PAHdb,][]{bau10,boe14b} to calculate the intrinsic emission spectrum for compact-symmetric PAH species over a wide range of sizes with average absorption photon energies of 6 and 9~eV. These authors show a clear inverse relationship between the 3.3/11.2 emission ratio and PAH size for both the coronene and ovalene families probing the range of sizes and structures thought to be prevalent in space. In addition, these authors demonstrate how a higher absorbed average photon energy increases the intrinsic 3.3/11.2 emission ratio for a given PAH size. Consequentially, the theoretical and observed 3.3/11.2 emission ratios can be compared to obtain an estimate for the PAH size dominating the observed PAH emission. 

This approach has been used to study the PAH population in other spatially resolved Galactic MIR bright sources. For the reflection nebulae NGC~7023 and NGC~2023, \citet{cro16} and \citet{Knight:21} reported that the average PAH size reaches a minimum at the PDR front and increases towards the illuminating star. In addition, NGC~2023 has greater average PAH sizes than those found for NGC~7023 \citep{Knight:21}. Both results indicate that PAH size depends on the radiation field intensity \citep{Knight:21}. We note that \citet{Knight:21} did not detect significant variation in the average PAH size found in the face-on PDR beyond the Orion Bar.  

\begin{figure}
\begin{center}
\includegraphics[clip,trim =0cm 0cm 0cm
 0cm,width=8.5cm]{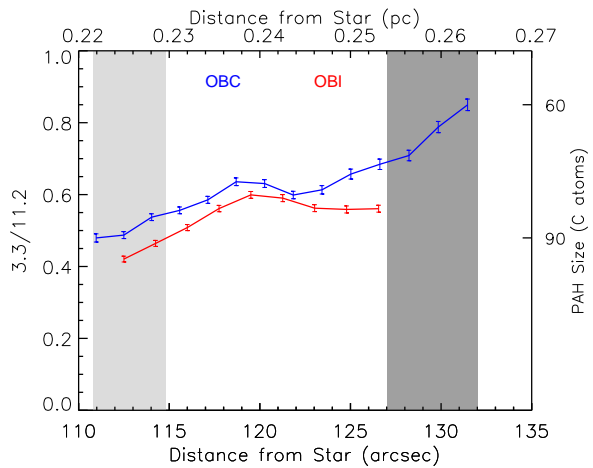}
\end{center}
\caption{The 3.3/11.2 PAH ratio across the edge--on PDR in the Orion Bar Combined (blue) and Orion Bar Ionized (red) slits.  The corresponding dominating PAH size for four 3.3/11.2 values is shown on the right y-axis (see Section~\ref{PAH size} for details). The dark and light grey shaded regions correspond to the Orion Bar PDR front and the ionization front (see Figure~\ref{orion_line_profiles1}). }
\label{orion_pah_size}
\end{figure}

In the case of the Orion Nebula, the average photon energy from $\theta^{1}$ Ori C absorbed by PAHs is 8.1~eV \citep{Knight:21}. We compare our observed 3.3/11.2 ratios in the edge--on PDR, the Orion Bar, with the 9~eV model of \cite{ric12} to obtain the average PAH size (Figure~\ref{orion_pah_size}). Both slits exhibit a similar 3.3/11.2 emission ratio cross cut: an overall increasing 3.3/11.2 ratio and thus a decreasing PAH size starting from the IF going deeper into the PDR with a slight bump slightly behind the PAH emission peak at $\sim$~118--121$^{\prime\prime}$. We find an average PAH size of $\sim$~90 carbon atoms at the IF, $\sim$~70 carbon atoms at the PAH emission peak, decreasing to 60 carbon atoms at the PDR front in the OBC aperture. Hence, the average PAH size in the edge-on PDR increases with increasing strength of the FUV radiation field, consistent with the results in the reflection nebulae \citep{Knight:21}. This suggests that intense UV fields lead to increased photo-processing of PAHs destroying the smallest PAHs. While the derived PAH size near the IF is sufficiently large for the formation of fullerenes, the tell-tale signature of fullerenes at 18.9 \mum\, is (unfortunately) not covered by our observations.

\subsection{Ionic PAH bands}
\label{spatial seq}

Recent observations indicate subtle differences between the PAH emission bands arising from cationic PAHs (i.e. the 6.2, 7.7, 8.6, and 11.0 \mum\, bands). In this section, we report similar behaviour for the Orion Bar. 

\cite{whe13} and \cite{sto14} reported a break-down of the tight correlation between the 6.2 and 7.7 \mum\, PAH bands on small spatial scales towards the giant star-forming region N66 in the Large Magellanic Cloud and W49A in the Milky Way respectively. \cite{pee17} found the 8.6 and 11.0~$\mu$m emission features share a similar spatial morphology peaking closer to the illuminating source in NGC~2023 than the 6.2 and 7.7~$\mu$m emission, which also have a similar spatial morphology. These authors define a spatial sequence of the PAH bands characterised by the distance of their peak emission from the illuminating source. They further argue that the 7.7 $\mu$m complex traces at least two PAH sub populations, one co-spatial with the 8.6 and 11.0 \mum\, emission and the other co-spatial with the 11.2 \mum\, emission attributed to neutral PAHs. The tight correlation between the 6.2 and 7.7~$\mu$m bands then suggests that at least two PAH sub populations also contribute to the 6.2~$\mu$m band. \citet{Sidhu:20} and \citet{Sidhu:21} further advocates for the distinction between the 6.2 and 7.7 \mum\, bands on one hand and the 8.6 and 11.0~$\mu$m bands on the other based on a Principal Component Analysis (PCA) of the main band intensities within NGC~2023 and NGC~7023.  
These authors discuss two origins for these subsets: i) a distinct charge (balance) of the PAHs contributing to the emission with the 6.2 and 7.7 \mum\, carriers being less ionized than the 8.6 and 11.0 \mum\, carriers, and ii) a contribution of the VSGs and PAH clusters to the 6 to 9 \mum\, PAH emission. 

A spatial sequence or stratification of the cationic PAH bands is also observed within our data set. This is most easily discerned when considering the 6.2/11.2, 7.7/11.2, 8.6/11.2, and 11.0/11.2 PAH ratios (see panel a of Figure~\ref{orion_line_profiles3}). In particular, the relative values of these emission ratios at the dust emission peak (at 100$^{\prime\prime}$) and at the PAH emission peak (at 117$^{\prime\prime}$) changes significantly. Ordering these from high to low values, we find the following sequence: 11.0/11.2, 8.6/11.2, 7.7/11.2, and 6.2/11.2 with the 11.0 and 8.6~$\mu$m ratios having a significantly higher value in the \HII\, region (PDR). 
In addition, the 8.6/7.7, 8.6/6.2, 11.0/6.2, and 11.0/7.7 ratios as well as the 11.0/8.6 ratio are strongest in the \HII\, region (PDR; see panel b in Figure~\ref{orion_line_profiles3} and panel a in Figure~\ref{orion_line_profiles5}). 
This is consistent with the spatial sequence found in NGC~2023 \citep[i.e. the 11.0 \mum\, PAH emission peaks closest to the star, then the 8.6 \mum\, PAH emission, followed by 7.7 and 6.2~$\mu$m PAH emission;][]{pee17}.  

We also observe significant variation in the 6.2/7.7 ratio which remains roughly constant across the edge--on PDR but reaches a minimum at the dust emission peak in the \HII\, region (PDR) for the OBC aperture (panel b in Figure~\ref{orion_line_profiles3}), consistent with results from \cite{whe13} and \cite{sto14}. \cite{sto14} suggests this behaviour may arise from the different vibrational assignments of these features  \citep[the 6.2~$\mu$m band is attributed to C--C stretching whereas the 7.7~$\mu$m band is a combination of the C--C stretching and C--H in plane bending modes;][]{all89} with the C--H in plane bending mode dominating within the \HII\, region (PDR). As the 8.6~$\mu$m band is attributed {\it solely} to C--H in plane bending modes, \cite{sto14} argue that, in this case, the 8.6/7.7 ratio is expected to exhibit an increase within the ionized cavity. Such an increase was detected in their observations. 
We also detect a rise in the 8.6/7.7 ratio in the OBC aperture moving from the edge--on PDR towards the \HII\, region (PDR, panel b in Figure~\ref{orion_line_profiles3}). 
This is further reflected in the weaker correlations between the 6.2 and 7.7~$\mu$m bands in the \HII\, region (PDR) while, conversely, the 7.7 and 8.6~$\mu$m bands have a stronger correlation in this region relative to the edge--on PDR (Figure~\ref{orion_corr1}, panels a to c). 
Hence, although the PAH emission at stellar distances of more than 136$^{\prime\prime}$ (i.e. in the face--on PDR beyond the Orion Bar) agrees better with the PAH emission arising from within the \HII\, region (PDR, face--on) than that from the edge--on PDR (Section~\ref{corr_morph}), both face--on PDRs exhibit distinct behaviour in their PAH emission. 

Notably, we do not find a similar trend in the cross cut of the OBI aperture for the 6.2/7.7 ratio (panel f in Figure~\ref{orion_line_profiles3}). Similarly, despite the fluctuations of the 8.6/7.7 ratio in the OBI aperture, they are not in sync with changes in the environment.

\subsection{Dust Continuum, Silicate and Plateau Emission}
\label{dust and plateau}

As mentioned in Section~\ref{cont}, the underlying dust continuum has a much steeper rise towards longer wavelengths within the \HII\, region (PDR) compared beyond the IF \citep[see also][]{sal16}. We find this change in continuum slope to occur at roughly the position of the edge--on IF. 
The increase in the slope of the underlying continuum is indicative of the irradiation of large dust grains within the \HII\, region along the line of sight \citep{ces00}. 
These authors further report that the 9.7~$\mu$m amorphous silicate emission peaks closest to the Trapezium cluster with appreciable emission near $\theta^{2}$ Ori A behind the PDR front. 
We only detect silicate emission in the \HII\, region (PDR) peaking at the dust emission peak and dropping very sharply towards the IF.  This is consistent with the results of \cite{ces00}, indicating these silicates do not become hot enough to emit within the relatively sheltered PDR environments. Furthermore, the lack of silicate emission behind the IF is consistent with the dust in the \HII\, region being heated by trapped Lyman $\alpha$ photons \citep{sal16}.

\begin{figure}
\begin{center}
\resizebox{\hsize}{!}{%
\includegraphics[clip,trim =0cm 1.9cm 0cm 0cm,width=7.5cm]{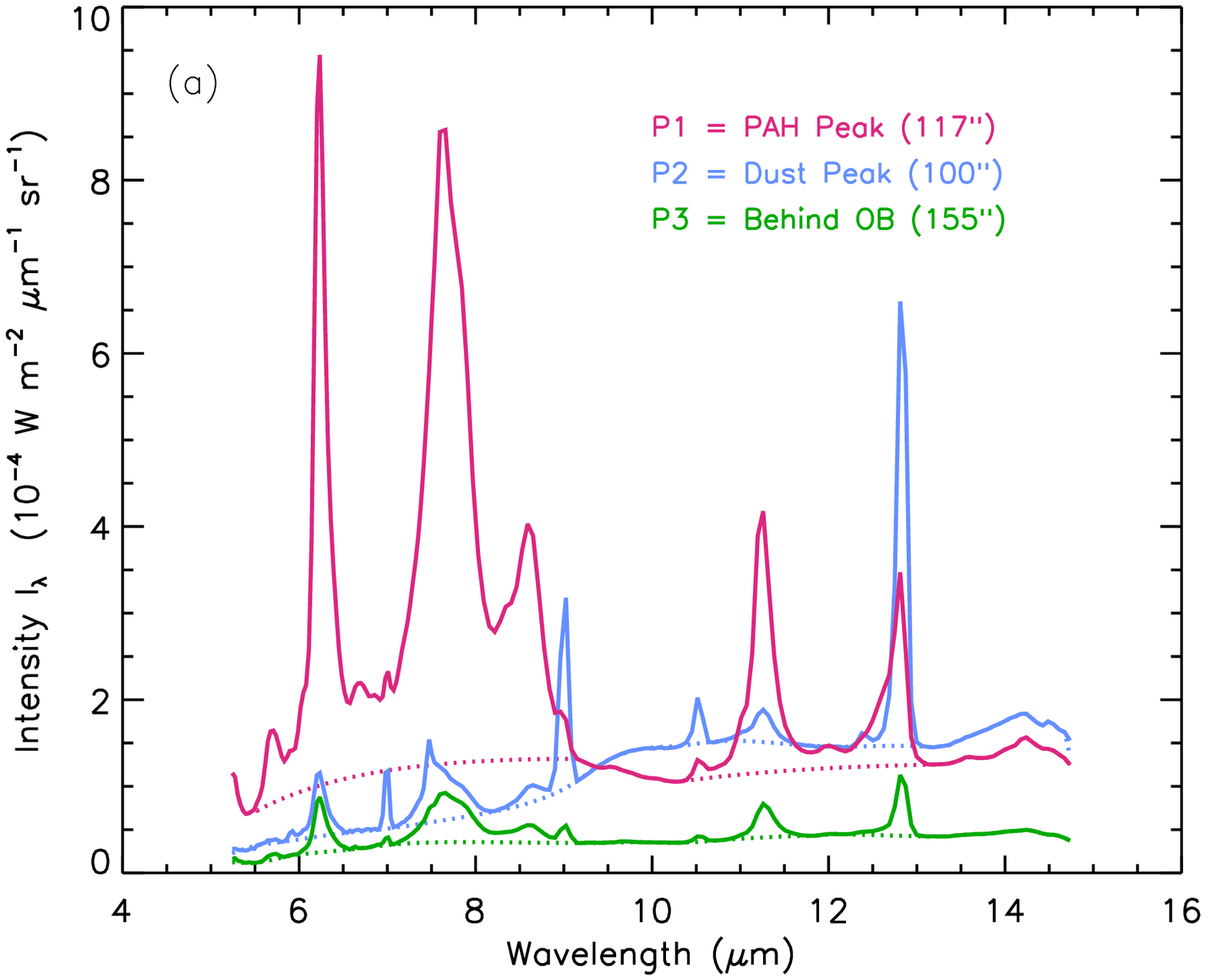}
}
\resizebox{\hsize}{!}{%
\includegraphics[clip,trim =0cm 0cm 0cm 0cm,width=7.5cm]{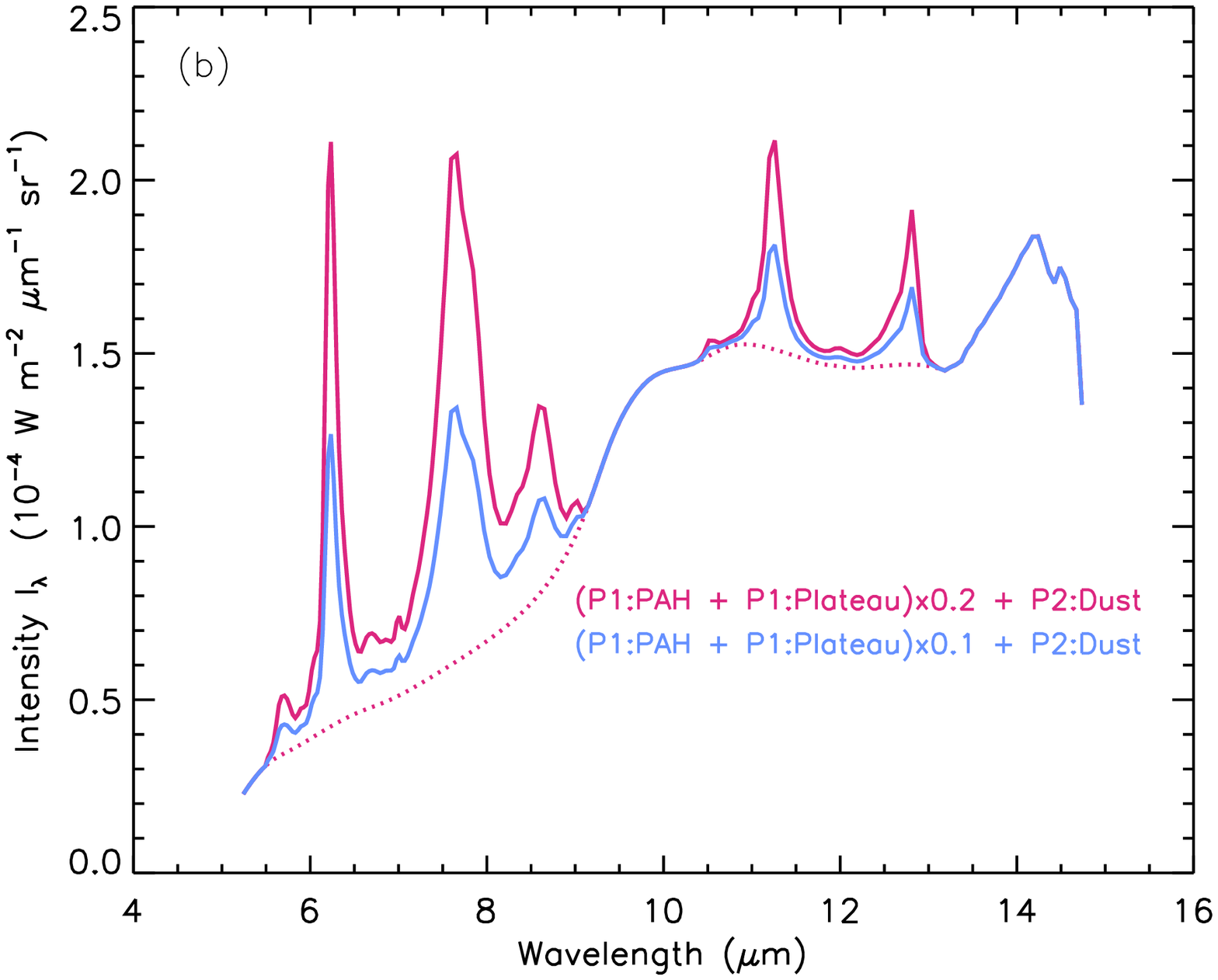}
}
\end{center}
\caption{{\it Top} A comparison of the IR emission (solid lines) and GS continuum (dotted lines) at three different locations within the OBC aperture: the PAH emission peak at $\sim$~117$^{\prime\prime}$ (P1), the dust peak at $\sim$~100$^{\prime\prime}$ (P2), and a position far behind the PDR front at $\sim$~155$^{\prime\prime}$ (P3). {\it Bottom} Spectral composites (solid lines) consisting of different fractions (10 and 20\%) of the combined PAH and plateau emission at the PAH emission peak position (P1) added to the peak dust emission at dust emission peak position (P2) and GS continuum (dotted line). }
\label{orion_pdr_v_hii}
\end{figure}

The spectral components associated with the plateaus are very weak or non-existent in front of the IF (panels b and e in Figure~\ref{orion_line_profiles1}). In particular, the 5--10~$\mu$m plateau in both apertures and the 10--13~$\mu$m plateau in the OBC aperture are not detected in front of the IF while the 10--13~$\mu$m plateau only shows marginal emission (above 3~$\sigma$) at the ``PAH bump'' in front of the IF in the OBI aperture. The very weak or lack of plateau emission coincides with an increasing strength of the dust continuum emission. This thus raises the question whether the increasing prominence of the dust emission drowns out the plateau emission or whether the plateau carriers do not survive in front of the IF. Therefore, we investigate the relative contribution of the dust continuum and plateau emission components at two distinct locations within the OBC aperture, namely the PAH emission peak at $\sim$~117$^{\prime\prime}$ (P1) and the dust peak at $\sim$~100$^{\prime\prime}$ (P2, Figure~\ref{orion_pdr_v_hii}). As the dust continuum emission in the \HII\, region (PDR) only rises steeply towards longer wavelengths, the 5--9~$\mu$m range (and thus the 5--10~$\mu$m plateau) is significantly less influenced compared to the 10--13~$\mu$m plateau (see also Figure~\ref{orion_spectra}). 
Notably the dust continuum emission at P1 is actually stronger in the 5--10~$\mu$m range than that at P2 thus dispelling the notion that the non-detection of the 5--10~$\mu$m plateau into the \HII\, region (PDR) is due to the dominance of the dust emission. In the 10--13~$\mu$m range, we find that the dust continuum emission has similar strength at P1 and P2. Hence, like the 5--10~$\mu$m plateau, the disappearance of the 10--13~$\mu$m plateau cannot be attributed to a drastic change in dust emission between the edge--on PDR and the \HII\, region.  In addition, the detection of plateau emission in front of the IF may be influenced by the presence of silicate emission as the latter may lead to increased uncertainty in the continuum determination. However, when adding 10\% or 20\% of the plateau and PAH emission at P1 (corresponding to the minimum relative strength of the main PAH bands in the \HII\ region (PDR) and to the average strength of the plateau emission in the face--on PDR beyond the Orion Bar) to the dust continuum emission at P2, the plateau emission is still detectable (Figure~\ref{orion_spectra}). We therefore conclude that the lack or weak detection of plateau emission is significant. Similar weak to no detections are seen for the 8 $\mu$m bump and the G8.2 and G7.8~$\mu$m components (see panel d in Figure~\ref{orion_line_profiles1}).

In Figure~\ref{orion_line_profiles5}, we show the cross cut of the ratio of the 5--10 and 10-13~$\mu$m plateaus to the 11.2~$\mu$m PAH band (tracing neutral PAHs) and to the sum of the 6.2, 7.7, and 8.6~$\mu$m PAH bands (which traces ionic PAHs). We find that the relative strength of the 10-13~$\mu$m plateau emission is highest just behind the edge--on PDR front (at $\sim$ 132$^{\prime\prime}$  which is thus displaced from and located behind the PAH emission peak (at $\sim$ 117$^{\prime\prime}$). From the edge--on PDR front towards the IF, we find these ratios drop steadily while across the face--on PDR behind the Orion Bar, they level off to about 80\%. Relative to the ionic PAH bands, the 5--10~$\mu$m plateau emission peaks at roughly the same position as the 10-13~$\mu$m plateau but with a sharp decrease beyond the peak (in the face--on PDR behind the Orion Bar) and a more gradual decrease in the edge--on PDR. In contrast, relative to the neutral PAH band, the 5--10~$\mu$m plateau emission peaks at the PAH emission peak only showing a slight decrease towards the location where the other plateau ratios peak, after which it drops equally sharp as seen for the 5--10~$\mu$m plateau relative to the ionic bands. 

The plateaus are typically assigned to loosely bound PAH clusters \citep[e.g.][]{all89,bre89a,pee17} while the G8.2 and G7.8 components may also arise from very large, irregularly--shaped PAHs \citep[in addition to or instead of PAH clusters,][]{pee17}. 
These crosscuts indicate that the carriers of the plateaus and the G7.8 and G8.2 components are more photo-chemically labile than the stable free-flying PAHs which can survive harsher radiation fields closer to the star \citep[e.g.][]{and15}. The very small contribution of the plateau emission to the total emission observed in the \HII\, region (PDR) with respect to the face--on PDR behind the Orion Bar and the edge--on PDR can be attributed to the stronger UV-field impinging on the face--on PDR in the \HII\, region (PDR). The removal or suppression of these plateau features with proximity to the illuminating source is a clear indicator of the photochemical evolution of carbonaceous species in this PDR environment.

\subsection{Environmental Diagnostics}

The PAH population, and thus the PAH emission characteristics, is influenced by the physical conditions of their host environment. In order to quantify these dependencies, we derive the physical conditions in Section~\ref{PDR calc}. We find that the G7.6/G7.8 PAH ratio is an excellent tracer for the radiation field strength while the 6.2/11.2 PAH ratio probes the PAH ionization parameter (Section~\ref{ PDR pahs}). 

\label{PDR}
\subsubsection{Deriving PDR Conditions}
\label{PDR calc}

We determine the FUV radiation field strength, G$_{0}$, across our apertures following the method employed by \cite{gal08}. Briefly, these authors measured the very small grain (VSG) continuum emission from 10 to 16 $\mu$m to determine G$_{0}$ using the relationship G$_{0}$ $\propto$ I$_{\textrm{cont}}^{1/1.3}$ and the absolute value of G$_{0}~=~4~\times~10^{4}$ at the ionization front taken from \cite{tau94}.

\begin{figure*}
\begin{center}
\resizebox{\hsize}{!}{%
\includegraphics[clip,trim =0cm 0cm 0cm 0cm,width=8.5cm]{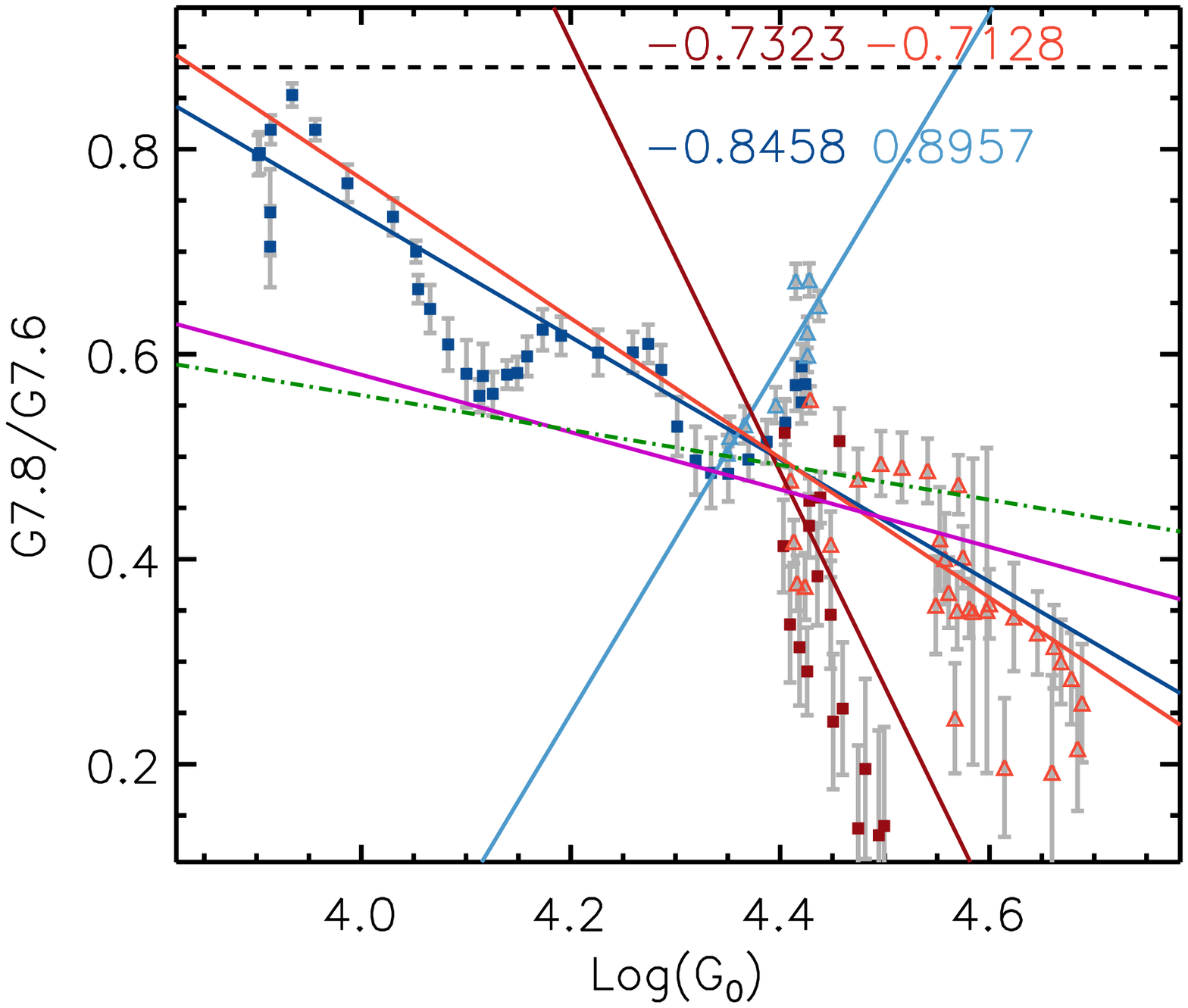}
\includegraphics[clip,trim =0cm 0cm 0cm 0cm,width=8.5cm]{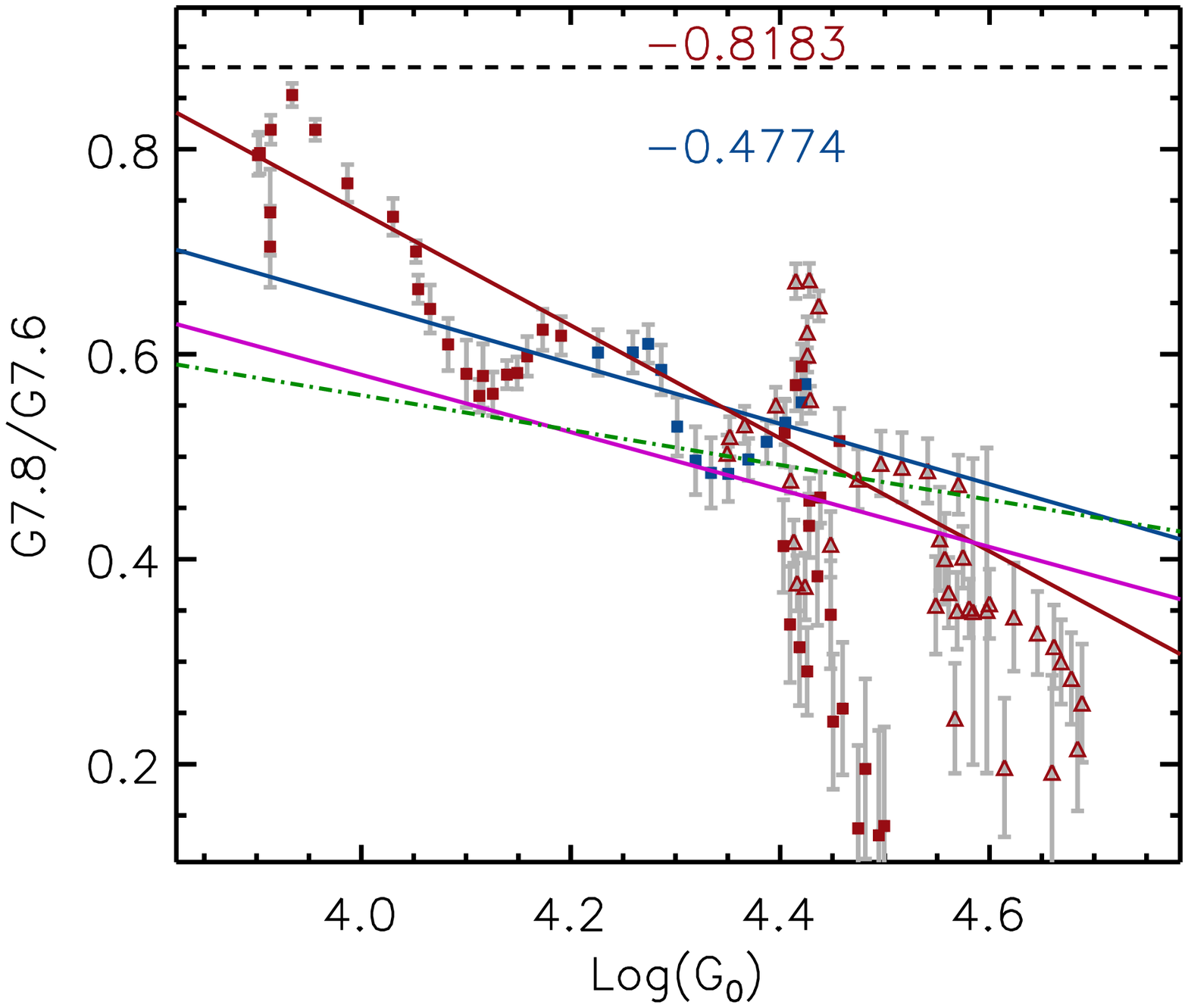}
}
\end{center}
\caption{G$_0$ versus G7.8/G7.6: Orion Bar combined (squares) and Orion Bar ionized (triangles). {\it Left} Pixels located beyond the IF and pixels in the \HII\, region (PDR) are represented in dark blue and red  respectively for the OBC aperture and in light blue and light red respectively for the OBI aperture (as depicted in panel a of Figure~\ref{orion_cont_10_14}).{\it Right} Pixels are color-coded such that the OBC edge--on PDR is shown in dark blue while all face--on PDRs and the OBI edge--on PDR are shown in dark red (as depicted in Figure~\ref{orion_cont_10_14} panel c).}
Correlation coefficients for each respective region are given in the same color as listed above. Weighted linear fits are shown as solid lines for each region in their respective color. The linear fits derived in \citet{sto17} are given as a green dot-dashed line and magenta solid line where they do and do not include the Ophiuchus diffuse cloud pointings respectively. The maximum G7.8/G7.6 ratio of 0.88 found in the outskirts of W49A by \citet{sto17} is shown as a black horizontal dashed line. 
\label{orion_G78_G76_G0}
\end{figure*} 

First, we measured the integrated strength of the dust continuum emission from 10--13.2~$\mu$m. To connect this 10--13.2~$\mu$m continuum flux with the 10-16~$\mu$m VSG continuum of \cite{gal08}, we use ISO-SWS spectra positioned across the Orion Bar (Table~\ref{table:sofia}). For these spectra, we determined the ratio of the integrated strength of the underlying dust continuum from 10--13.2 and 10--16 $\mu$m to be an average of 2.2 $\pm$ 0.1 over the five SWS pointings, indicating that these continua are effectively proportional. For the absolute calibration, we use the observation-based value of 2.6 $\times$ 10$^{4}$ at the edge--on ionization front in the Bar as reported by \citet{mar98} instead of the model-based value of 4 $\times$ 10$^{4}$ as reported by \citet{tau94}.  Thus, we calculate the G$_{0}$ cross cut for both apertures using the following scaling relationship:

\begin{equation}
G_{0}(r) = G_{0}(\textrm{IF}) \left( \frac{I_{\textrm{10--13 cont}}(r)}{I_{\textrm{10--13 cont}}(\textrm{IF})} \right)^{1/1.3}   
\label{G0_cont_relate}
\end{equation}

\noindent where IF refers to the position of the ionization front and r refers to an arbitrary position in either aperture. This results in G$_{0}$ ranging from 0.7--3.1~$\times$~10$^{4}$ in the OBC aperture and ranging from 2.1--4.7~$\times$~10$^{4}$ in the OBI aperture. In both apertures, G$_{0}$ peaks in the \HII\, region (PDR). The derived G$_{0}$ cross cut is shown in Figures~\ref{orion_line_profiles1} and~\ref{orion_line_profiles3}.
We obtain G$_{0}$ values that are about a factor of $\sim$~1.5 lower than those of \citet{gal08} due to the different absolute calibration.

Next, we derive an estimate for the average gas density, n$_{\textrm{H}}$, using the UV extinction fit to the total PAH emission cross cut in the OBC aperture (Section~\ref{UV_fit}). The fit traces the exponential drop in the edge--on PDR from 117$^{\prime\prime}$ to 136$^{\prime\prime}$, and thus up to a distance $s$ from the PAH emission peak of 1.18~$\times$~10$^{17}$~cm. This corresponds to $ F_{\mathrm{PAH}}(s)/F_{\mathrm{PAH}}(\mathrm{peak}) $ = 0.195 from the UV extinction fit as defined in equation~\ref{uv_ext_eq}, or an optical depth of $\tau_{\textrm{UV}}$~=~1.63. 
We use a conversion factor of $\tau_{\textrm{UV}}$~=~1.8~A$_{\textrm{V}}$ from the model 2 of \cite{fla80} and \cite{rob81} along with the standard N$_{\textrm{H}}$/A$_{\textrm{V}}$~=~1.9~$\times$~10$^{21}$~cm$^{-2}$/mag \citep{boh78} and solve for the average gas density, $n_{\textrm{H}}$, across the edge--on PDR as

\begin{equation}
n_{\textrm{H}} = \frac{N_{\textrm{H}}}{A_{\textrm{V}}} \frac{A_{\textrm{V}}}{\tau_{\textrm{UV}}} \frac{\tau_{\textrm{UV}}}{s}.     
\end{equation}

\noindent We derive an average gas density of 1.44~$\times$~10$^4$~cm$^{-3}$ within the Orion Bar. We note that there is a systematic uncertainty associated with this derivation as the UV extinction curve is shallower for Orion \citep[$R_V \sim 5.5$,][]{Lee68}. However, our derivation makes it consistent with PDR models.

Subsequently, we derive the PAH ionization parameter, $\gamma$~=~G$_{0}$~T$^{0.5}$~/~n$_{\textrm{e}}$, where n$_{\textrm{e}}$ is the electron density and T the gas temperature \citep{bak94}. We use the same method as described in \cite{gal08} within the OBC edge--on PDR. We take the value for gas temperature at the Orion Bar PDR front of 500~K as derived in \cite{tau94}. 
The average gas density is converted to electron density using the assumption that all free electrons result from the photo-ionization of carbon and all gas--phase carbon is ionized, n$_{\textrm{e}}$ $\simeq$ (C/H) n$_{\textrm{H}}$ $\simeq$ 1.6 $\times$ 10$^{-4}$ n$_{\textrm{H}}$, where 1.6~$\times$~10$^{-4}$ is the interstellar gas-phase carbon abundance \citep{sofia04}. Under the assumption that the electron density and gas temperature remains constant within the Orion Bar, we derive $\gamma$ across both apertures. We find $\gamma$  within the OBC edge--on PDR ranges from $\sim$~5--8.5~$\times$~10$^{3}$. For reference, \cite{gal08} determined the Orion Bar $\gamma$ range from $\sim$~1.5--4~$\times$~10$^{3}$, below our estimates (by a factor of $\sim$0.46).  This can be accounted for in terms of the different absolute calibration for G$_{0}$ and the difference in n$_{\textrm{H}}$ (\cite{gal08} used the density reported by \cite{tau94}, which is a factor of $\sim$~3.5 higher than our value).

\subsubsection{PAHs as PDR Tracers}
\label{ PDR pahs}

First, we compare our Orion Bar data with the previously established relationship of \cite{sto17} between G$_{0}$ and G7.8/G7.6 (Figure~\ref{orion_G78_G76_G0}). We observe a strong anti--correlation  between these two parameters with G7.8/G7.6 being highest in low G$_{0}$ environments, consistent with \cite{sto17}. In the left panel of this figure, we compare these parameters separating the spectra before and after the IF in two groups (as illustrated by the shaded regions in panel (a) of Figure~\ref{orion_cont_10_14}) for the OBC and OBI apertures separately. We find a strong anti-correlation in the OBC PDR spectra behind the edge--on IF, a slightly weaker correlation in the \HII\, region (PDR) spectra of both apertures, and a strong positive correlation in the OBI edge--on PDR spectra. The latter, however, only probes a very small range in G$_{0}$.

\begin{figure}
\begin{center}
\includegraphics[clip,trim =0cm 0cm 0cm 0cm,width=8.5cm]{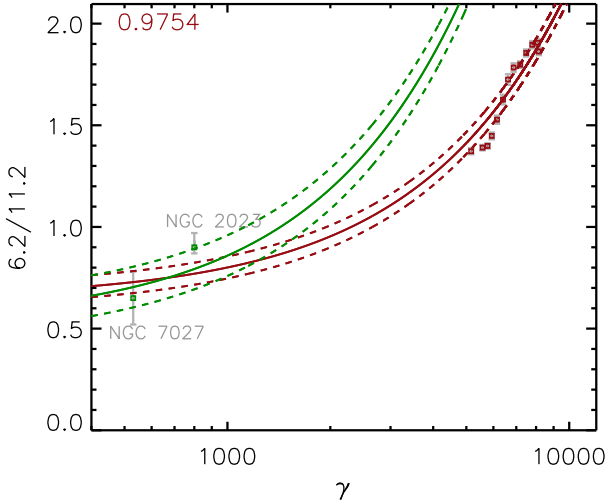}
\end{center}
\caption{ The PAH ionization parameter $\gamma$ versus the 6.2/11.2 PAH ratio for the OBC edge--on PDR. 
A weighted linear fit and their 1 $\sigma$ fit uncertainties (in offset) are shown as solid and dashed lines respectively and the correlation coefficient is given in red. The linear fit derived by \citet{gal08} is shown as a solid green line with dashed green lines representing the 1 $\sigma$ deviations (in offset). Data points for the reflection nebula NGC~2023 and the planetary nebula NGC~7027 are shown in green.}
\label{orion_gamma_62_112}
\end{figure}

Compared to the results of \cite{sto17}, our linear fits are significantly steeper. This may be attributed to i) the much smaller range of G$_{0}$ values covered here and ii) the fact that we only probe high UV field regions relative to the range given in \cite{sto17}. In fact, if we only consider the \HII\, region observations of \cite{sto17}, the slope of their relation will increase substantially. Additionally, it should be noted these authors only consider the global, spatially integrated, values of G$_{0}$ and G7.8/G7.6 in each of the sources they included in their relationship, which resulted in considerable uncertainties in G$_{0}$.

We discussed in Section~\ref{PDR morph} that the grouping sperating the PDR behind the Orion Bar IF versus the \HII\, region (PDR) fails to take the underlying differences in PDR morphology into account. In order to understand the effects of PDR morphology on the G7.8/7.6 ratio versus G$_{0}$, we use a different grouping of our data, i.e. the grouping with the OBC edge--on PDR in one group and all of the face--on and the OBI edge--on PDR in the second group (Figure~\ref{orion_G78_G76_G0}, right panel; the grouping is illustrated in panel c of Figure~\ref{orion_cont_10_14}). In the OBC edge--on PDR, we find a moderate anti-correlation but more interestingly, the linear fit to these data points agrees within the uncertainties with the \cite{sto17} relationship that excludes the diffuse ISM. Additionally, we find that the grouping including the face--on PDRs and the OBI edge--on PDR has a very strong anti-correlation between G7.8/7.6 ratio and G$_{0}$ with a steeper slope than the \citet{sto17} relationships. This suggests that an edge--on and face--on PDR morphology may generally also yield a different linear relationship between G7.8/G7.6 and G$_{0}$. Despite the differences between these studies and between face--on and edge--on PDRs, it is clear that the G7.8/G7.6 has the potential to become a useful tracer of the FUV radiation field for a wide variety of PDRs. \\

We proceed with investigating the dependence of the PAH emission characteristics on the PAH ionization parameter. To this end, we first investigate the relationship between the PAH ionization parameter $\gamma$ and the 6.2/11.2 ratio in the OBC edge--on PDR and compare with the results reported by \cite{gal08}. In Figure~\ref{orion_gamma_62_112}, we show a fit to this relationship 
making use of data points from \cite{gal08} for NGC~2023 and NGC~7027 to constrain this relationships at low $\gamma$ ($<$ 10$^{3}$). 
This relationship is expressed as follows 
\begin{equation}
I_{6.2}/I_{11.2} = 0.00015 \  \gamma + (0.65 \pm 0.05). 
\label{62_112_v_gamma}
\end{equation}
Our derived relationship is a factor of $\sim$~2 lower in slope in comparison with that of \cite{gal08} due to the different $\gamma$-values (see Section~\ref{PDR calc}). 

\begin{figure}
\begin{center}
\includegraphics[clip,trim =1cm .8cm 1cm 1cm,width=7.5cm]{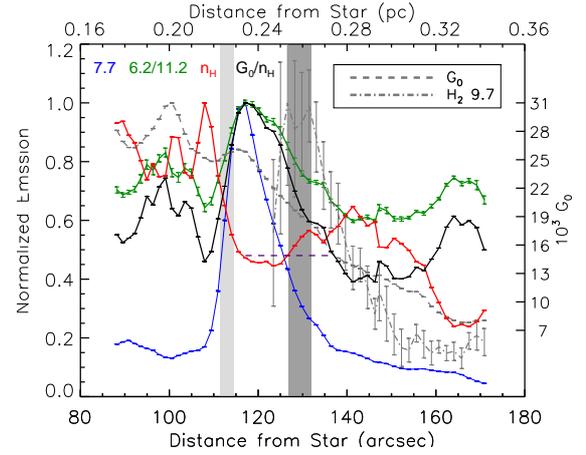}
\end{center}
\caption{The derived physical conditions across the OBC aperture. Cross cuts are normalized to the peak values for each parameter. The horizontal indigo dashed line from 117--136$^{\prime\prime}$ indicates the average density, $n_{\textrm{H}}$, of 1.44~$\times$~10$^{4}$ cm$^{-3}$ derived from the UV extinction fit in the OBC edge--on PDR (Section~\ref{PDR calc}) normalized to the peak gas density in the aperture. The dark and light grey shaded regions correspond to the Orion Bar PDR front and the IF (see Figure~\ref{orion_line_profiles1}). G$_{0}$ cross cuts values are shown on the right y-axis in units of 10$^{3}$ Habings (see Section~\ref{PDR calc} for derivation).  The 7.7 $\mu$m and 6.2/11.2 cross cuts are shown for reference. }
\label{orion_combined_PDR_profile}
\end{figure}

We can now derive how the gas density, $n_{\textrm{H}}$, varies with distance from the illuminating source. Indeed, using the relationship between the 6.2/11.2 and $\gamma$ calibrated on the edge--on PDR (equation~\ref{62_112_v_gamma}) and the observed 6.2/11.2 emission ratios, we solve for $\gamma$ for the remainder of the data, namely the face--on PDRs in both apertures and the OBI edge--on PDR. 
Based on these calculated $\gamma$ values and the derived G$_{0}$ values (equation~\ref{G0_cont_relate}, Section~\ref{PDR calc}) and assuming a constant gas temperature of 500~K, we derive the corresponding gas density, $n_{\textrm{H}}$, cross cuts (Figure~\ref{orion_combined_PDR_profile}). The derived gas density $n_{\textrm{H}}$ shows a maximum of 3.00~$\times$~10$^{4}$ cm$^{-3}$ in front of the IF, decreasing down to $\sim$~1.5~$\times$~10$^{4}$ cm$^{-3}$ throughout the edge--on PDR, followed by a slight rise behind the edge--on PDR front and subsequent drop to a minimum of 7.16~$\times$~10$^{3}$ cm$^{-3}$ at $\sim$~165$^{\prime\prime}$. Note that this derived gas density depends on G$_{0}$ and the (inverse of the) 6.2/11.2 emission ratio, evident in comparing the respective cross cuts. Additionally, we note that the resulting G$_{0}$/n$_{\textrm{H}}$ cross cut follows a very similar trend as the 6.2/11.2 cross cut primarily due to how the gas densities were calibrated on the $\gamma$ vs 6.2/11.2 relationship and the assumption of a constant gas temperature.

\begin{figure*}
\begin{center}
\resizebox{\hsize}{!}{%
\includegraphics[clip,trim =0cm 0cm 0cm 0cm,width=8.5cm]{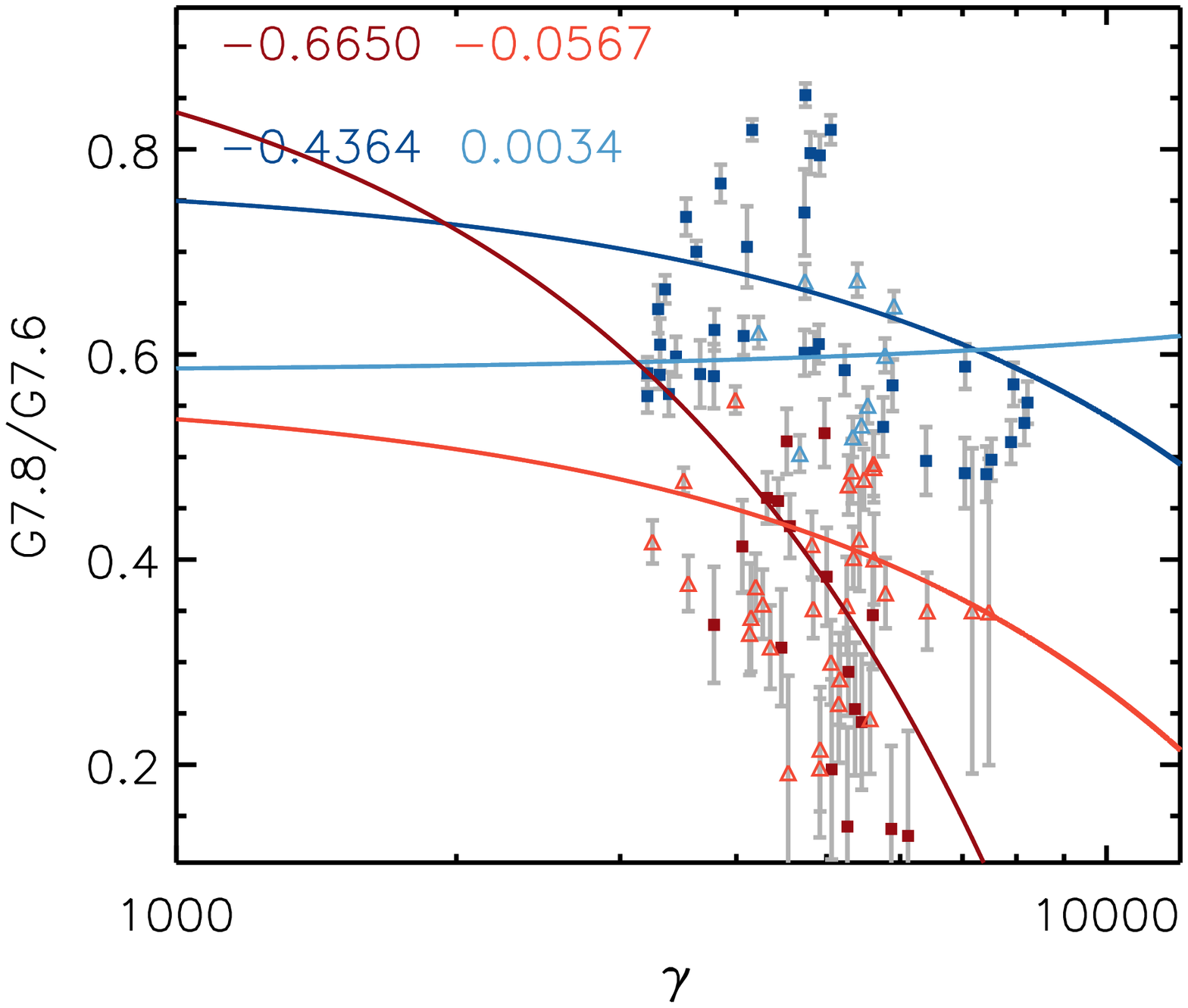}
\includegraphics[clip,trim =0cm 0cm 0cm 0cm,width=8.5cm]{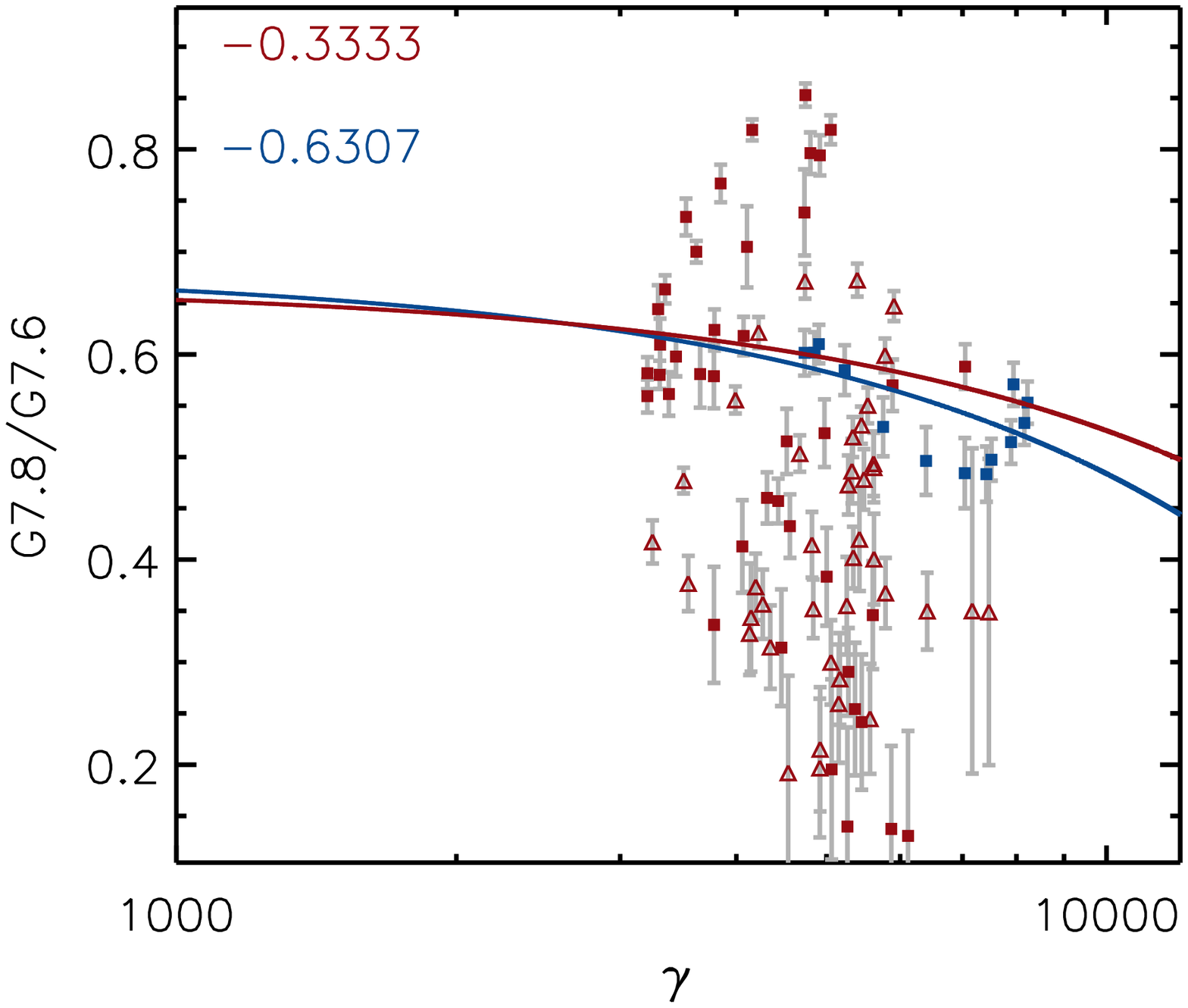}
}
\end{center}
\caption{The PAH ionization parameter $\gamma$ versus the G7.8/G7.6 ratio: Orion Bar combined (squares) and Orion Bar ionized (triangles). {\it Left} Pixels located beyond the IF and pixels in the \HII\, region (PDR) are represented in dark blue and red  respectively for the OBC aperture and in light blue and light red respectively for the OBI aperture (as depicted in panel a of Figure~\ref{orion_cont_10_14})  {\it Right} Pixels are color-coded such that the OBC edge--on PDR is shown in dark blue while all face--on PDRs and the OBI edge--on PDR is shown in dark red (as depicted in Figure~\ref{orion_cont_10_14} panel c). 
 Correlation coefficients for each respective region are given in their respective color. Weighted linear fits are shown as solid lines for each region in their respective color.}
\label{orion_G78_G76_gamma}
\end{figure*} 

We re-calculate the PAH ionization parameter, $\gamma$, employing a variable gas density and investigate the relation between $\gamma$ and the G7.8/G7.6 ratio (Figure~\ref{orion_G78_G76_gamma}). In the left panel, we use the original division of pixels before and behind the edge--on IF for each aperture separately (as illustrated by the shaded regions in panel (a) of Figure~\ref{orion_cont_10_14}). No clear correlation is found: only datapoints from the \HII\, region (PDR) in the OBC aperture (in dark red) show a moderate anti-correlation. In the right panel, we consider the grouping where the OBC edge--on spectra are in one group with the rest of the spectra in the other group (as illustrated by the shaded regions in panel (a) of Figure~\ref{orion_cont_10_14}). In this case, the OBC edge--on spectra show a moderate anti-correlation while the remainder group is significantly weaker. 

As G7.8/G7.6 is a strong tracer of G$_{0}$ (see Figure~\ref{orion_G78_G76_G0}), it is unsurprising to find some resemblance of a relationship between G7.8/G7.6 and $\gamma$ as the latter is directly proportional to G$_{0}$. However, the inclusion of a variable gas density to calculate $\gamma$ seems to weaken this relationship. This suggests G7.8/G7.6 is not an ideal tracer for PAH ionization, whereas ratios involving cationic to neutral PAH emission features, such as 6.2/11.2, are much better suited to this role.  In addition, this suggests that variations in the G7.8/G7.6 ratio originates in a different photo-chemical processing of the carriers responsible for these two components.

\section{Conclusion}
\label{conclusion}

In this paper, we investigate the characteristics of the PAH emission features across the Orion Bar through the use of {\it Spitzer} IRS~SL spectroscopic observations consisting of two apertures that cross the Orion Bar at different locations along with {\it SOFIA} FLITECAM imaging observations. We make use of the spline decomposition method to separate the PAH emission features from the underlying continuum components. We measure the fluxes on the various emission components found within both spectral apertures including the PAH features and related components, the atomic recombination lines, the H$_{2}$ lines, and the underlying dust continuum. Variations in these spectral components are found using cross cuts taken with respect to distance from the primary illuminating source in the Orion Nebula. Correlations between the PAH related features are considered based on the relative position to the ionization front of the Orion Bar as well as the different PDR morphologies present in each aperture, these being edge--on and face--on. Finally, we compare specific PAH emission ratios with the derived cross cuts of the FUV radiation field and the PAH ionization parameter. Our major findings are as follows:

\begin{itemize}
    \item All of the PAH-related emission has a strong peak located within the Orion Bar. Variations in PAH features become more prominent away from the this peak behind the PDR front and into \HII\, region (PDR) in front of the Bar. Additional maxima or mimima in the PAH emission cross cuts and PAH ratio cross cuts typically coincide with the G$_{0}$ peaks as defined by the dust continuum emission in each aperture. 
    \item We quantify the effect of the extinction of UV photons into the edge--on PDR and thus the decrease in available energy for PAH excitation. We show that the excess emission found behind the edge--on PDR front can be attributed to a face--on PDR. We derive an average gas density within the edge--on PDR of 1.44~$\times$~10$^{4}$ cm$^{-3}$.
    \item Grouping the spectra based on PDR morphology yields much tighter correlations between the PAH emission features in comparison to grouping based on relative position with respect to the ionization front in general. PAH emission correlations in many cases demonstrate two distinct trends that are attributed to the edge--on PDR of the Bar and the face--on PDRs located in front of the ionization front and behind the PDR front. Hence, the PDR viewing angle influences the observed PAH characteristics. 
    \item The PAH emission within the edge--on PDRs in both apertures behaves differently confirming the Bar is not a uniform structure.
    \item The average PAH size in the edge--on PDR increases with increasing strength of the FUV radiation field. This suggests that intense UV fields lead to increased photo-processing of PAHs, destroying the smallest PAHs.
    \item Subtle differences are observed between the bands assigned to cationic PAHs (at 6.2, 7.7, 8.6, and 11.0 $\mu$m). There is a spatial sequence evident in the relative intensities of these features with respect to proximity to the illuminating source, namely 11.0, 8.6, 7.7, and 6.2 $\mu$m. We also report a deviation from the well--known tight relationship between the 6.2 and 7.7~$\mu$m bands within the \HII\, region (PDR) where the UV radiation field strength is strongest. 
    \item The carriers of the PAH plateaus and the G7.8 and G8.2 components are more photo-chemically labile than the stable free-flying PAHs responsible for the main PAH bands. This is indicative of photoprocessing driving the chemical evolution of aromatic species throughout the PDR environment. Conversely, a broad emission component attributed to silicates only becomes prominent at the highest UV radiation fields within the \HII\, region (PDR) and becomes invisible into the PDR. 
     \item Overall, we confirm the anti--correlation between the G7.8/G7.6 and G$_{0}$ reported by  \cite{sto17}. 
     Using a grouping based on PDR morphology shows that the OBC edge--on PDR displays a similar linear relationship as \cite{sto17} whereas the face--on PDRs show a steeper relationship. A weaker correlation of G7.8/G7.6 with the PAH ionization parameter indicates that the carriers of the G7.6 and G7.8 components experience different photo-chemical processing.
     \item We replicate the linear relationship found between 6.2/11.2 and the PAH ionization parameter of \cite{gal08} within the edge--on PDR. Using the derived relationship and the observed 6.2/11.2 emission ratios, we derive the gas density with distance from the illuminating source. 
     
\end{itemize}

To summarize, the Orion Bar PDR has clearly earned its reputation as the prototypical edge--on PDR. We show clear stratification of the relative intensities of the PAH emission features within the different layers of the Orion Bar yielding insight into the physical and chemical structure of this well--studied environment. However, the murkier face--on PDRs on both sides of the Bar also deserve closer scrutiny as few observed PDRs are as unambiguous as the Orion Bar. Novel JWST observations of the Orion Bar will allow us to further advance our understanding of astronomical PAHs and their relationship with their PDR environments. 

\section*{Data Availability}

The data underlying this article will be shared on reasonable request to the corresponding author.

\section*{Acknowledgements}

C.K. acknowledges support from an Ontario Graduate Scholarship (OGS). E.P. acknowledges support from an NSERC Discovery Grant and a SOFIA grant. Studies of interstellar PAHs at Leiden Observatory are supported through a Spinoza award. This work is based [in part] on observations made with the Spitzer Space Telescope, which is operated by the Jet Propulsion Laboratory, California Institute of Technology under a contract with NASA.  This work is further based [in part] on observations made with the NASA-DLR Stratospheric Observatory for Infrared Astronomy (SOFIA). SOFIA is jointly operated by the Universities Space Research Association, Inc. (USRA), under NASA contract NAS2-97001, and the Deutsches SOFIA Institut (DSI) under DLR contract 50 OK 0901 to the University of Stuttgart. 




\bibliographystyle{mnras}
\bibliography{mainbib}



\appendix

\begin{table}
\begin{center}
\caption{\label{table:2}{}Orion 7 to 9 Decomposition Parameters}

\begin{tabular}{ p{1.5cm} p{2.9cm}  p{2.2 cm} }
\hline\hline
 Feature ($\mu$m) & $<$ Peak Position ($\mu$m) $>$$^{1}$ & $<$ FWHM ($\mu$m) $>$$^{1}$ \\

\hline\\[-5pt]
Pfund~$\alpha$ & 7.481~$\pm$~0.006 & 0.06~$\pm$~0.01 \\
GPAH 7.6 & 7.59~$\pm$~0.02 & 0.44~$\pm$~0.01 \\
GPAH 7.8 & 7.86~$\pm$~0.04 & 0.40~$\pm$~0.03 \\
GPAH 8.2 & 8.29~$\pm$~0.02 & 0.263~$\pm$~0.02 \\
GPAH 8.6 & 8.61~$\pm$~0.01  & 0.34~$\pm$~0.01 \\
 Ar III  8.99 & 9.004~$\pm$~0.008 & 0.122~$\pm$~0.009 \\ [2pt]
\hline\hline

\end{tabular}

\end{center}
$^{1}$ Combined averages of fits in spectra of all three slits.
\end{table}

\begin{table*}
\begin{center}
\caption{\label{ch3 app table:1}{}Orion Bar IRS~SL and FLITECAM Emission Components}
\begin{tabular}{ c c c }
\hline\hline
 Feature Tag$^{1}$ &  Emission Description & Peak Location$^{2}$  \\

\hline\\[-2pt]

\multicolumn{3}{c}{\bf Atomic Lines}\\[2pt]
{[}\ArII\,{]}  &  [\ArII\,] 6.98~\mum\ recombination line & \HII\, region (PDR)\\
\HI\, 7.46 & \HI\, 7.45~$\mu$m recombination line &  \HII\, region (PDR)\\
{[}\ArIII\,{]} & [\ArIII\,] 8.99~\mum\, recombination line &  \HII\, region (PDR) \\
{[}\SIV\,{]} & [\SIV\,] 10.5~\mum\ recombination line &  \HII\, region (PDR) \\
\HI\, 12.37 & \HI\, 12.37~$\mu$m recombination line &  \HII\, region (PDR) \\
{[}\NeII\,{]} & [\NeII\,] 12.8~\mum\ recombination line &  \HII\, region (PDR) \\
{[} \OI\,{]}  & [\OI\,] 6300~\AA\ emission line &  IF\\
\hline\\[-10pt]
\multicolumn{3}{c}{\bf Dust Emission}\\[2pt]
cont. 10.2 & dust continuum emission at  10.2~\mum\, &  \HII\, region (PDR) \\
cont. 13.2 & dust continuum emission at  13.2~\mum\, &  \HII\, region (PDR) \\
cont.10--13 & integrated dust continuum emission from 10--13.2~\mum\, &  \HII\, region (PDR)\\
\hline\\[-10pt]
\multicolumn{3}{c}{\bf PAH--Related Emission}\\[2pt]
3.3 (FC) & PAH 3.3~\mum\, band$^{3}$  &  edge--on PDR (Orion Bar) \\
6.2 & PAH 6.2~\mum\, band  &  edge--on PDR (Orion Bar) \\
7.7 & PAH 7.7~\mum\, band  &  edge--on PDR (Orion Bar) \\
8.6 & PAH 8.6~\mum\, band  &  edge--on PDR (Orion Bar) \\
G7.6 & PAH G7.6~\mum\, Gaussian sub--component  &  edge--on PDR (Orion Bar) \\
G7.8 & PAH G7.8~\mum\, Gaussian sub--component  &  edge--on PDR (Orion Bar) \\
G8.2 & PAH G8.2~\mum\, Gaussian sub--component  &  edge--on PDR (Orion Bar) \\
G8.6 & PAH G8.2~\mum\, Gaussian sub--component  &  edge--on PDR (Orion Bar) \\
G11.0 & PAH 11.0~\mum\, band  &  edge--on PDR (Orion Bar) \\
11.2 & PAH 11.2~\mum\, band  &  edge--on PDR (Orion Bar) \\
12.7 & PAH 12.7~\mum\, band  &  edge--on PDR (Orion Bar) \\
8~bump & 8~\mum\, bump PAH plateau   &  edge--on PDR (Orion Bar) \\
5--10 plat & 5--10~\mum\, PAH plateau  &  edge--on PDR (Orion Bar) \\
10--13 plat & 10--13~\mum\, PAH plateau  &  edge--on PDR (Orion Bar) \\
\hline\\[-10pt]
\multicolumn{3}{c}{\bf Molecular Hydrogen Lines}\\[2pt]
H$_{2}$ 9.7 & H$_{2}$ 9.7~\mum\, emission line  &  Orion Bar PDR Front\\

\hline\hline

\end{tabular}
\end{center}
$^1$ Shorthand used to refer to individual features in Figures~\ref{orion_line_profiles1}, ~\ref{orion_line_profiles3}, and~\ref{orion_line_profiles5}, and throughout the paper.
$^{2}$ Peak of the emission component (see Section~\ref{proj}).
$^{3}$ Observed with SOFIA FLITECAM photometry \citep{Knight:21}.
\end{table*}

\section{ Decomposition of 7--9~$\mu$m region}
\label{Fitparameters}

We decomposed the GS subtracted spectra in the 7--9.2~$\mu$m region with 6 Gaussians. These include the 7--9~\mum\, PAH components G7.6, G7.8, G8.2, and G8.6~\mum\, \citep{pee17,sto17}, and the atomic lines \HI\, 7.45~\mum\, and [\ArIII] 8.99~\mum. As in \cite{pee02}, we chose Gaussian components to fit the 7.6 and 7.8~$\mu$m peaks of the 7.7~\mum\, complex, the 8.6~\mum\ PAH band, and a fourth Gaussian component at 8.2~\mum\, to obtain a good fit in the 7--9~$\mu$m region. We also note that the 8~\mum\, bump found between the LS and GS spline continua is incorporated into these Gaussian components, hence the similarities between the cross cuts of the 8~\mum\, bump and the G7.8 and G8.2~\mum\, components in particular (Figure~\ref{orion_line_profiles1} d and h). This fitting procedure was first run with the starting parameters of all 6 Gaussians allowed to vary in peak position in a window of 0.2~$\mu$m and in FWHM in a window of 0.1~$\mu$m. Subsequently, the average peak positions for each feature within each aperture was obtained and fixed (Table~\ref{table:2}). We determine the average FWHM for each of these Gaussian fits and fix these parameters as well.

\section{Orion Bar Spitzer IRS~SL Cross Cut Data}
\label{ob irs components}

In Table~\ref{table:2}, we summarize the prominent emission components found in the Orion Bar Spitzer IRS~SL spectra as discussed in Section~\ref{results}. To facilitate parsing the large amount of cross cuts shown in Figures~\ref{orion_line_profiles1} and~\ref{orion_line_profiles3}, we use the same nomenclature used to refer to each feature. We organize these components based on the different types of emission found as well as where they spatially peak within each aperture relative to the Orion Bar ionization front. Furthermore, in Tables~\ref{table:3} and~\ref{table:4}, we provide the normalization factors for each cross cut shown in this work.

\renewcommand{\arraystretch}{1.5}
\begin{table}
\caption{\label{table:3}{}Orion emission cross cut normalization factors}
\begin{center}
\begin{tabular}{ p{3cm} p{1.5 cm}  p{1.5 cm} }
\hline\hline
Emission Feature ($\mu$m) &  OBC$^{1,3}$  &  OBI$^{2,3}$  \\

\hline\\[-5pt]
{[}\ArII\,{]} & 1.25 (-5) & 8.19 (-6) \\
\HI\, 7.46 & 2.67 (-6) & 1.94 (-6)\\
{[}\ArIII\,{]} & 2.43 (-5) & 2.09 (-5)\\
{[}\SIV\,{]} & 1.34 (-5) & 2.39 (-5)\\
\HI\, 12.37 & 9.10 (-7) & 5.02 (-7)\\
{[}\NeII\,{]} & 6.85 (-5) & 6.11 (-5)\\
cont. 10.2 & 1.47 (-4) & 1.88 (-4) \\
cont. 13.2 & 1.19 (-4)  &  1.25 (-4) \\
cont.10--13 & 4.69 (-4) & 5.40 (-4) \\
3.3 & 3.90 (-5) & 3.59 (-5) \\
6.2 & 1.33 (-4) & 1.05 (-4) \\
7.7 & 2.75 (-4) & 2.10 (-4) \\
8.6 & 6.20 (-5) & 4.85 (-5)\\
G7.6 & 2.19 (-4) & 1.63 (-4) \\
G7.8 & 1.21 (-4) & 1.05 (-4) \\
G8.2 & 2.54 (-5) & 2.18 (-5)\\
G8.6 & 8.53 (-5) & 7.05 (-5)\\
11.0 & 5.69 (-6) & 4.30 (-6)\\
11.2 & 7.05 (-5) & 6.74 (-5)\\
12.7 & 3.48 (-5) & 3.03 (-5)\\
8~bump  & 1.22 (-4) & 1.05 (-4)\\
5--10~plat  & 2.07 (-4) & 1.71 (-4) \\ 
10--13~plat  & 3.80 (-5) & 3.41 (-5)\\
5--10~plat  & 2.07 (-4) & 1.71 (-4)\\
H$_{2}$ 9.7 & 1.04 (-6) & N/A \\
\hline\\[-10pt]

\end{tabular}

\end{center}
$^1$ Combined aperture of Orion Bar/ Orion Bar Neutral;
$^2$ Orion Bar Ionized;
$^3$ Normalization factors in units W~m$^{-2}$~sr$^{-1}$ (i.e multiply by these values to get the original values) in the format: 3 significant digits (order of magnitude). 
\end{table}

\renewcommand{\arraystretch}{1.5}
\begin{table}
\caption{\label{table:4}{}Orion emission ratio cross cut normalization factors}
\begin{center}
\begin{tabular}{ p{3.5cm} p{1.7 cm}  p{1.7 cm} }
\hline\hline
Emission Ratio ($\mu$m) &  OBC$^{1,3}$  &  OBI$^{2,3}$  \\

\hline\\[-5pt]

6.2/7.7 & 0.542 & 0.507 \\
8.6/7.7  & 0.263 & 0.267 \\
8.6/6.2  & 0.680 & 0.601 \\
6.2/11.2  & 1.91 & 1.79 \\
11.0/11.2  & 0.152 & 0.187 \\
7.7/11.2 & 4.18 & 3.88 \\
8.6/11.2 & 1.08 & 0.983 \\
12.7/7.7 & 0.341 & 0.336 \\ 
12.7/11.2 & 1.22 & 0.879 \\
12.7/6.2 & 0.854 & 0.768 \\
12.7/8.6 & 1.41 & 1.39 \\
11.0/12.7 & 0.176 & 0.398 \\
11.0/6.2 & 0.104 & 0.139 \\
11.0/7.7 & 0.0409 & 0.0582 \\
11.0/8.6 & 0.172 & 0.239 \\
G7.8/G7.6 & 0.853 & 0.672 \\
G8.2/G7.8 & 0.287 & 0.275 \\
G8.2/G8.6  & 0.481 & 0.358 \\
G8.6/G7.6 & 0.481 & 0.454 \\
G8.2/G7.6  & 0.167 & 0.138\\
G8.6/G7.8  & 1.59 & 1.35\\
6.2/(G7.6 + G7.8)  & 0.467 & 0.453 \\
7.7/G7.6 & 1.47 & 1.37 \\
7.7/G7.8 & 9.68 & 6.08 \\
7.7/(G7.6 +G7.8) & 1.12 & 0.982 \\
3.3/6.2  & 0.663 & 0.622 \\
3.3/7.7 & 0.315 &  0.276 \\
3.3/11.2  & 1.11  & 0.932 \\
3.3/11.0  & 16.4 & 8.66 \\
(6.2+7.7+8.6)/11.2  & 6.84 & 6.61 \\
8~Bump/7.7 & 0.658 & 0.560 \\
5--10~plat/(6.2+7.7+8.6) & 0.448 & N/A \\
10--13~plat/(6.2+7.7+8.6) & 0.181 & N/A \\
5--10~plat/11.2 & 2.96 & N/A \\
10--13~plat/11.2 & 0.756 & N/A \\

\hline\\[-10pt]

\end{tabular}

\end{center}
$^1$ Combined aperture of Orion Bar/ Orion Bar Neutral;
$^2$ Orion Bar Ionized;
$^3$  Multiply emission ratio by respective normalization factors to get original ratio.
\end{table}

\section{Additional Emission Ratio Cross Cuts}
\label{appab}

\begin{figure*}
\begin{center}
    \centering
    \begin{tabular}{cc}

\resizebox{.5\hsize}{!}{%
\includegraphics[clip,trim =0cm 1.5cm 0cm 0cm,width=7.5cm]{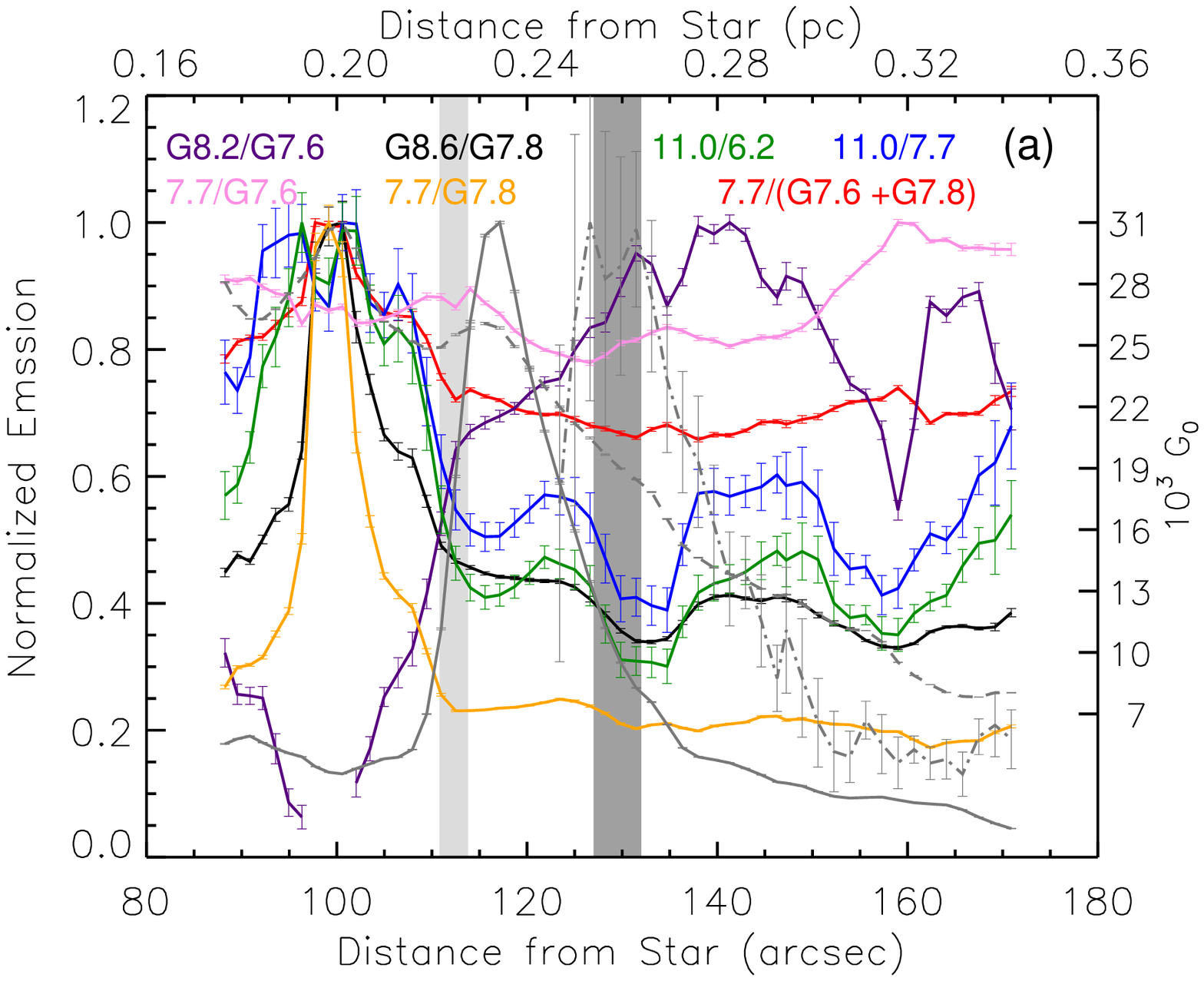}} & 
\resizebox{.5\hsize}{!}{%
\includegraphics[clip,trim =0cm 1.5cm 0cm 0.0cm,width=7.5cm]{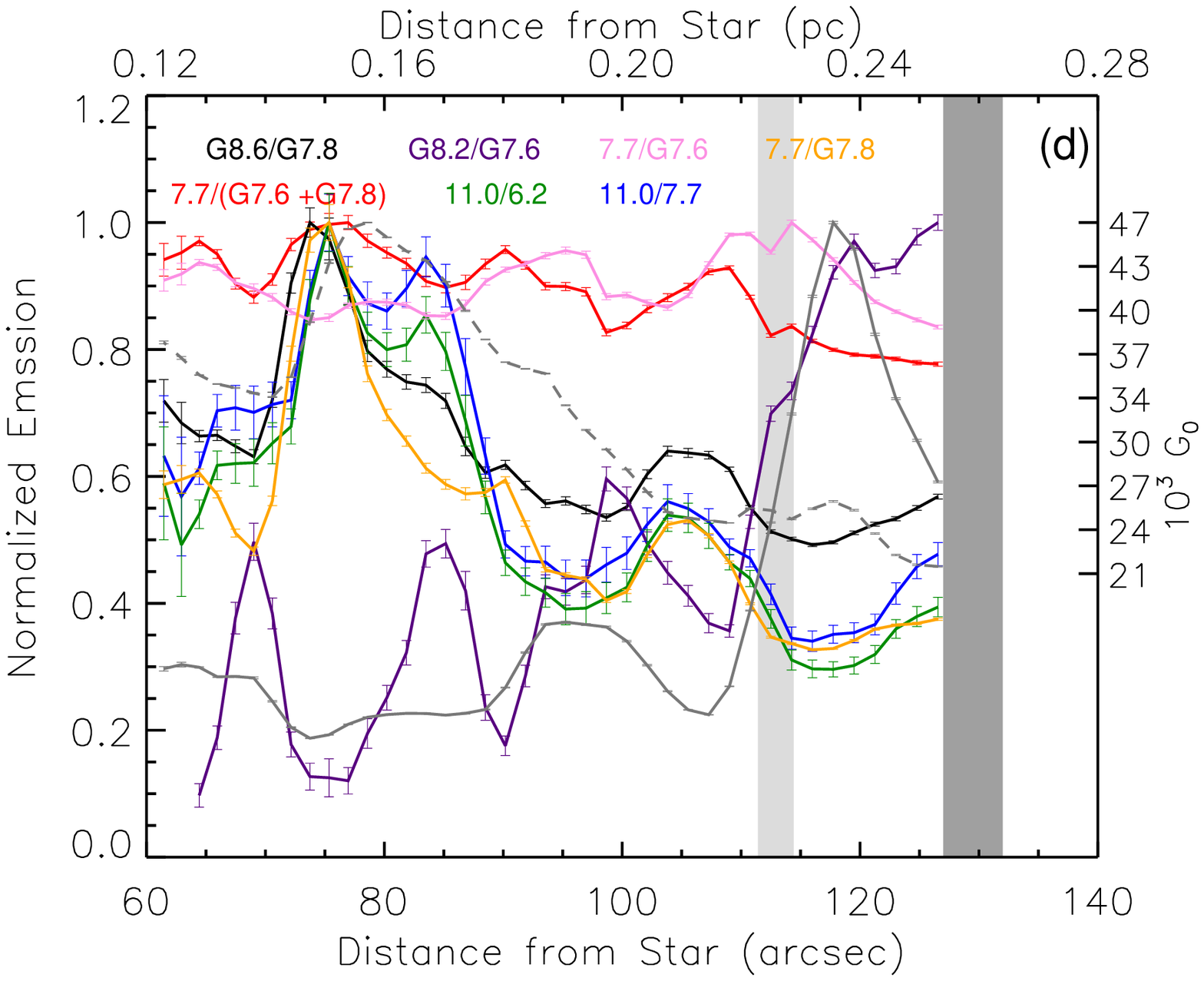}}\\
\resizebox{.5\hsize}{!}{%
\includegraphics[clip,trim =0cm 0cm 0cm 2.2cm,width=7.5cm]{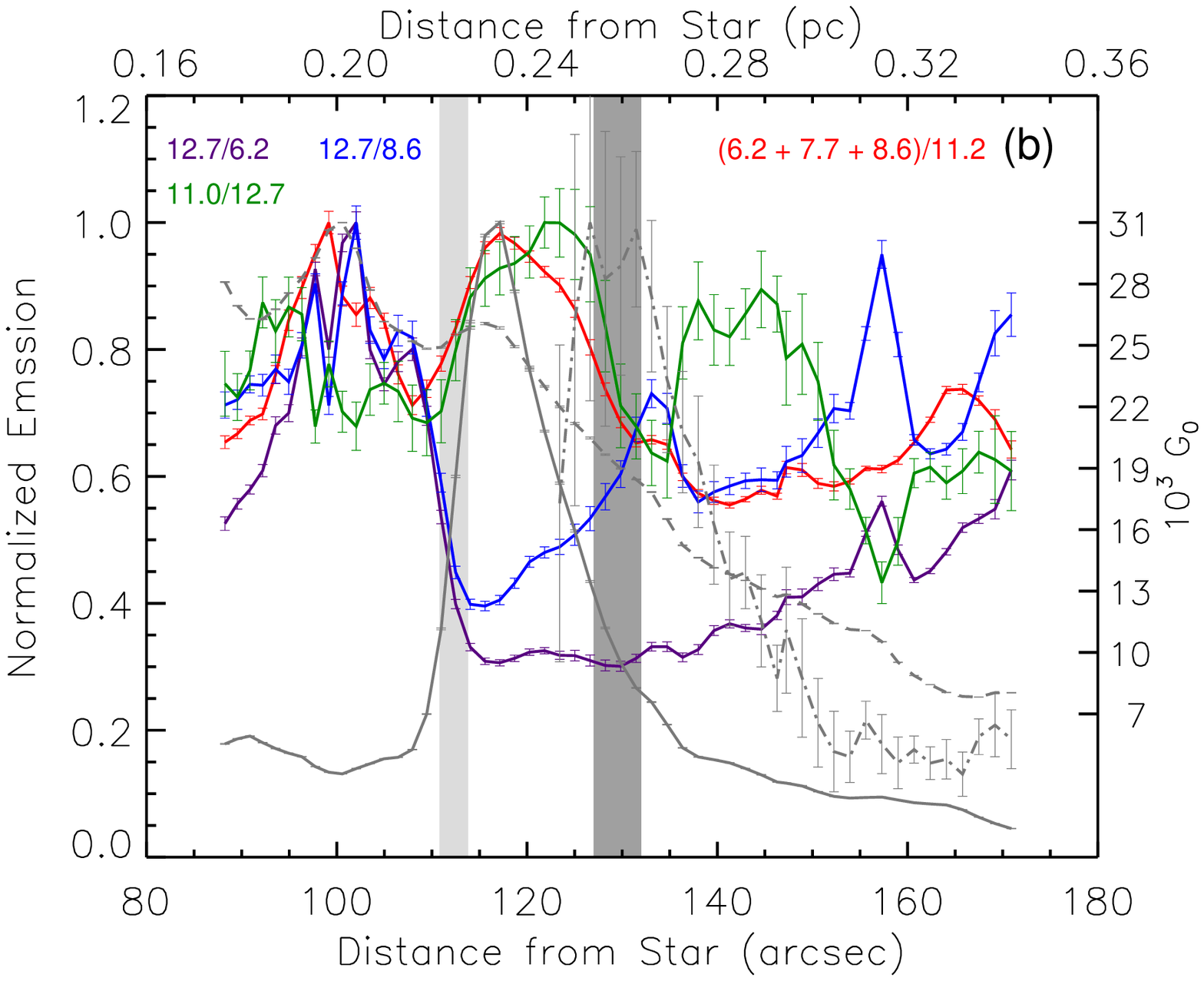}} & 
\resizebox{.5\hsize}{!}{%
\includegraphics[clip,trim =0cm 0cm 0cm  2.2cm,width=7.5cm]{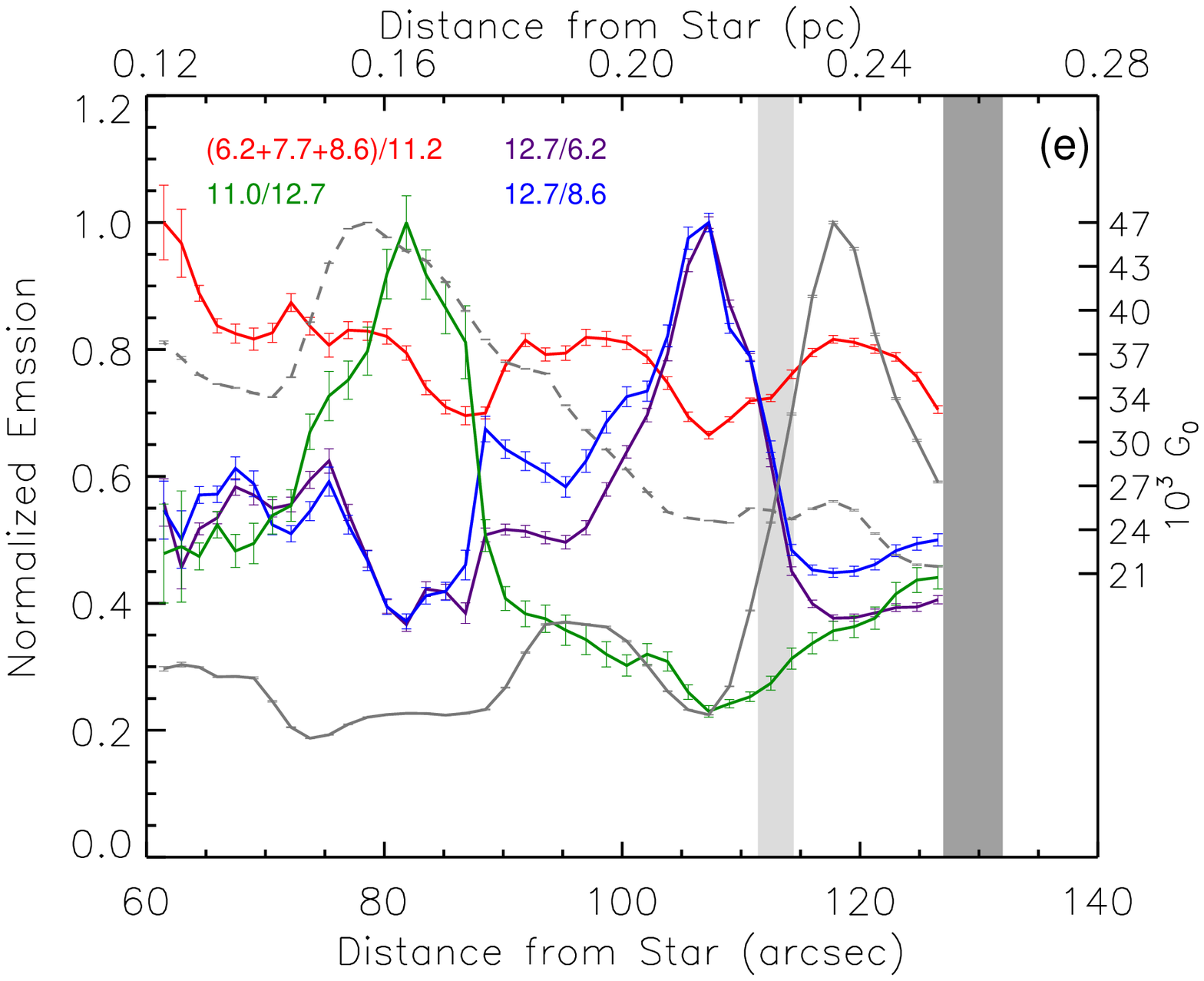}}\\
\resizebox{.5\hsize}{!}{%
\includegraphics[clip,trim =0cm 0cm 0cm 2.2cm,width=7.5cm]{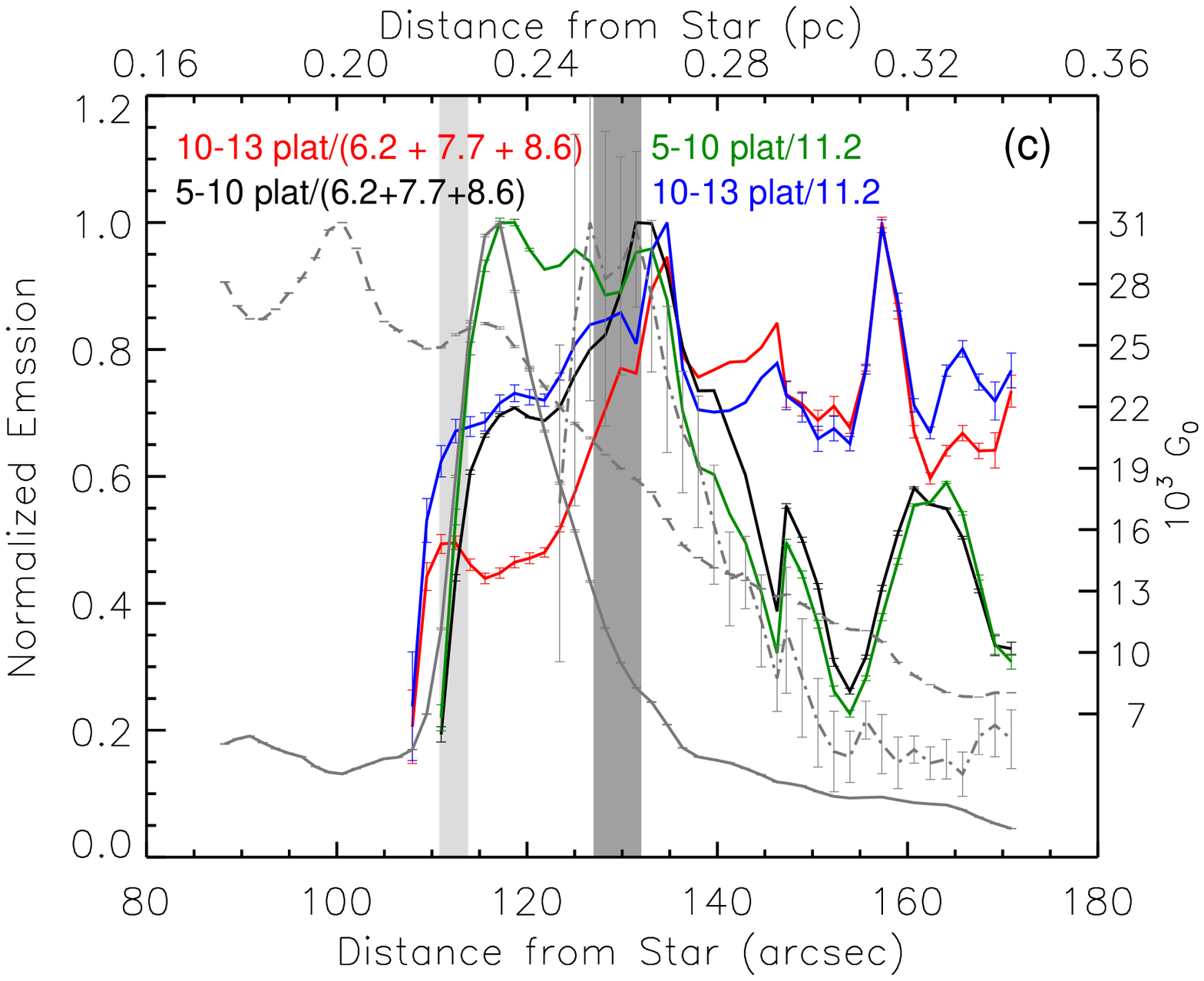}} & \\
\end{tabular}
\end{center}
\caption{Orion Bar combined (OBC, left) and Orion Bar ionized (OBI, right) emission ratio cross cuts normalized to the peak values for each ratio. The grey shaded regions correspond to the Orion Bar PDR front and the IF (see Figure~\ref{orion_line_profiles1}).  
G$_{0}$ cross cuts values are shown on the right y-axis in units of 10$^{3}$ Habings (see Section~\ref{PDR} for derivation). The PAH 7.7 $\mu$m emission, 9.7~$\mu$m H$_{2}$ line, and G$_{0}$ cross cuts are shown in grey as a solid, dot--dashed and dashed line respectively. Error bars for each emission ratio are given in the same color as the associated cross cut.}
\label{orion_line_profiles5}
\end{figure*}

In this section, we present a selection of supplementary emission ratio cross cuts within both Orion Bar apertures described in the text (Figure~\ref{orion_line_profiles5}).

\section{Correlation plots}
\label{corr_add}

\begin{figure*}
\begin{center}
\resizebox{\hsize}{!}{%
\includegraphics[clip,trim =0cm 0cm 0cm 0cm,width=0.33\textwidth]{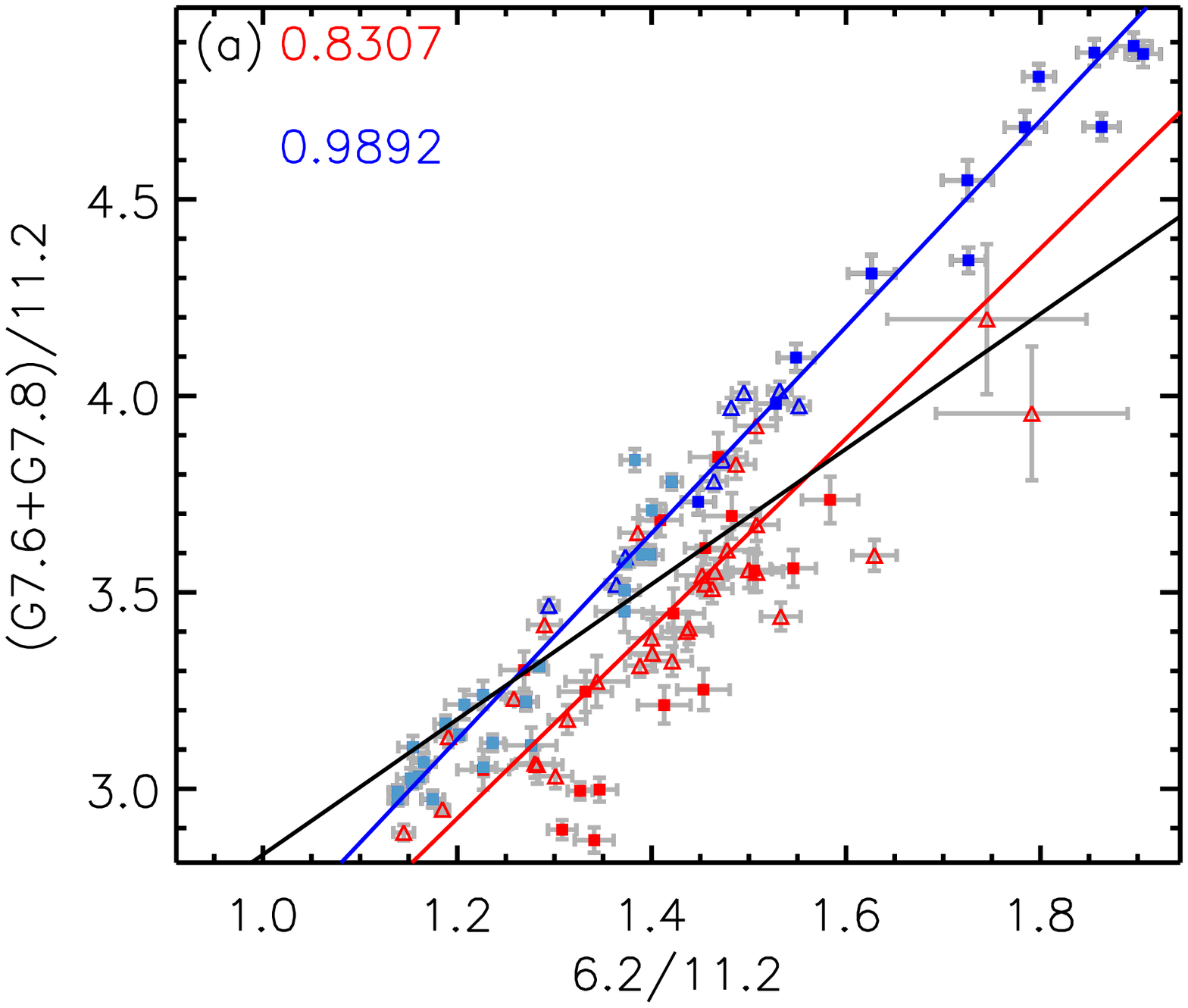}
\includegraphics[clip,trim =0cm 0cm 0cm 0cm,width=0.33\textwidth]{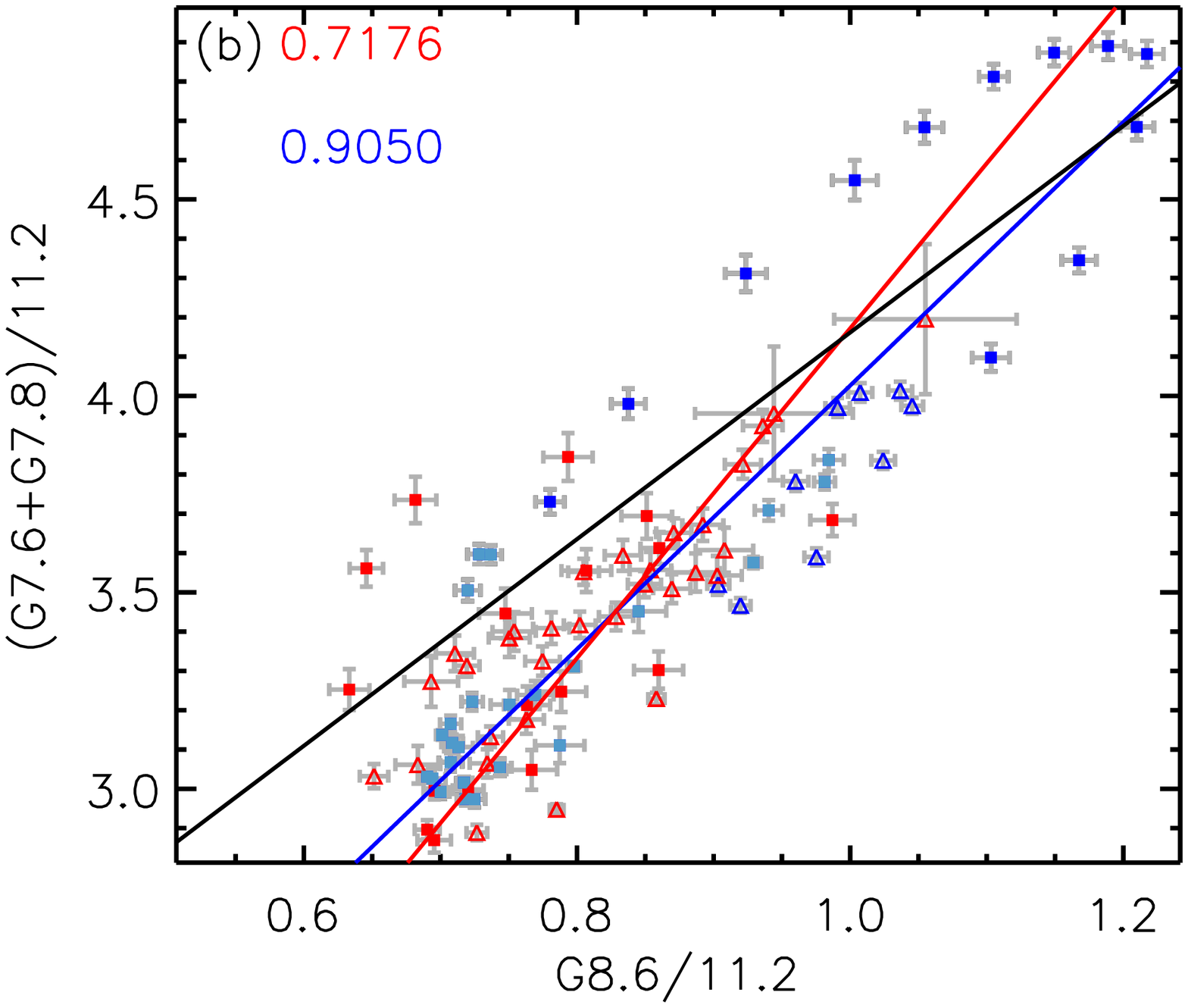}
\includegraphics[clip,trim =0cm 0cm 0cm 0cm,width=0.33\textwidth]{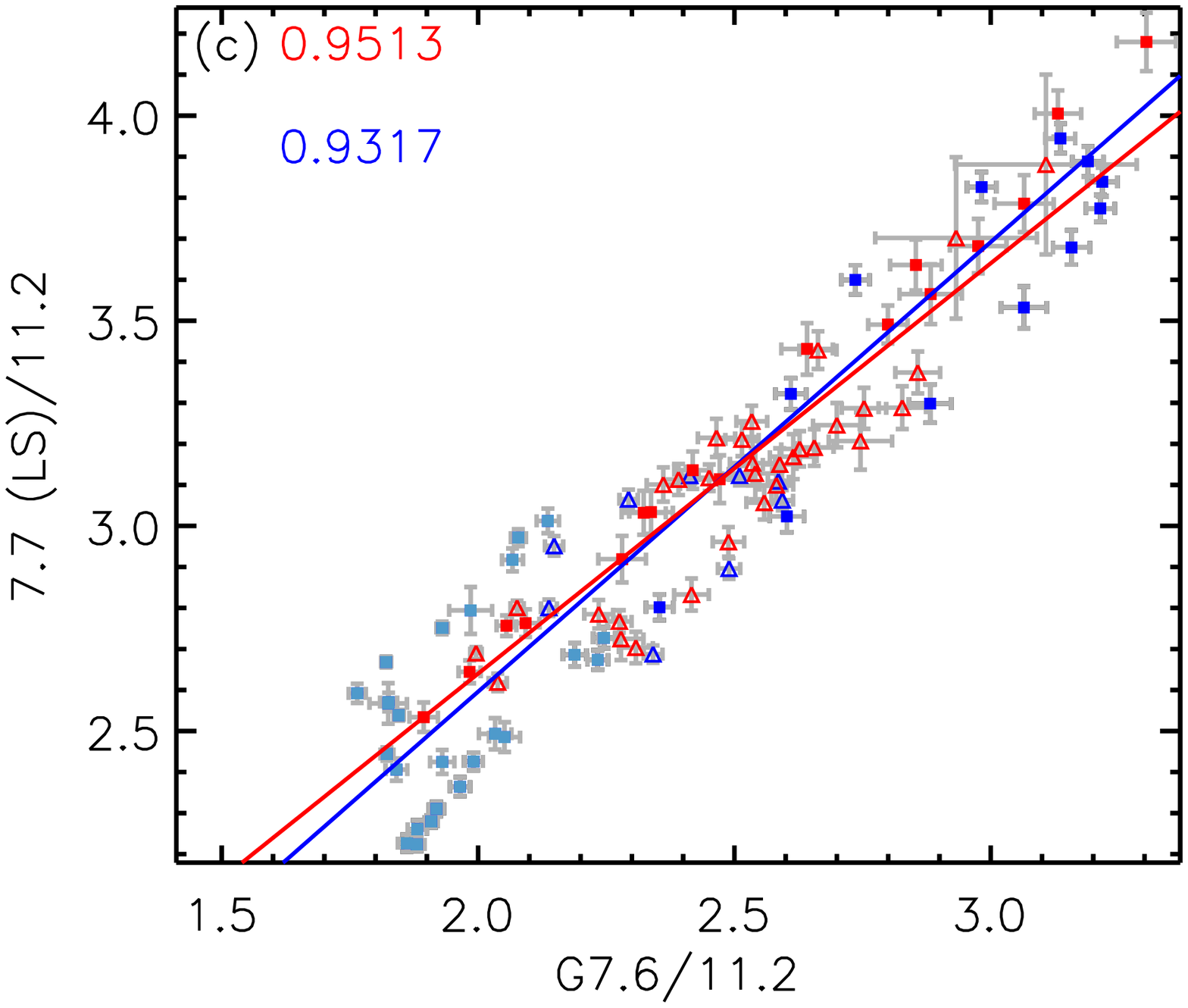}}
\resizebox{\hsize}{!}{%
\includegraphics[clip,trim =0cm 0cm 0cm 0cm,width=0.33\textwidth]{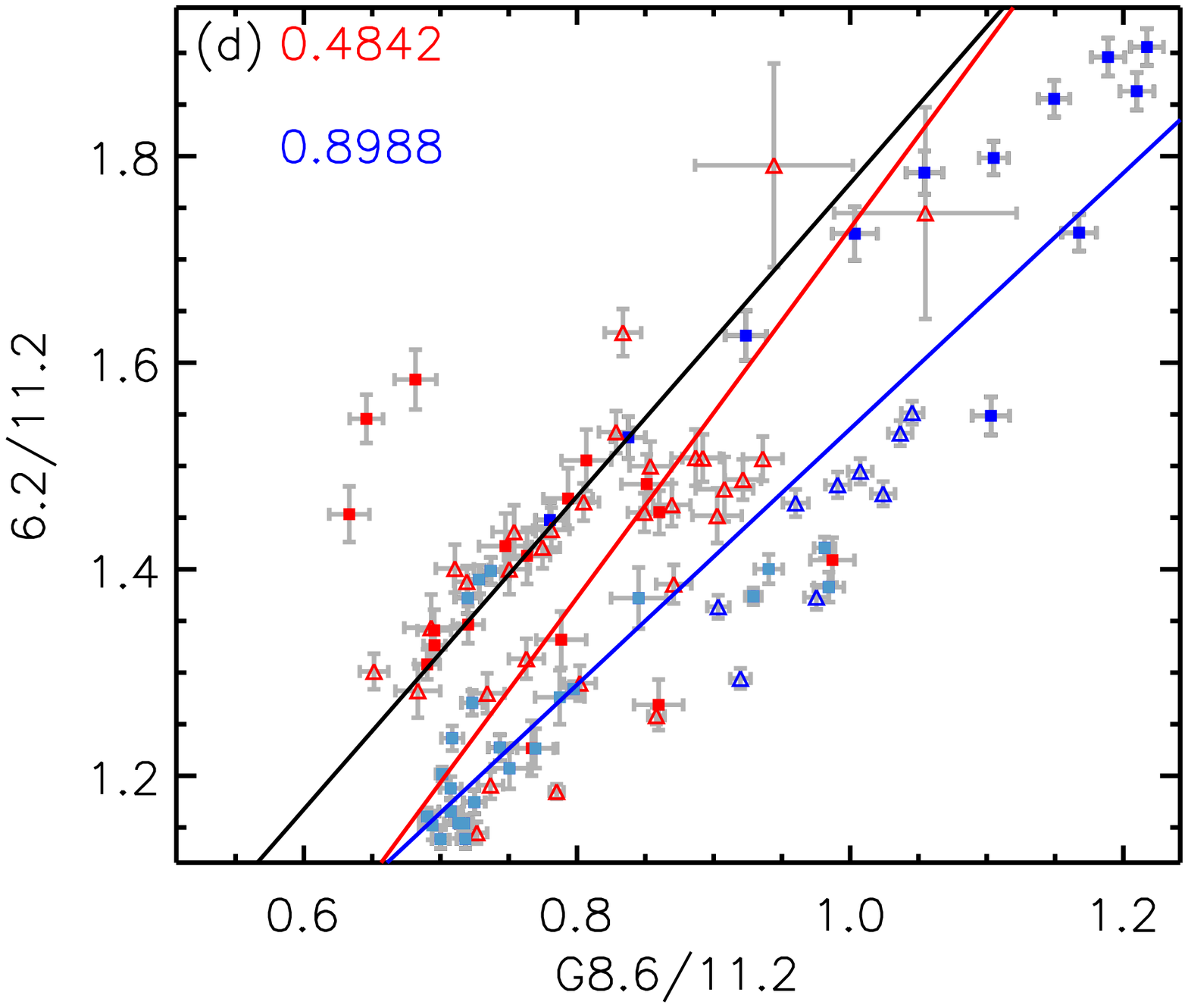}
\includegraphics[clip,trim =0cm 0cm 0cm 0cm,width=0.33\textwidth]{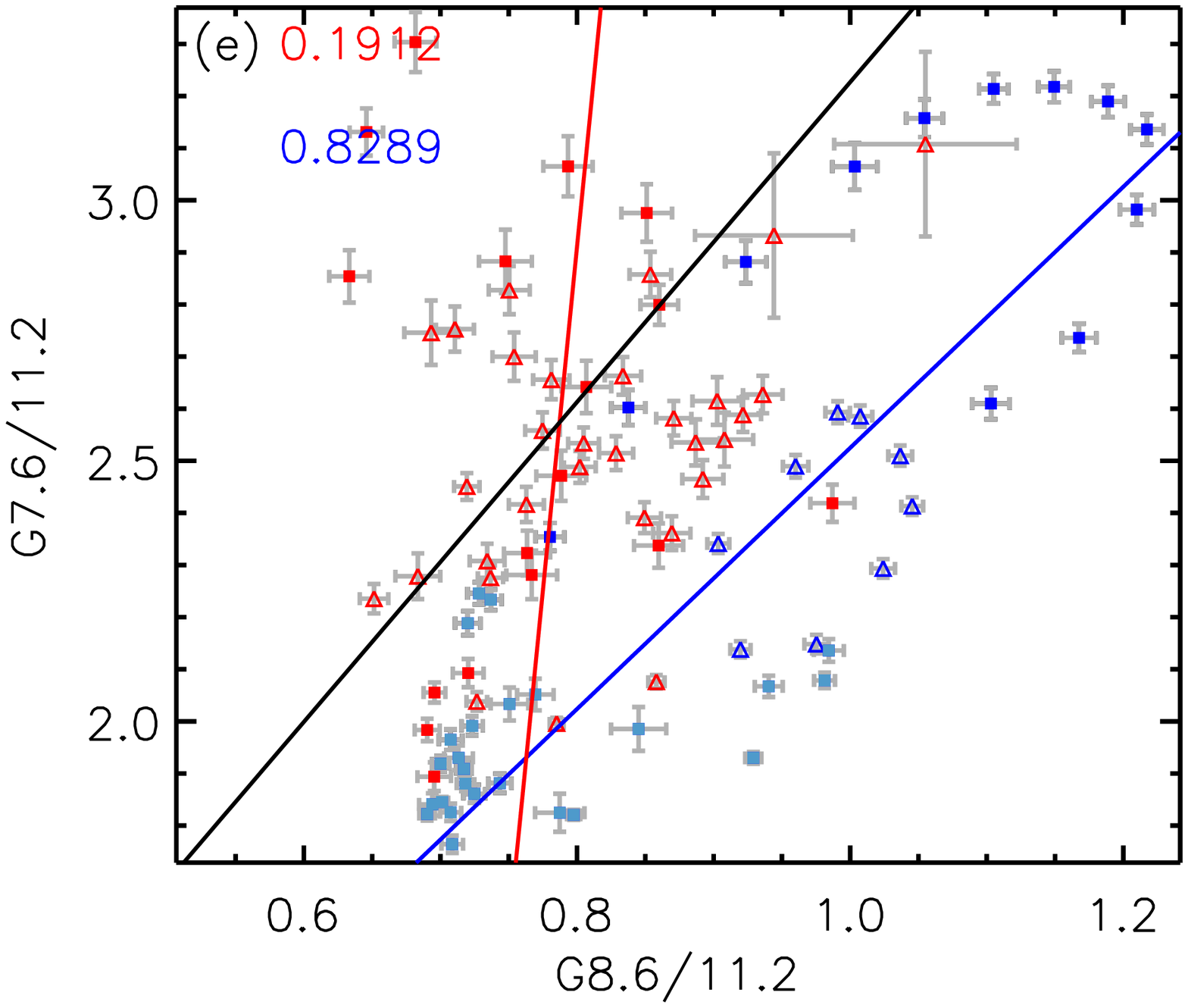}
\includegraphics[clip,trim =0cm 0cm 0cm 0cm,width=0.33\textwidth]{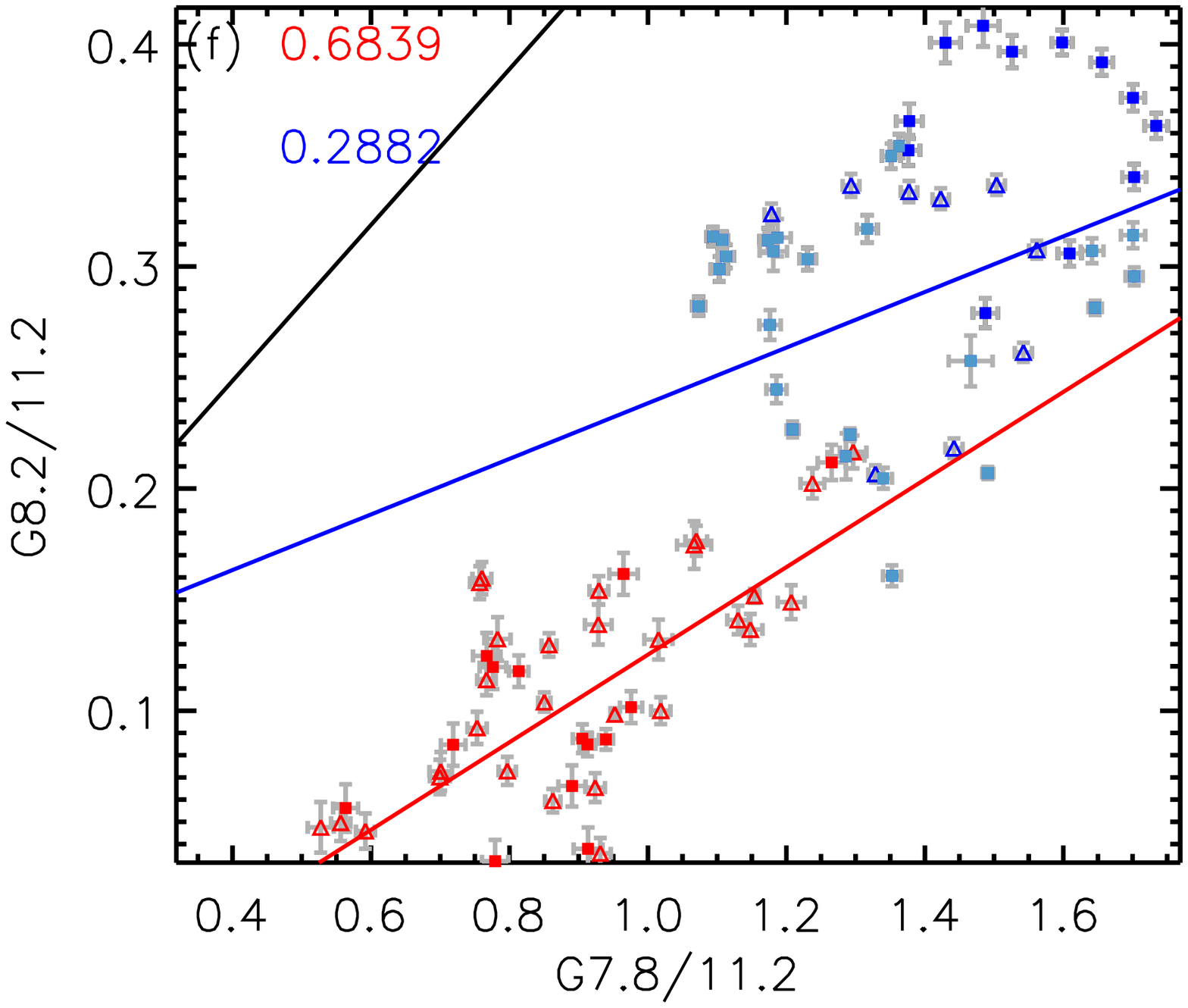}}

\includegraphics[clip,trim =0cm 0cm 0cm 0cm,width=0.33\textwidth]{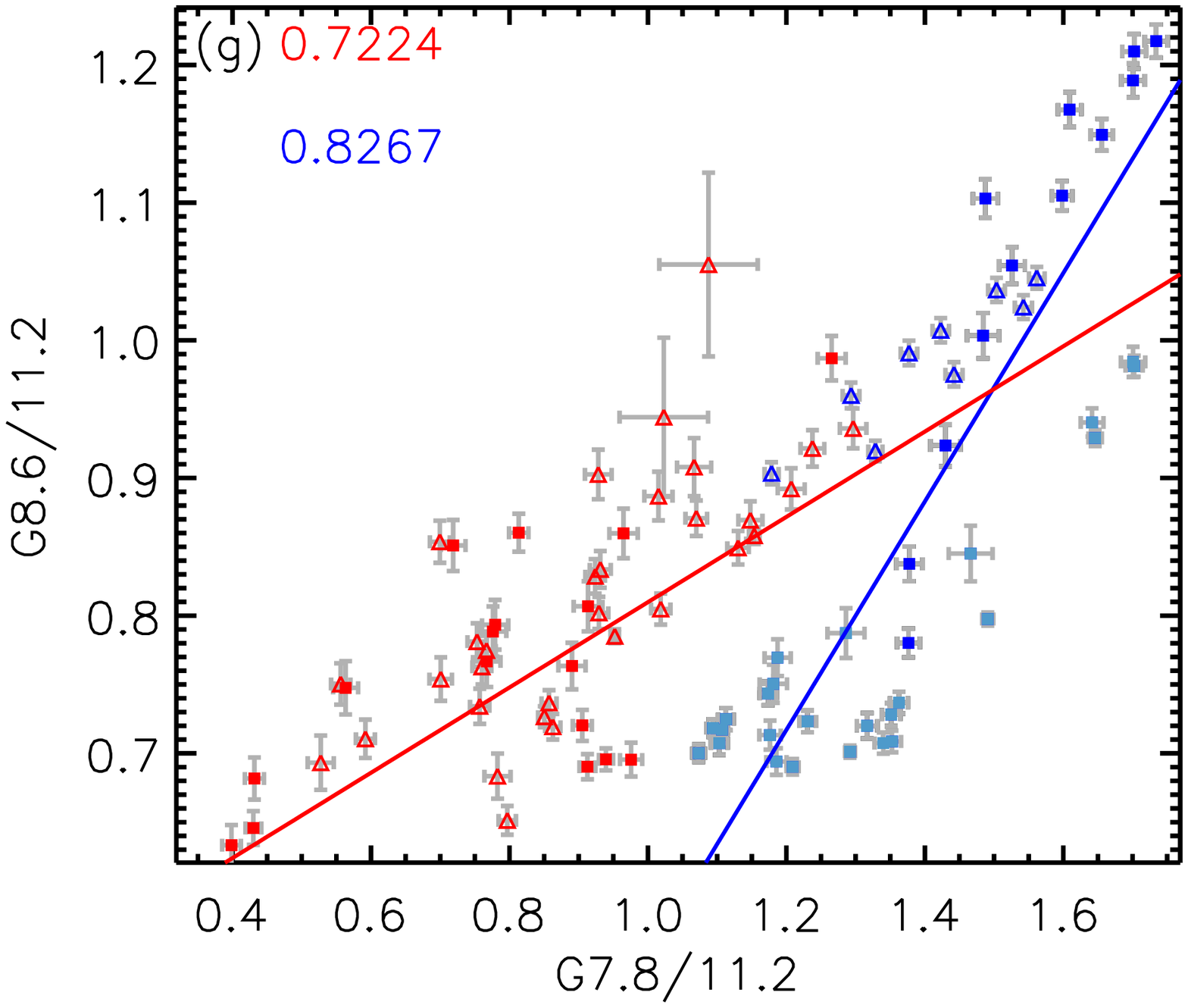}
\includegraphics[clip,trim =0cm 0cm 0cm 0cm,width=0.33\textwidth]{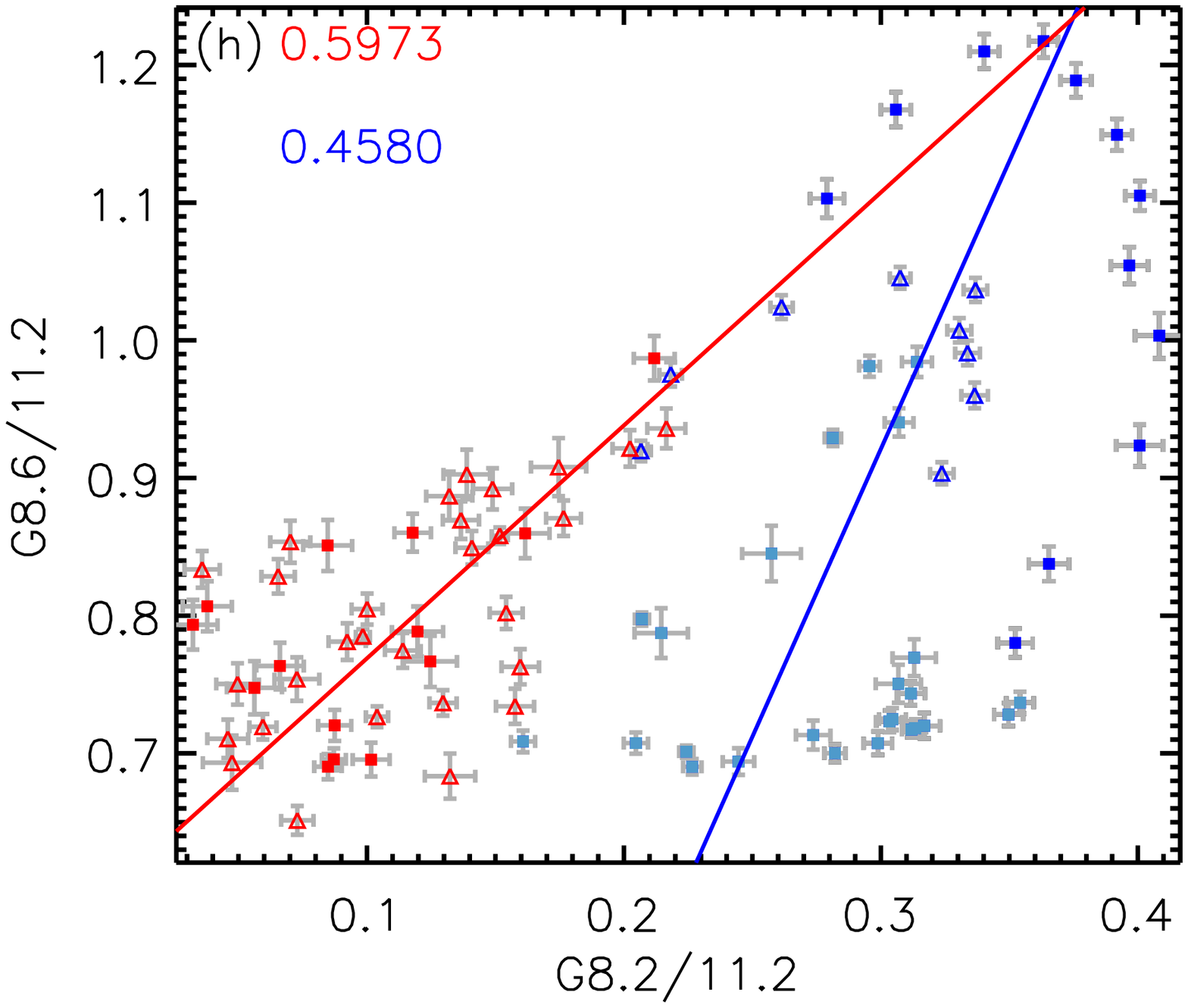}
\resizebox{\hsize}{!}{%
\includegraphics[clip,trim =0cm 0cm 0cm 0cm,width=0.33\textwidth]{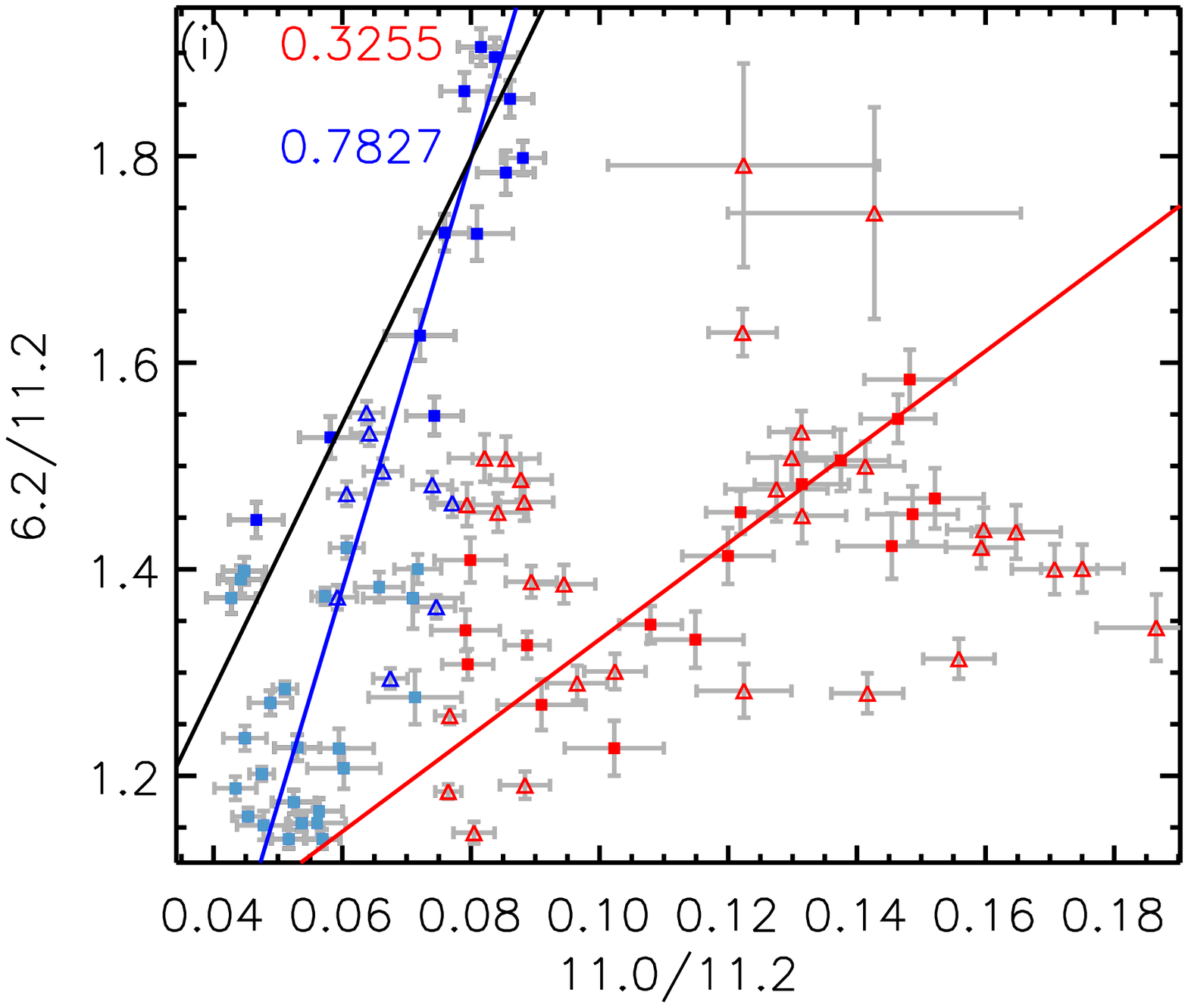}
\includegraphics[clip,trim =0cm 0cm 0cm 0cm,width=0.33\textwidth]{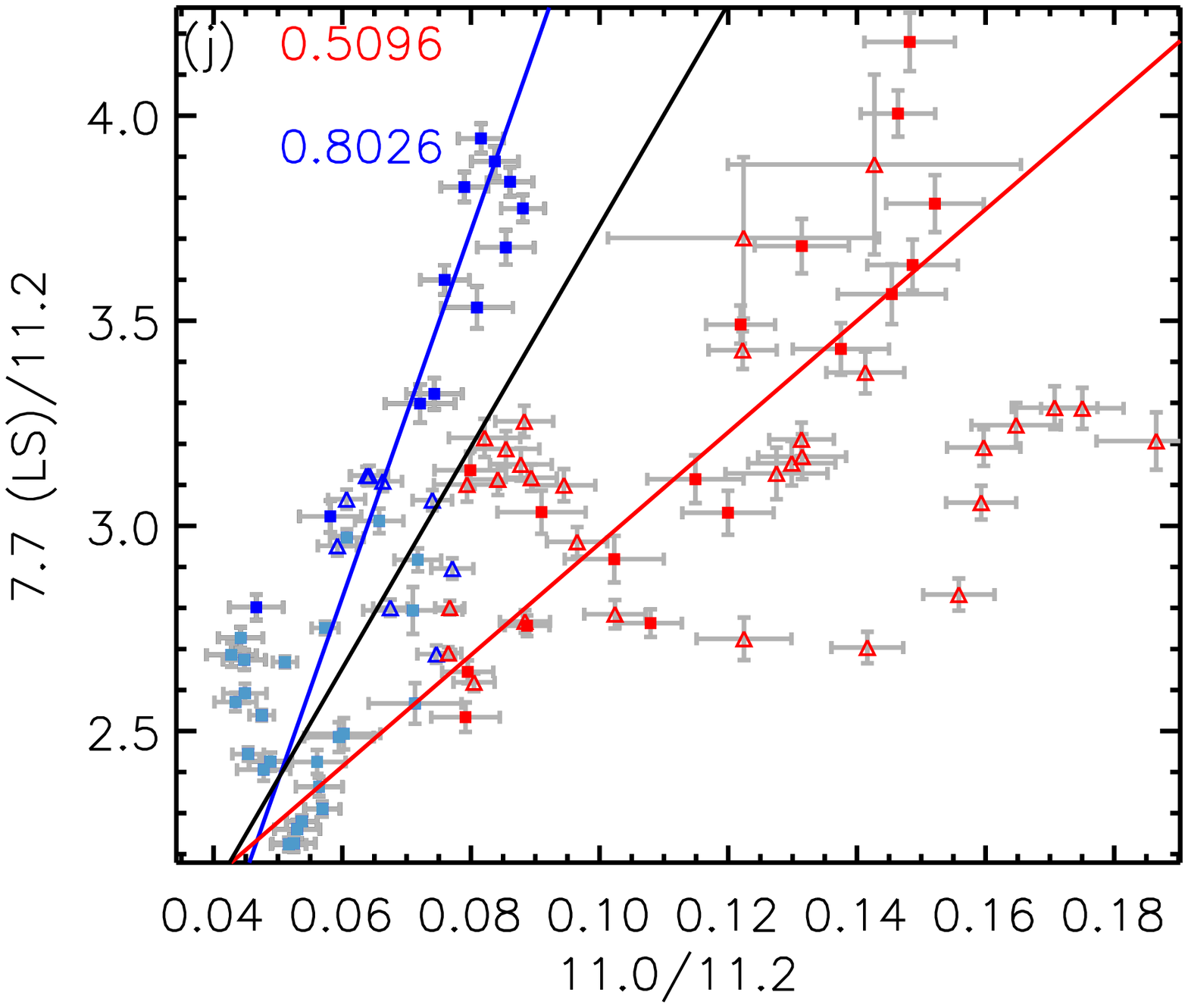}
\includegraphics[clip,trim =0cm 0cm 0cm 0cm,width=0.33\textwidth]{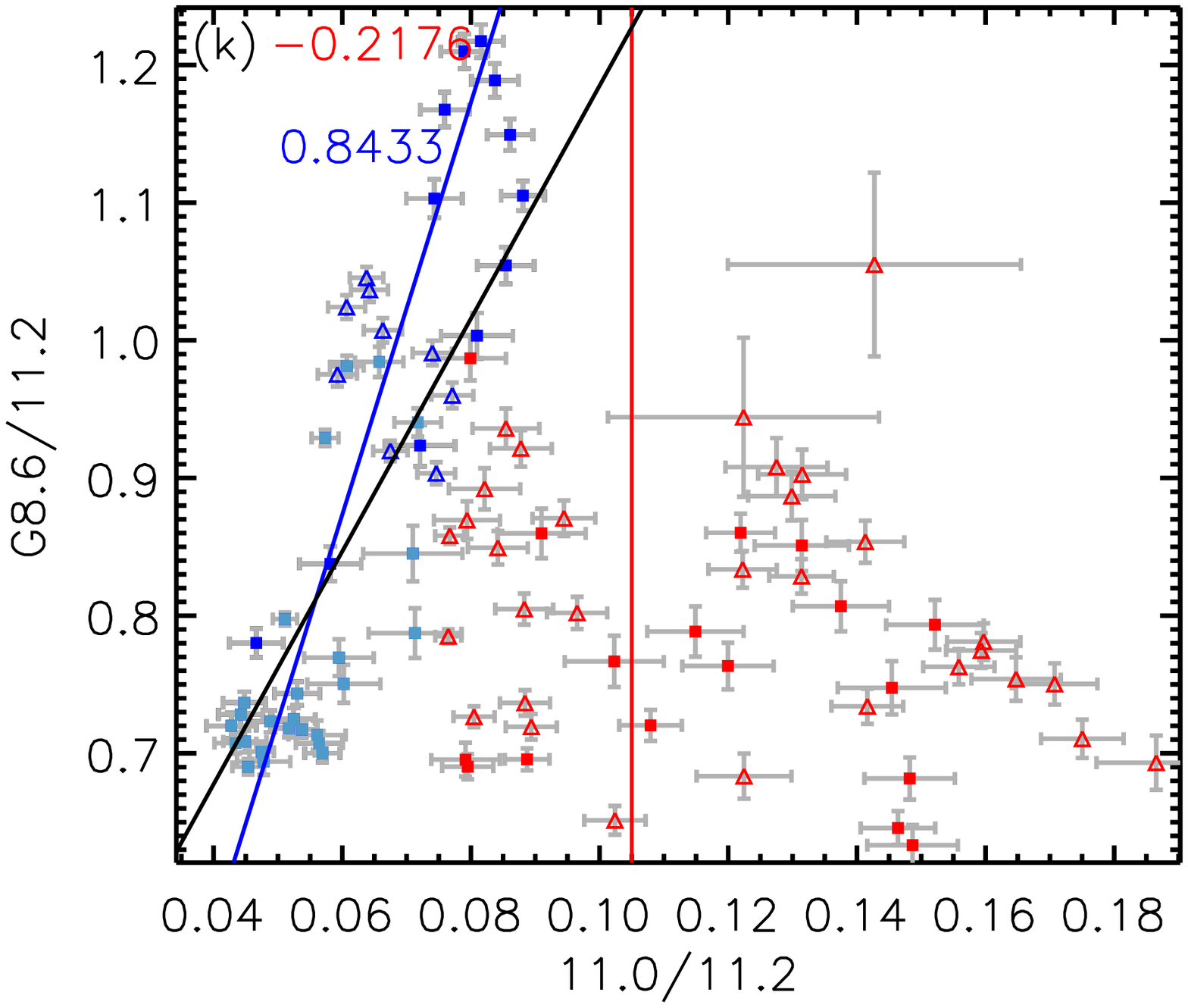}
}
\end{center}
\caption{Correlation plots within the Orion Bar Combined (OBC, squares) and the Orion Bar Ionized (OBI, triangles): ratios from within the Orion Bar PDR are shown in blue, ratios from the PDR spectra behind the Orion Bar PDR front in light blue ( > 131.5$^{\prime\prime}$ from the Trapezium), and ratios from the \HII\, region (PDR) in red (see shaded areas in panel (a) of Figure~\ref{orion_cont_10_14}). Correlation coefficients for the PDR behind the Orion Bar IF (i.e both blue and light blue data points) and the \HII\, region (PDR; red data points) are given in blue and red respectively. Weighted linear fits are shown as solid lines for each respective region given in the same colors as the correlation coefficients. The black lines correspond to the respective correlation fits found for NGC~2023 in \citet{pee17}.}
\label{orion_corrD1}
\end{figure*}

\begin{figure*}
\begin{center}
\resizebox{\hsize}{!}{%
\includegraphics[clip,trim =0cm 0cm 0cm
0cm,width=0.33\textwidth]{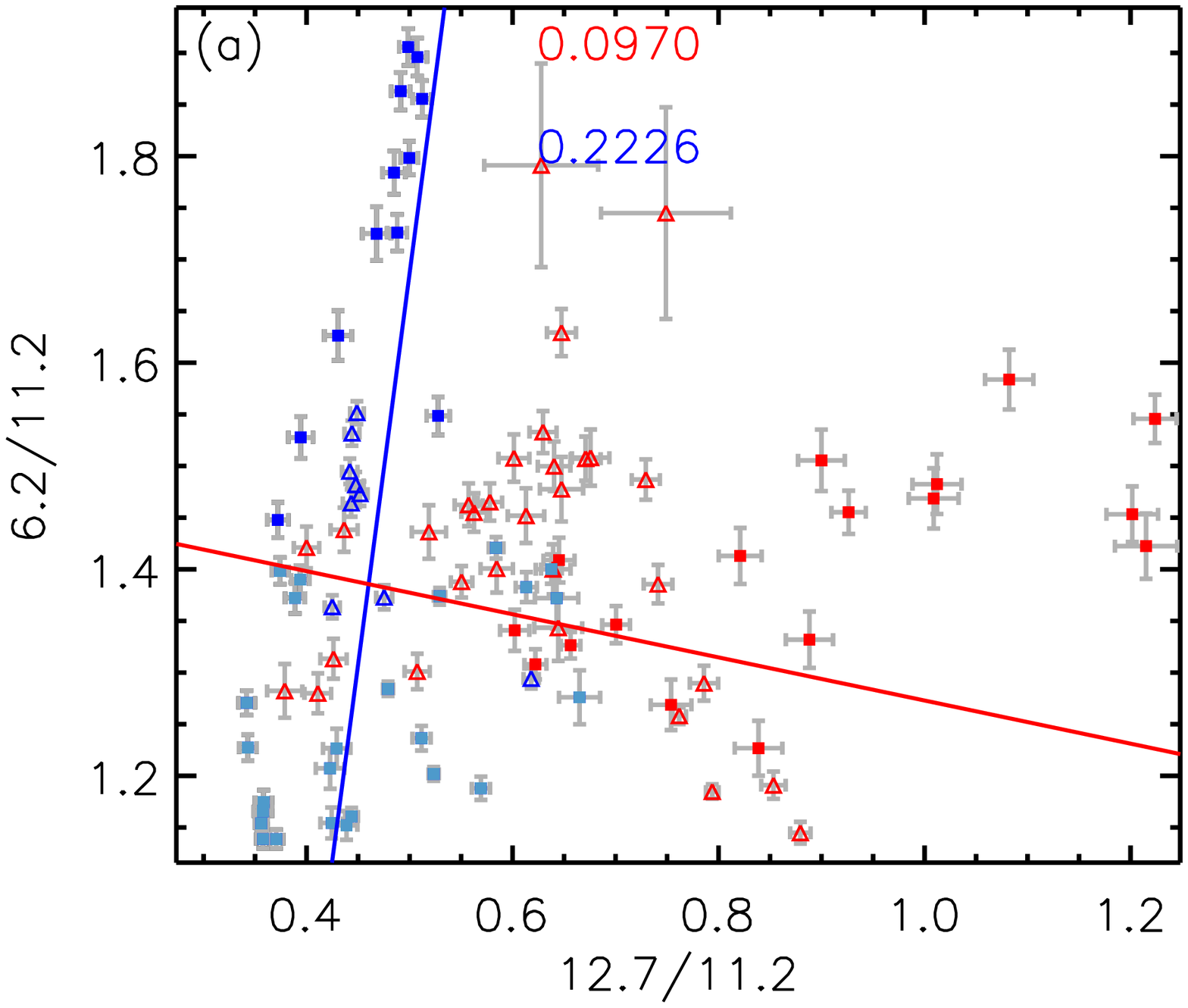}
\includegraphics[clip,trim =0cm 0cm 0cm 0cm,width=0.33\textwidth]{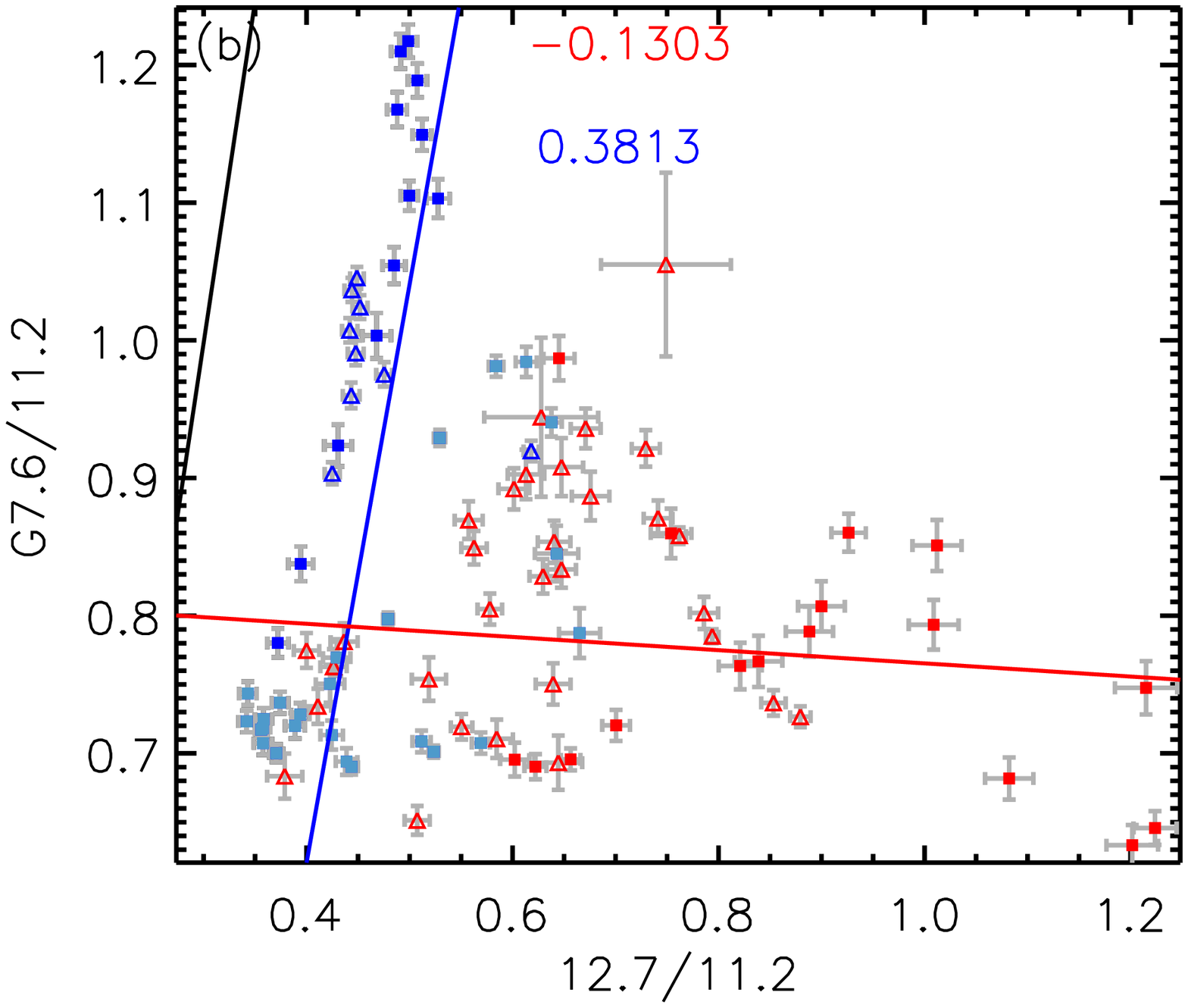}
\includegraphics[clip,trim =0cm 0cm 0cm 0cm,width=0.33\textwidth]{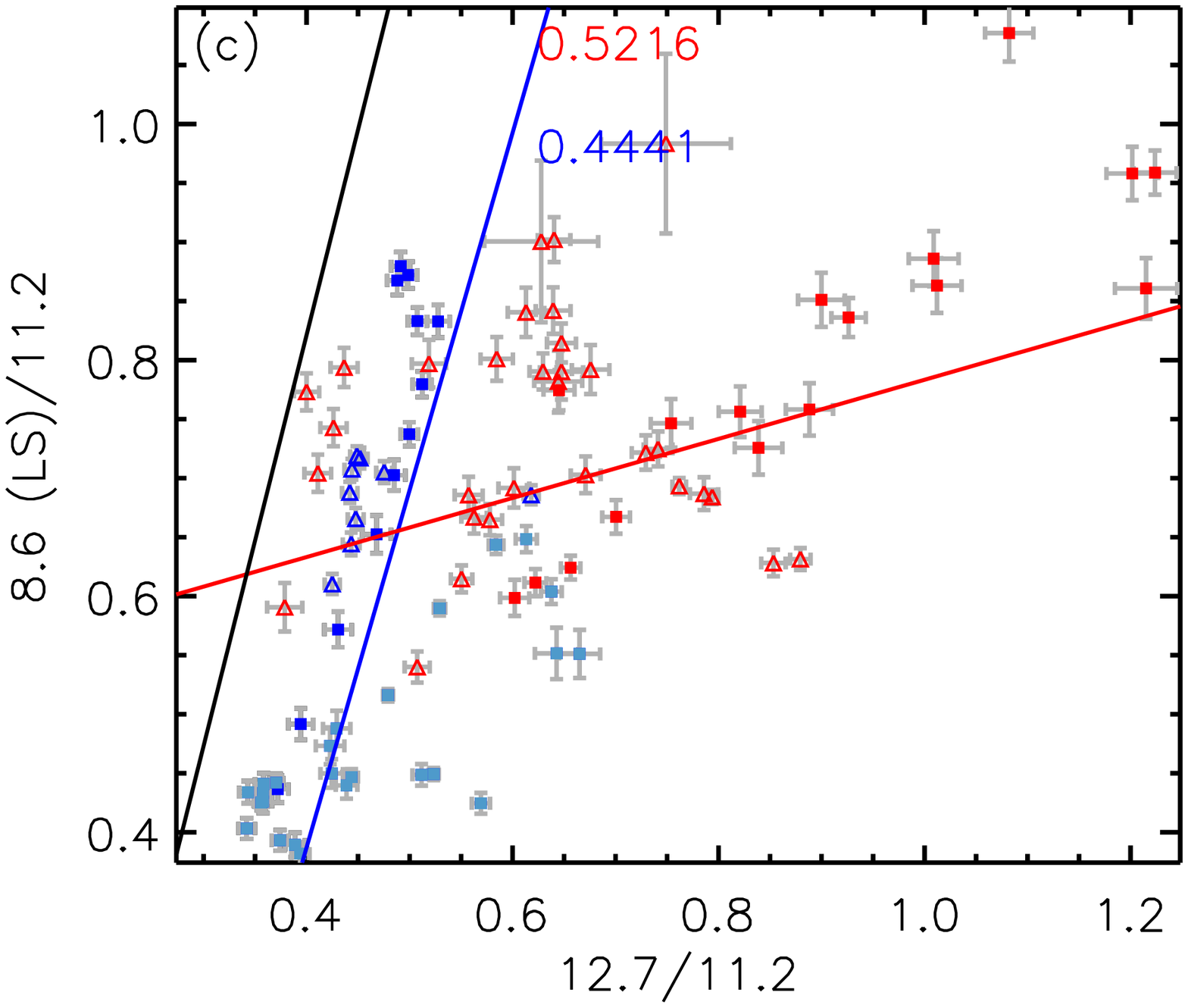}}
\resizebox{\hsize}{!}{%
\includegraphics[clip,trim =0cm 0cm 0cm  0cm,width=0.25\textwidth]{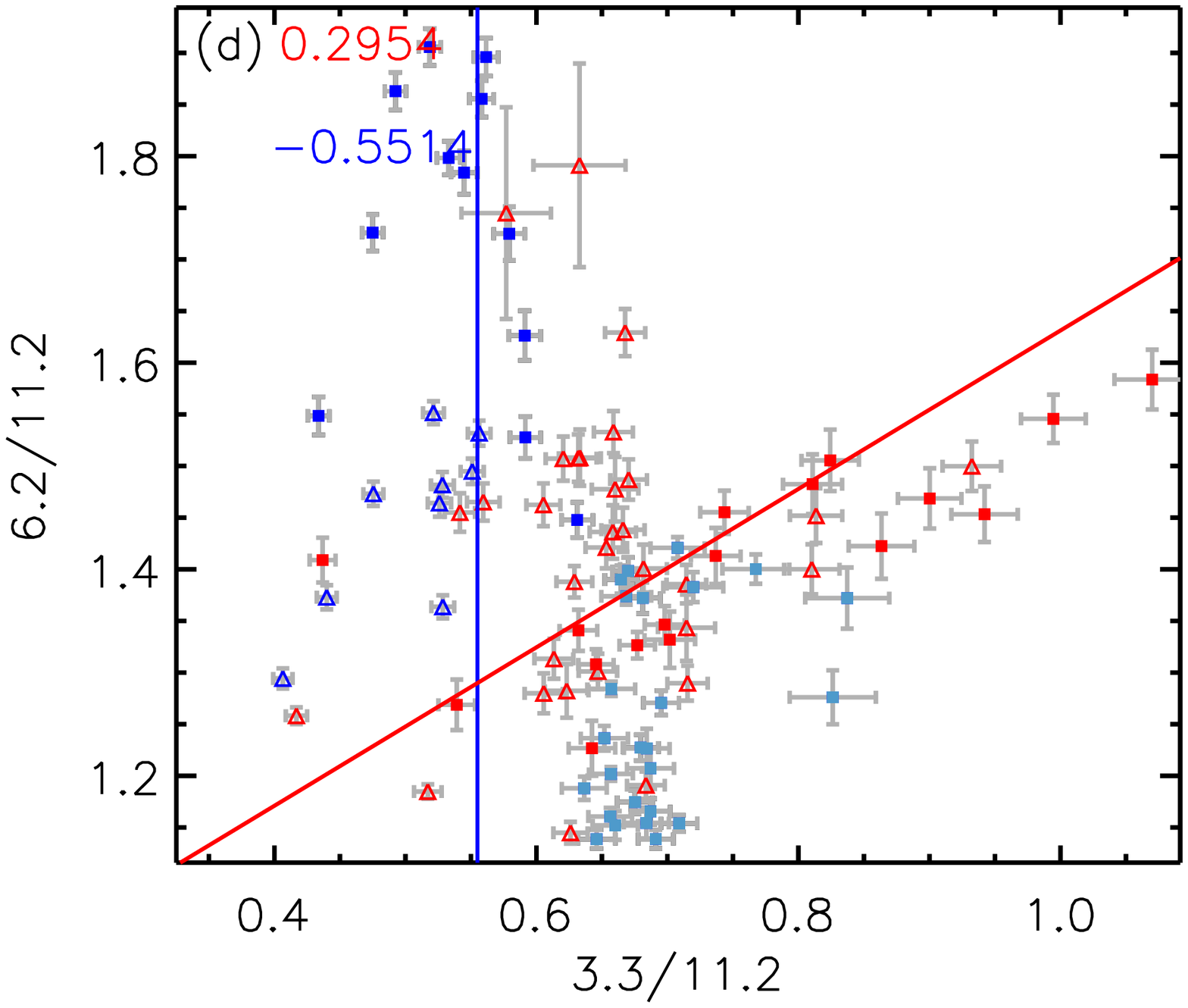}
\includegraphics[clip,trim =0cm 0cm 0cm 0cm,width=0.25\textwidth]{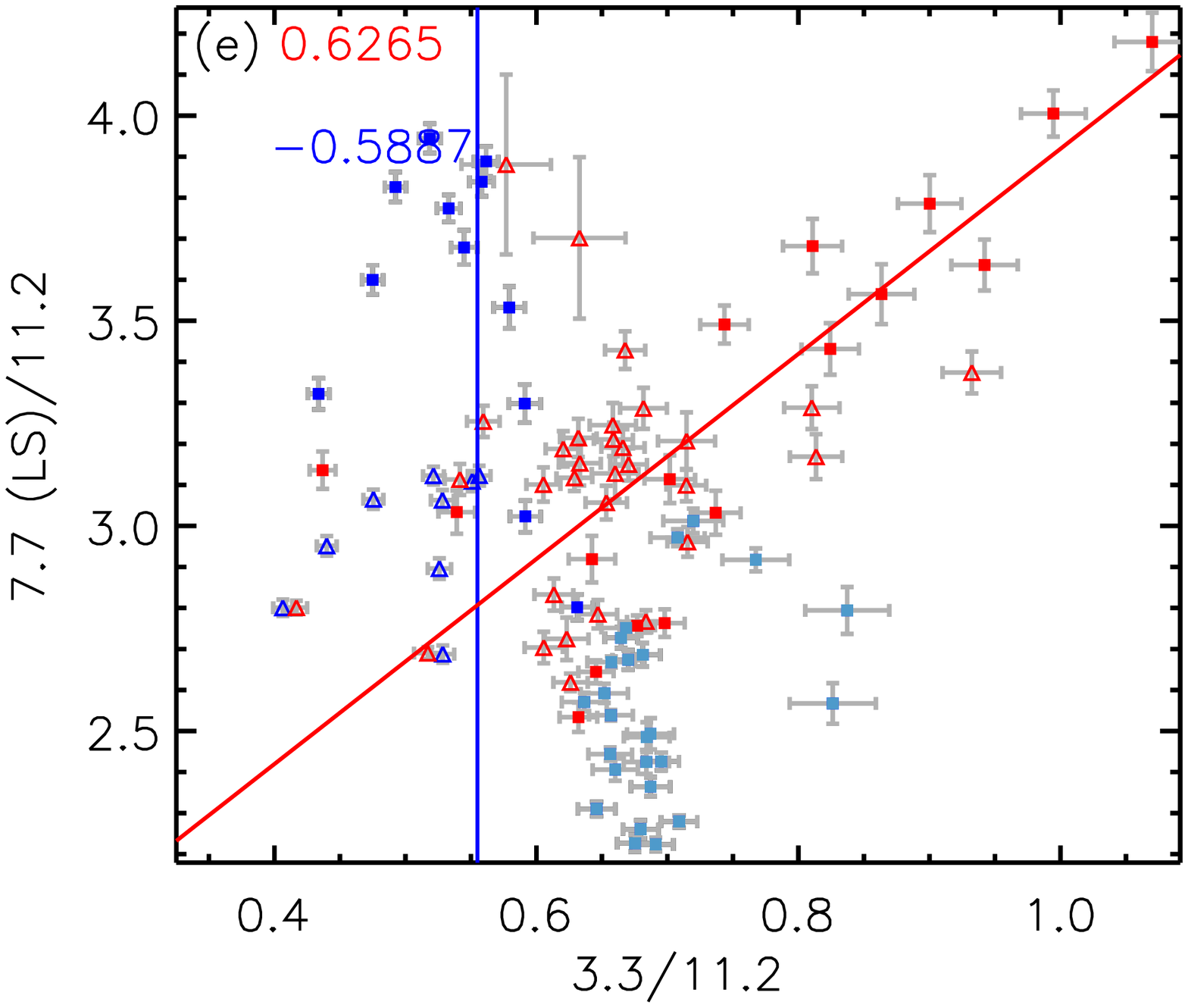}
\includegraphics[clip,trim =0cm 0cm 0cm 0cm,width=0.25\textwidth]{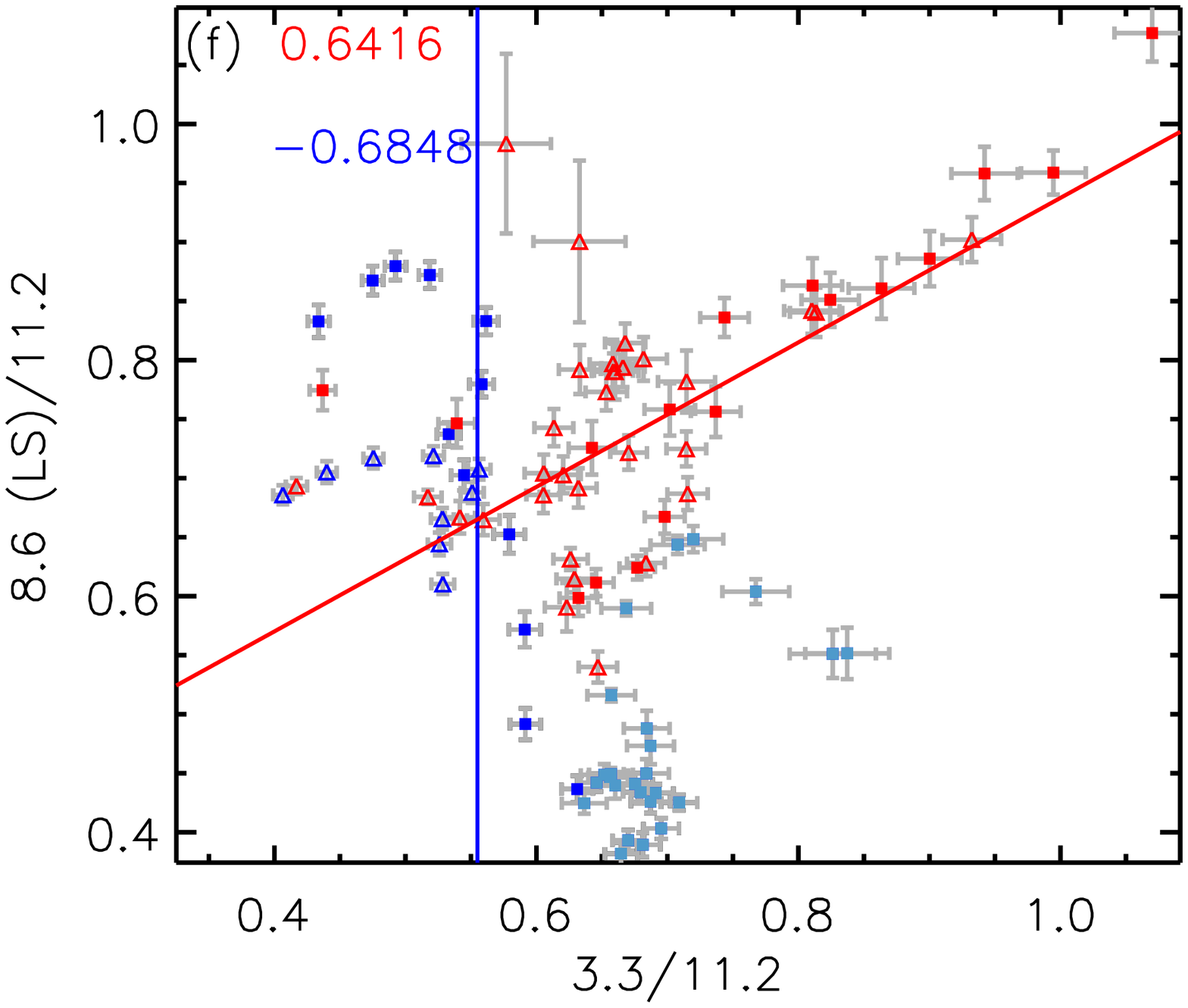}}
\resizebox{\hsize}{!}{%
\includegraphics[clip,trim =0cm 0cm 0cm 0cm,width=0.25\textwidth]{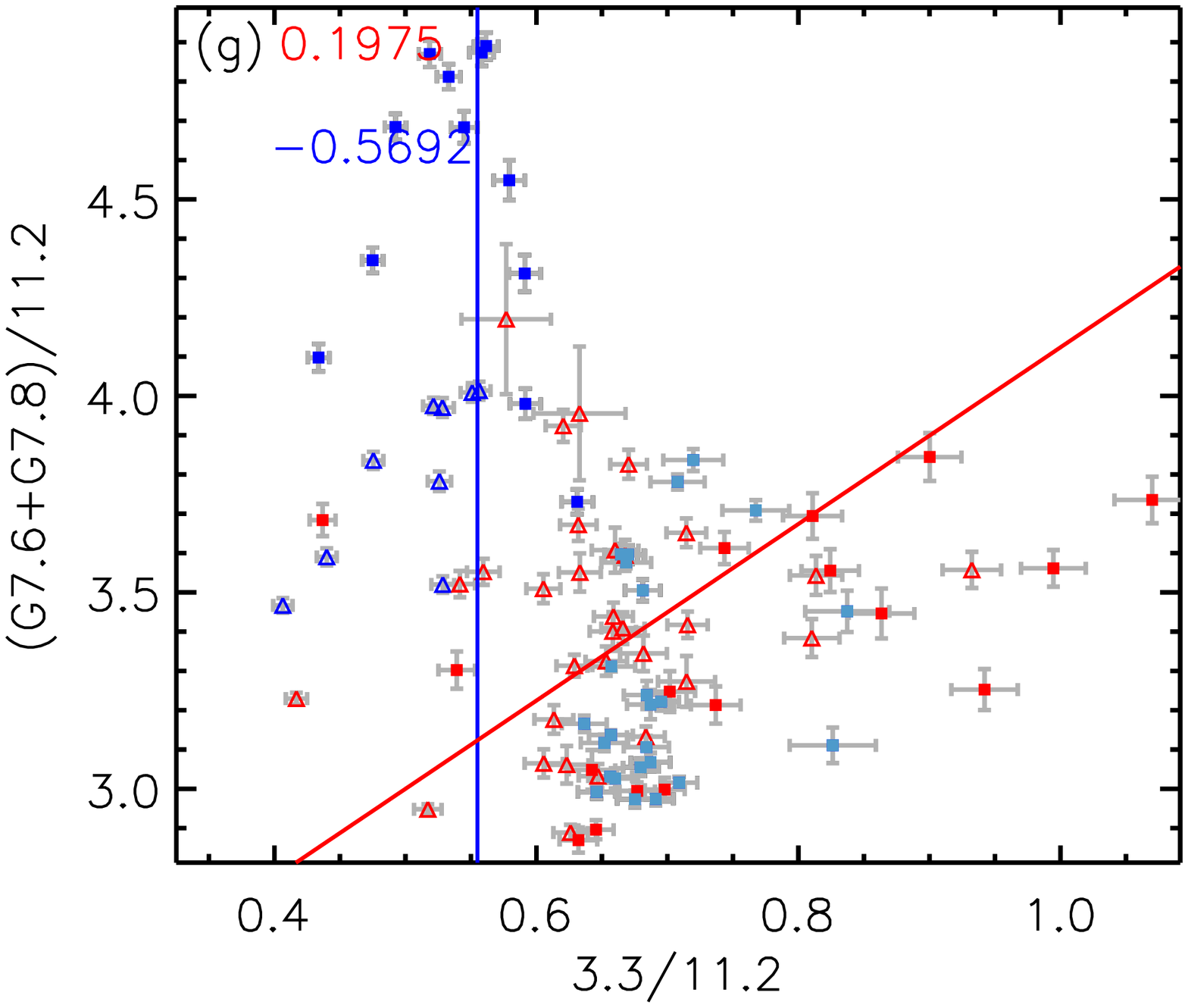}
\includegraphics[clip,trim =0cm 0cm 0cm 0cm,width=0.25\textwidth]{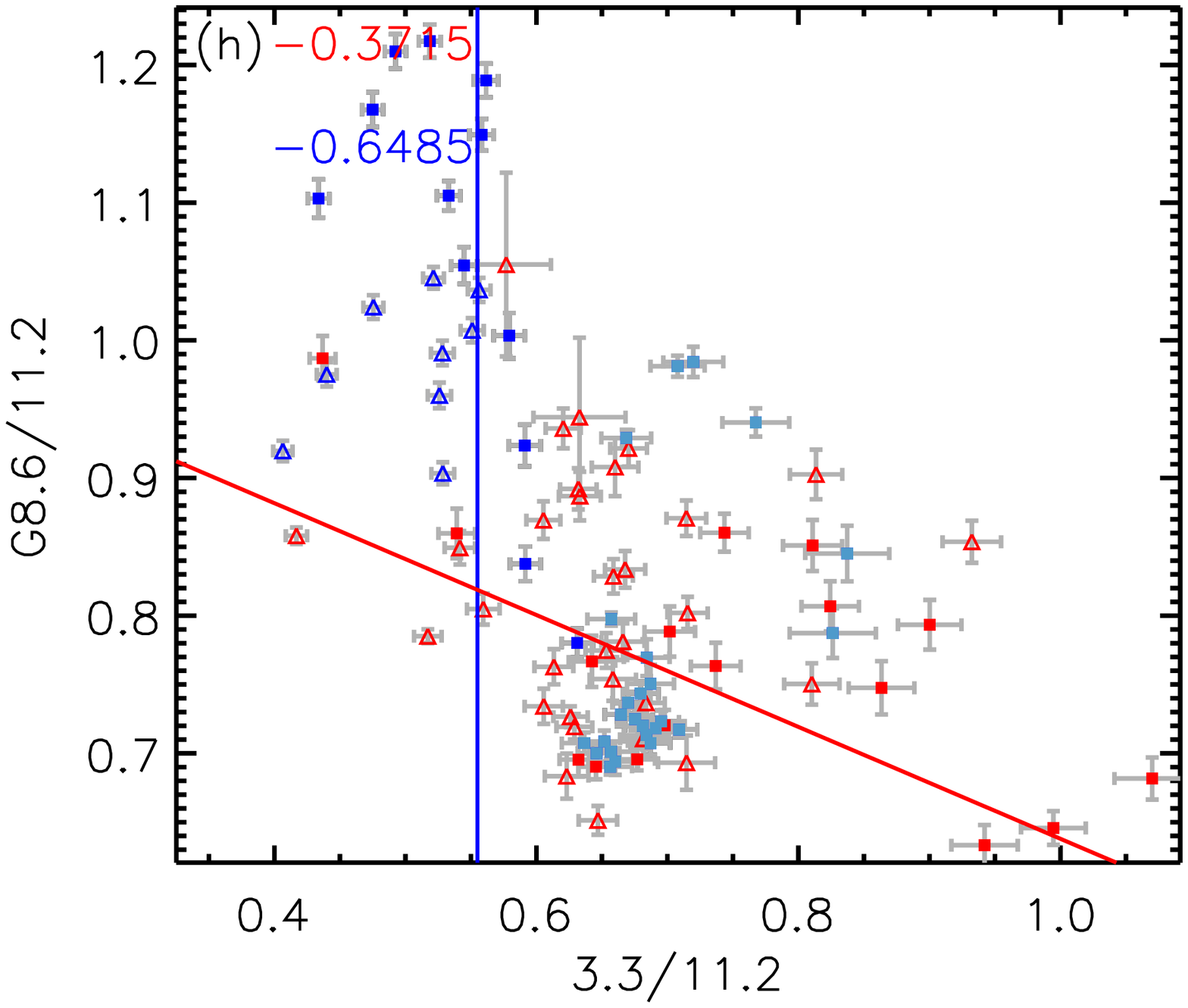}
\includegraphics[clip,trim =0cm 0cm 0cm
0cm,width=0.25\textwidth]{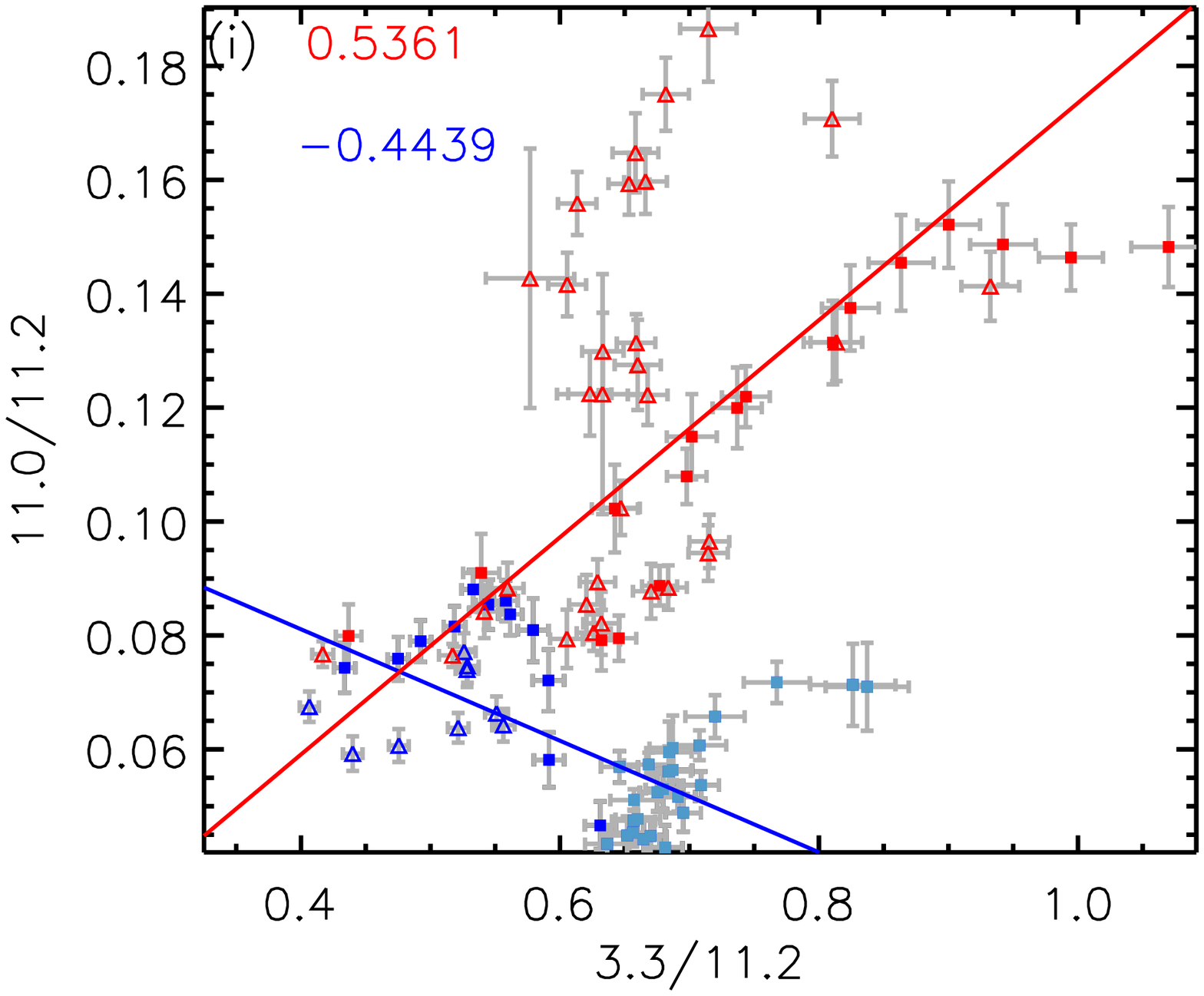}}
\resizebox{\hsize}{!}{%
\includegraphics[clip,trim =0cm 0cm 0cm 0cm,width=0.25\textwidth]{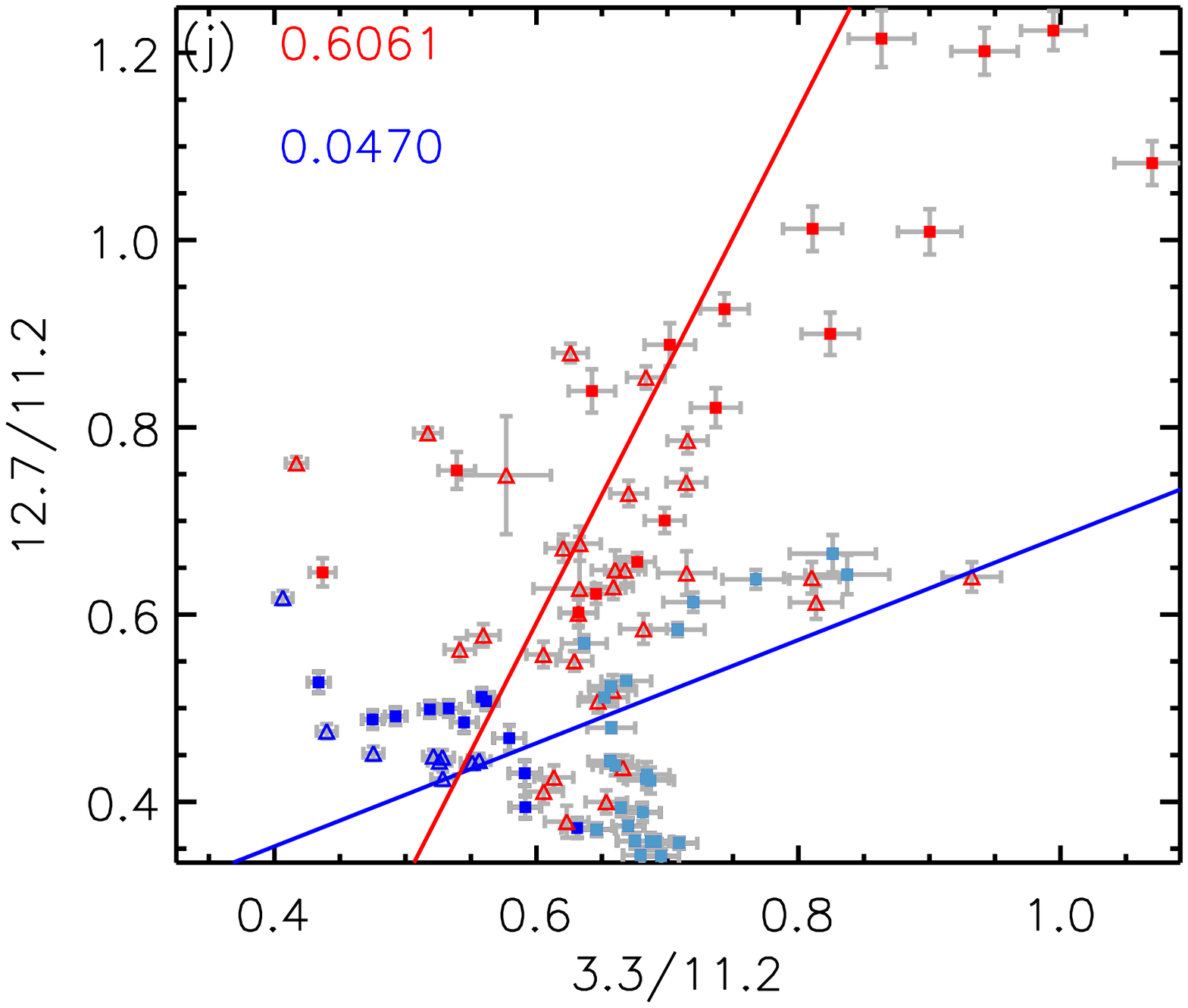}
\includegraphics[clip,trim =0cm 0cm 0cm 0cm,width=0.25\textwidth]{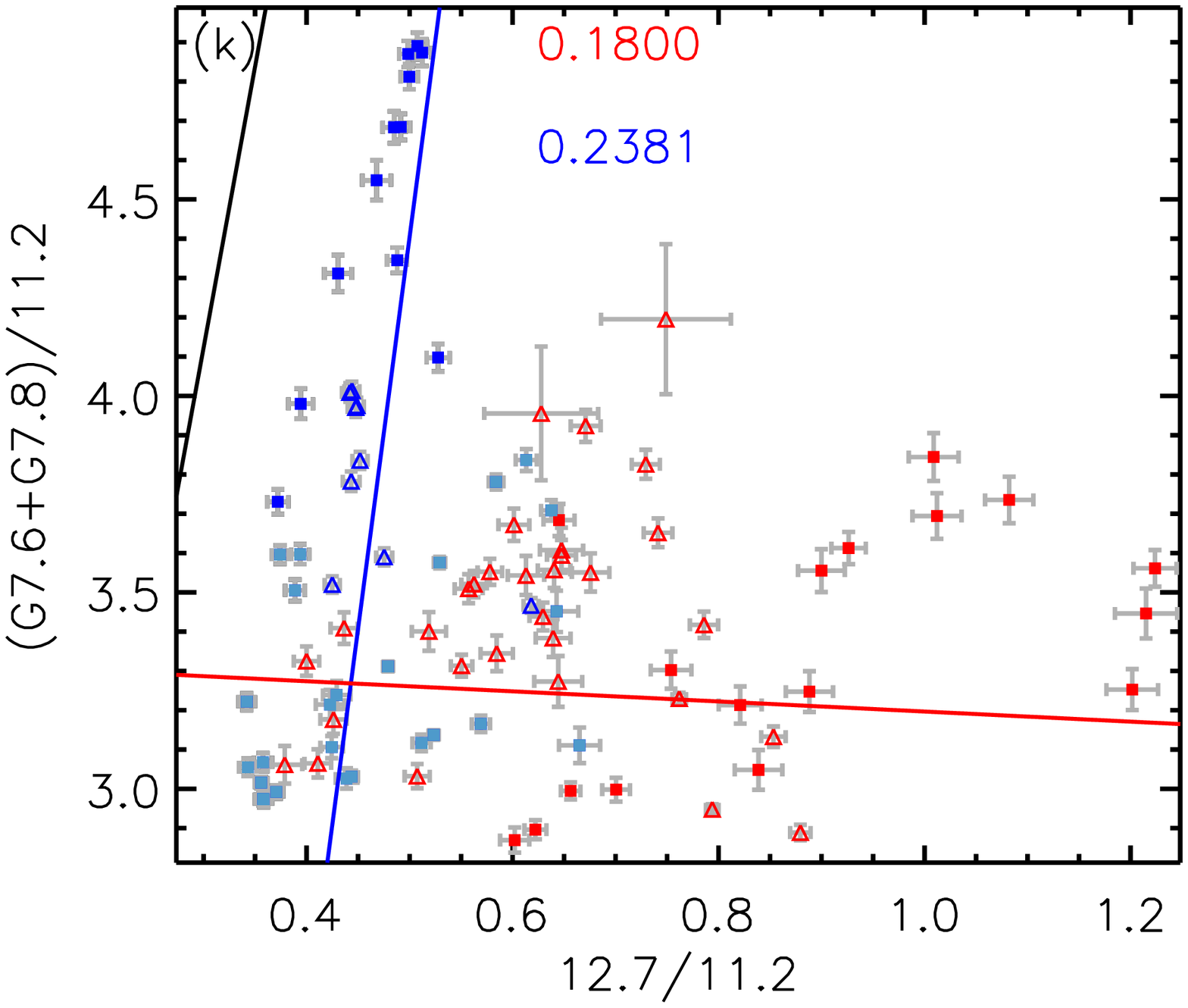}
\includegraphics[clip,trim =0cm 0cm 0cm
0cm,width=0.25\textwidth]{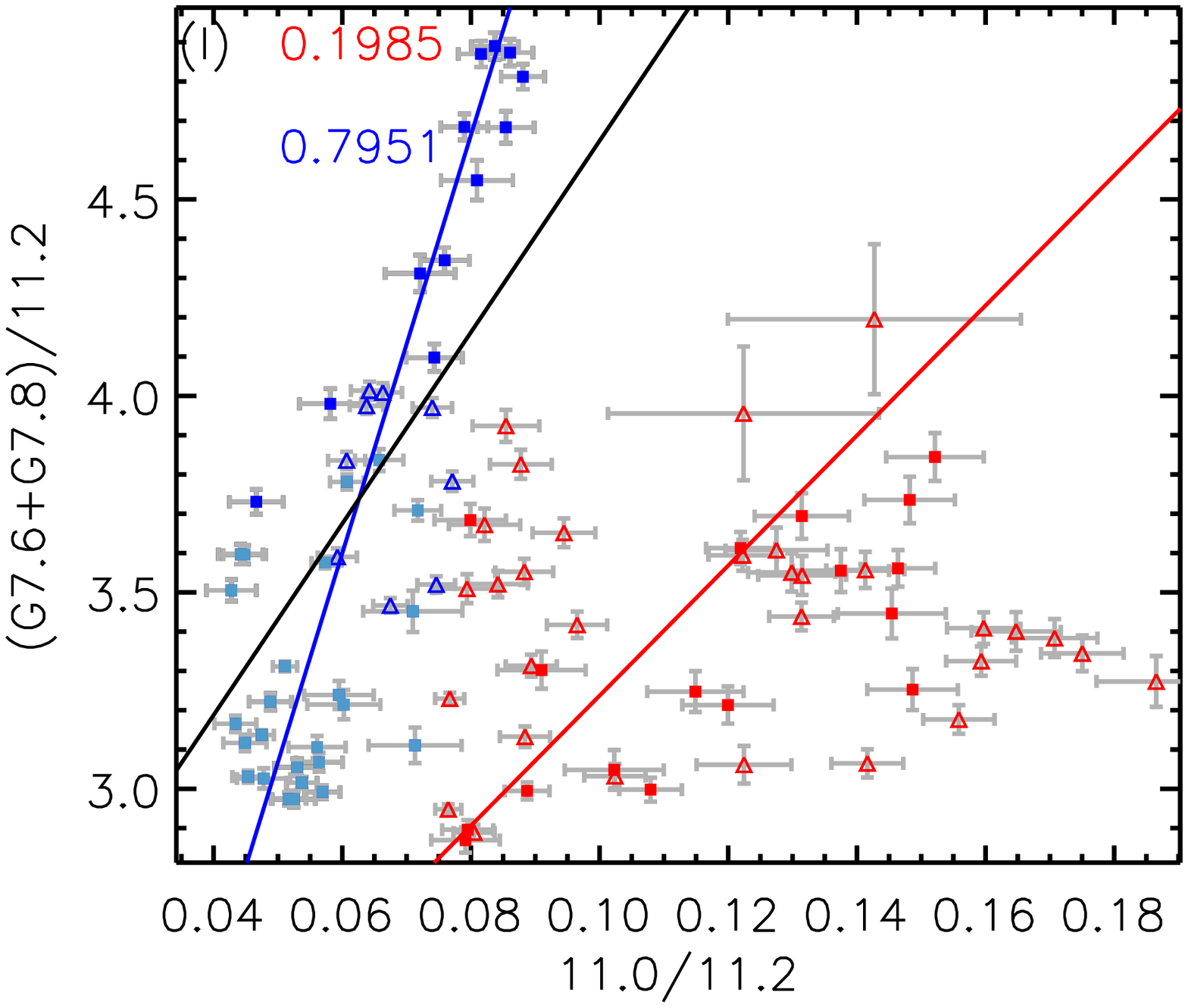}}
\end{center}
\caption{Correlation plots within the Orion Bar Combined (OBC, squares) and the Orion Bar Ionized (OBI, triangles): ratios from within the Orion Bar PDR are shown in blue, ratios from the PDR spectra behind the Orion Bar PDR front in light blue ( > 131.5$^{\prime\prime}$ from the Trapezium), and ratios from the \HII\, region (PDR) in red (see shaded areas in panel (a) of Figure~\ref{orion_cont_10_14}). Correlation coefficients for the PDR behind the Orion Bar IF (i.e both blue and light blue data points) and the \HII\, region (PDR; red data points) are given in blue and red respectively. Weighted linear fits are shown as solid lines for each respective region given in the same colors as the correlation coefficients. The black lines correspond to the respective correlation fits found for NGC~2023 in \citet{pee17}.}
\label{orion_corrD2}
\end{figure*}

To supplement Section~\ref{corr}, we present a selection of additional correlation plots within both apertures (Figures~\ref{orion_corrD1}, ~\ref{orion_corrD2}) and discuss correlations involving the Gaussian components. As in Section~\ref{corr}, we use our first grouping which is based on the relative position to the IF: 1) the \HII \, region (PDR, shown in red) and 2) the PDR spectra behind the Orion Bar IF as described in Sections~\ref{ob} and~\ref{proj} and illustrated by the shaded regions in panel (a) of Figure~\ref{orion_cont_10_14}.

The 6.2~$\mu$m band correlates much stronger in both groups with the sum of the G7.6 and G7.8~$\mu$m (Figure~\ref{orion_corrD1} (a)) than with the 7.7~$\mu$m band (Figure~\ref{orion_corr1} (a)). This is consistent with our findings for the cross cuts of these features (Section~\ref{proj}). The 7.7~$\mu$m strongly correlates with the G7.6~$\mu$m component, reflecting the dominance of the G7.6 component to the 7.7 $\mu$m complex (Figure~\ref{orion_corrD1} (c)). 
The Gaussian components show more variation between each other. The G7.6 and G8.6~$\mu$m components correlate within the PDR behind the Orion Bar IF but show significantly greater scatter within the \HII\, region (PDR; Figure~\ref{orion_corrD1} (e)), the G7.8 and G8.2~$\mu$m components are only moderately correlated in the \HII\, region (PDR; Figure~\ref{orion_corrD1} (f)), and the G7.8 and G8.6 components are correlated in both the \HII\, region (PDR) and the PDR behind the Orion Bar IF (Figure~\ref{orion_corrD1} (g)). In addition, correlations involving the G7.8 and G8.2 components exhibit a clear separation in values associated with the PDR behind the Orion Bar IF and \HII\, region (PDR; Figure~\ref{orion_corrD1} (f, g, h)).

\bsp	
\label{lastpage}
\end{document}